\documentclass{article}
\usepackage{graphicx}  % standard LaTeX graphics tool;
\usepackage{amsmath}   % For getting proper math eqns
\usepackage[compress]{cite}
\usepackage{amssymb}   % Used for getting various symbols eg. gtrsim, lesssim etc.
\usepackage{bm} % For bold math (esp of lower greek letters)
\usepackage{dcolumn}% Align table columns on decimal point
\usepackage{color}
\usepackage{mathrsfs}
\usepackage{amsfonts}
\usepackage{varioref}
\usepackage{textcomp}
\RequirePackage[colorlinks,citecolor=blue,urlcolor=magenta,linkcolor=blue]{hyperref}
\allowdisplaybreaks
\def\gr{general relativity}

\labelformat{section}{Section #1} 
\labelformat{subsection}{Section #1} 
\labelformat{subsubsection}{Section #1}
\labelformat{subsubsubsection}{Section #1}
\labelformat{equation}{Eq.~(#1)} 
\labelformat{figure}{Fig.~#1} 
\labelformat{subfigure}{Fig.~\thefigure#1} 
\labelformat{table}{Tab.~#1} 
\labelformat{appendix}{Appendix #1}
\title{Limits on stellar structures in Lovelock theories of gravity}
\author{Sumanta Chakraborty\footnote{sumantac.physics@gmail.com}$~^{1}$ and Naresh Dadhich\footnote{nkd@iucaa.in}$~^{2}$
\\
{$^{1}$\small{School of Mathematical and Computational Sciences}}
\\
{\small{and}}
\\
{\small{School of Physical Sciences}}
\\
{\small{Indian Association for the Cultivation of Science, Kolkata-700032, India}}
\\
{$^{2}$\small{IUCAA, Post Bag 4, Ganeshkhind, Pune 411007}}}
\begin{document}
  
\maketitle
%%%%%%%%%%%%%%%%%%%%%%%%%%%%%%%%%%%%%%%%%%%%%%%%%%%%%%%%%%%%%%%%%%%%%%%%%%%%%%%%%%%%%%%%%%%%%%%%%%%
%%%%%%%%%%%%%%%%%%%%%%%%%%%%%%%%%%%%%%%%%%%%%%%%%%%%%%%%%%%%%%%%%%%%%%%%%%%%%%%%%%%%%%%%%%%%%%%%%%%
%%%%%%%%%%%%%%%%%%%%%%%%%%%%%%%%%%%%%%%%%%%%%%%%%%%%%%%%%%%%%%%%%%%%%%%%%%%%%%%%%%%%%%%%%%%%%%%%%%%
\begin{abstract}
We study the bound on the compactness of a stellar object in pure Lovelock theories of arbitrary order in arbitrary spacetime dimensions, involving electromagnetic field. The bound we derive for a generic pure Lovelock theory, reproduces the known results in four dimensional Einstein gravity. Both the case of a charged shell and that of a charge sphere demonstrates that for a given spacetime dimension, stars in general relativity are more compact than the stars in pure Lovelock theories. In addition, as the strength of the Maxwell field increases, the stellar structures become more compact, i.e., the radius of the star decreases. In the context of four dimensional Einstein-Gauss-Bonnet gravity as well, an increase in the strength of the Gauss-Bonnet coupling (behaving as an effective electric charge), increases the compactness of the star. Implications are discussed.  
\end{abstract}
%%%%%%%%%%%%%%%%%%%%%%%%%%%%%%%%%%%%%%%%%%%%%%%%%%%%%%%%%%%%%%%%%%%%%%%%%%%%%%%%%%%%%%%%%%%%%%%%%%%
%%%%%%%%%%%%%%%%%%%%%%%%%%%%%%%%%%%%%%%%%%%%%%%%%%%%%%%%%%%%%%%%%%%%%%%%%%%%%%%%%%%%%%%%%%%%%%%%%%%
%%%%%%%%%%%%%%%%%%%%%%%%%%%%%%%%%%%%%%%%%%%%%%%%%%%%%%%%%%%%%%%%%%%%%%%%%%%%%%%%%%%%%%%%%%%%%%%%%%%
%\newpage
%%%%%%%%%%%%%%%%%%%%%%%%%%%%%%%%%%%%%%%%%%%%%%%%%%%%%%%%%%%%%%%%%%%%%%%%%%%%%%%%%%%%%%%%%%%%%%%%%%%
%%%%%%%%%%%%%%%%%%%%%%%%%%%%%%%%%%%%%%%%%%%%%%%%%%%%%%%%%%%%%%%%%%%%%%%%%%%%%%%%%%%%%%%%%%%%%%%%%%%
%%%%%%%%%%%%%%%%%%%%%%%%%%%%%%%%%%%%%%%%%%%%%%%%%%%%%%%%%%%%%%%%%%%%%%%%%%%%%%%%%%%%%%%%%%%%%%%%%%%
\section{Introduction and Motivation}

Existence of stable stellar objects is a fundamental requirement that any theory claiming to describe gravitational interaction around us must satisfy. The stability of such stellar objects can be arrived at by balancing the gravitational force against pressure of the fluid material forming the stellar structure. If the internal pressure of a star cannot balance the gravitational force, the star will collapse and will come to equilibrium at a new radius where these two forces are balanced. However, if the gravitational force is so strong that at no finite radius it can be balanced by the internal pressure, a black hole will form. This raises an interesting question, how small a star can be, so that its internal pressure can balance its tendency for gravitational collapse and stop it from becoming a black hole. Expressions for such limiting compactness ratio for a stellar object was derived by Buchdahl in the context of general relativity. He obtained the desired limit by considering an isotropic fluid star under very general conditions, which reads $(M/R) \leq 4/9$ \cite{Buchdahl:1959zz,Mak:2001gg,Stuchlik:2008xe,Andreasson:2007ck,Karageorgis:2007cy,Goswami:2015dma,Zarro:2009gd,Barraco:2002ds}. It turns out that even for anisotropic distribution of stellar matter, the same limit on the compactness of a stellar structure can be obtained \cite{2000JMP....41.4752H}. This suggests that there is a region between the compactness limit of a star and of a black hole, satisfying the relation $(1/2)>(M/R)>(4/9)$, within which any stellar object cannot be stable and eventually becomes a black hole. This should hold for all stellar objects irrespective of their composition, or, equivalently, the maximum compactness ratio of any stellar object should always be smaller than that of the black hole. This acts as an acid test for viability of any theory of gravity aiming to describe the world around us. At this stage it is worth mentioning that in Einstein gravity, the compactness limit can also be derived by prescribing that gravitational field energy must be less than or equal to half of the matter energy. The usefulness of this prescription stems from the fact that, it involves gravitational field energy as measured by an observer in the exterior of the stellar structure and hence makes no reference to the interior distribution, thereby making the analysis universal \cite{Dadhich:2019jyf,Dadhich:2010qh}. 

It turns out that most of the results discussed above modifies significantly for stellar structures inheriting Maxwell field. In particular, for a charged stellar object, the compactness limit depends on the assumptions about the nature of the charged object, unlike the neutral case discussed above. In particular, there can be more than one compactness limit depending upon the equation of state of the fluid and the charge distribution within the stellar structure. This is because, the Maxwell field brings in additional degrees of freedom in the theory, thus rendering the analysis to be an involved one \cite{Giuliani:2007zza,Boehmer:2007gq,Andreasson:2012dj}. Furthermore, in Einstein gravity stable stellar structure, surrounded by a charged shell, can exist provided the electric charge satisfies the following relation, $(Q^{2}/M^{2})<(9/8)$.

However all these discussions were mostly in the context of Einstein gravity in four spacetime dimensions (for similar scenario in five dimensional Einstein-Gauss-Bonnet gravity, see \cite{Wright:2015yda,Dadhich:2010qh}). Since it is natural to expect that the spacetime will inherit additional spacelike dimensions at a higher energy scale, it seems worthwhile to understand the fate of stellar structure in pure Lovelock theories. These being natural candidates to describe gravity in higher dimensions, while remaining ghost free. The interest in the pure Lovelock theories stem from the fact that in the past few years several interesting and attractive properties of these gravity theories have been discovered \cite{Dadhich:2015lra}. Some of these properties include --- (a) non-existence of non-trivial vacuum solution in all critical odd $d=2N+1$ dimensions, where $N$ is the order of the Lovelock polynomial \cite{Dadhich:2012cv,Camanho:2015hea}; (b) thermodynamic features of \gr\ continues to hold true in Lovelock gravity as well \cite{Chakraborty:2015wma,Chakraborty:2014rga}; (c) vacuum Einstein gravity in four spacetime dimensions becomes identical to pure Lovelock theory of order $N$ in $d=3N+1$ dimensions \cite{Chakraborty:2016qbw,Gannouji:2019gnb}; (d) just like the equipartition of gravitational and non-gravitational energy defines the horizon for pure Lovelock black holes \cite{Chakraborty:2015kva,Dadhich:1997ze}, gravitational energy being half of the non-gravitational energy provides the Buchdahl limit \cite{Dadhich:2019jyf,Dadhich:2016fku}. Motivated by these set of results \cite{Dadhich:2012pd,Dadhich:2015lra}, we will study the compactness limit for stable stellar structures in pure Lovelock theories in higher spacetime dimensions ($d\geq 2N+1$) with or without Maxwell field in this work (for earlier works regarding stellar structure in pure Lovelock theories, see \cite{Dadhich:2016fku,Dadhich:2016wtb,Molina:2016xeu}). 

Recently, there have also been some interest in the context of a four dimensional Einstein-Gauss-Bonnet gravity, derived from a higher dimensional setting by taking the limit $d\rightarrow 4$, such that $(d-4)\alpha$ is finite, where $\alpha$ is the Gauss-Bonnet coupling parameter \cite{Glavan:2019inb}. Even though there are several criticisms against this model, including the above limiting procedure being invalid (for a few other problems with this approach, see the subsequent discussions in this paper and \cite{Dadhich:2020ukj,Ai:2020peo,Gurses:2020ofy,Ge:2020tid}), we would like to explore whether it is possible to have stellar structure in this context as well. This will act as an interesting test for the techniques used to derive limits on the stellar structures in various other contexts, including the case of pure Lovelock theories considered here. Moreover, as pointed out in \cite{Dadhich:2020ukj}, the Gauss-Bonnet coupling parameter behaves as an effective charge for the system and hence is very much in tune with the stellar limit derived for pure Lovelock black holes with Maxwell field.  

The paper is organized as follows: We start by providing a brief introduction to the Lovelock theories of gravity and the associated gravitational field equations in static, spherically symmetric context in \ref{Intro_Lovelock}. Using these equations we derive the limit on the stellar structure involving Maxwell field for a charged shell as well as for a charged sphere within the realm of pure Lovelock gravity in \ref{charged_pure_Section}. As an illustration of the techniques involved, we have also analyzed the stellar structure and a possible bound on the same for four dimensional Einstein-Gauss-Bonnet gravity, where the redefined Gauss-Bonnet coupling parameter behaves as an effective charge, in \ref{section_4degb}. Finally we conclude with a review on the results obtained. Some relevant computations have been presented in \ref{AppA}. 

\emph{Notations and Conventions:} We will set the fundamental constants $c=1=\hbar$ and shall use the mostly positive signature convention. All the Roman indices ($a,b,c,\ldots$) denote the spacetime indices running from $0$ to $(d-1)$. 
%%%%%%%%%%%%%%%%%%%%%%%%%%%%%%%%%%%%%%%%%%%%%%%%%%%%%%%%%%%%%%%%%%%%%%%%%%%%%%%%%%%%%%%%%%%%%%%%%%%
%%%%%%%%%%%%%%%%%%%%%%%%%%%%%%%%%%%%%%%%%%%%%%%%%%%%%%%%%%%%%%%%%%%%%%%%%%%%%%%%%%%%%%%%%%%%%%%%%%%
%%%%%%%%%%%%%%%%%%%%%%%%%%%%%%%%%%%%%%%%%%%%%%%%%%%%%%%%%%%%%%%%%%%%%%%%%%%%%%%%%%%%%%%%%%%%%%%%%%%
\section{A brief introduction to Lovelock theories of gravity}\label{Intro_Lovelock}

In this section we will provide a very brief introduction to the Lovelock theories of gravity, which will become useful in the subsequent sections. The starting point is the action functional for the Lovelock class of theories, which is a polynomial in the Riemann curvature, constructed in such a manner, that the field equations derived from the same is second order. Expressed in an explicit manner, the action for Lovelock theories of gravity in $d$ spacetime dimensions takes the following form,
%%%%%%%%%%%%%%%%%%%%%%%%%%%%%%%%%%%%%%%%%%%%%%%%%%%%%%%%%
\begin{align}
\mathcal{A}=\int d^{d}x~\sqrt{-g}\sum _{i=0}^{N\leq(d/2)}\kappa_{i}\delta^{a_{1}b_{1}\cdots a_{i}b_{i}}_{c_{1}d_{1}\cdots c_{i}d_{i}}R^{c_{1}d_{1}}_{a_{1}b_{1}}\cdots R^{c_{i}d_{i}}_{a_{i}b_{i}}~.
\end{align}
%%%%%%%%%%%%%%%%%%%%%%%%%%%%%%%%%%%%%%%%%%%%%%%%%%%%%%%%%
In the above action $\delta^{a_{1}b_{1}\cdots a_{i}b_{i}}_{c_{1}d_{1}\cdots c_{i}d_{i}}$ is the completely antisymmetric determinant tensor and $R^{ab}_{cd}$ is the Riemann curvature tensor. From the completely antisymmetric nature of the determinant tensor it follows that, $d\geq 2N$ in order to have non-trivial contribution from the various terms in the action. For example, in four spacetime dimensions one can have the Einstein-Hilbert and the Gauss-Bonnet term. In the above action, $\kappa_{i}$ denotes the coupling constants, with mass dimension $[\kappa_{i}]=M^{d-2i}$ (since, we have set $c=1=\hbar$, $L=T=M^{-1}$). For example, $\kappa_{0}=-(\Lambda/8\pi G)$, $\kappa_{1}=(1/16\pi G)$ and $\kappa_{2}=\kappa_{1}\alpha$, where $\Lambda$ is the cosmological constant ($[\Lambda]=M^{2}$), $G$ is the gravitational constant ($[G]=M^{2-d}$) and $\alpha$ is the Gauss-Bonnet coupling ($[\alpha]=M^{-2}$, as $[\kappa_{2}]=M^{d-4}$).

Starting from the above action, describing the Lovelock theories of gravity, one can determine the associated field equations by variation of the action with respect to the metric. The field equations for $N$th order Lovelock gravity will involve $N$ curvature tensors. Since our aim in this work is to study the spherically symmetric stellar configuration in static equilibrium, we will present the field equations in the context of a static and spherically symmetric metric with isotropic perfect fluid source. The metric ansatz, compatible with the above symmetry, is taken to be,
%%%%%%%%%%%%%%%%%%%%%%%%%%%%%%%%%%%%%%%%%%%%%%%%%%%%%%%%%
\begin{align}\label{sph_symm_ansatz}
ds^{2}=-e^{\nu(r)}dt^{2}+e^{\lambda(r)}dr^{2}+r^{2}d\Omega_{d-2}^{2}~,
\end{align}
%%%%%%%%%%%%%%%%%%%%%%%%%%%%%%%%%%%%%%%%%%%%%%%%%%%%%%%%%
where $\nu(r)$ and $\lambda(r)$ are arbitrary functions of the radial coordinate. The temporal and radial part of the field equations arising out of the Lovelock action, presented above, take the following form, 
%%%%%%%%%%%%%%%%%%%%%%%%%%%%%%%%%%%%%%%%%%%%%%%%%%%%%%%%%
\begin{align}
\sum_{N}\frac{(d-2)!}{(d-2N-1)!}\kappa_{N}\frac{(1-e^{-\lambda})^{N-1}}{2r^{2N}}\left[-Nr\lambda'e^{-\lambda}-\left(d-2N-1\right)\left(1-e^{-\lambda}\right)\right]
&=\frac{1}{2}T^{t}_{t}~,
\label{gen_eq_love_01}
\\
\sum_{N}\frac{(d-2)!}{(d-2N-1)!}\kappa_{N}\frac{(1-e^{-\lambda})^{N-1}}{2r^{2N}}\left[Nr\nu'e^{-\lambda}-\left(d-2N-1\right)\left(1-e^{-\lambda}\right)\right]
&=\frac{1}{2}T^{r}_{r}~.
\label{gen_eq_love_02}
\end{align}
%%%%%%%%%%%%%%%%%%%%%%%%%%%%%%%%%%%%%%%%%%%%%%%%%%%%%%%%%
Here, $T^{t}_{t}$ and $T^{r}_{r}$ are the temporal and radial part of the energy momentum tensor and the summation is restricted such that $N\leq (d/2)$. As emphasized earlier, we have taken isotropic perfect fluid as the energy momentum tensor, sourcing the gravitational field. Thus with the present symmetry we can safely assume the energy momentum tensor to be, $T^{a}_{b}=\textrm{diag}(-\rho(r),p(r),p(r),p(r),\cdots)$, such that the conservation equation becomes,
%%%%%%%%%%%%%%%%%%%%%%%%%%%%%%%%%%%%%%%%%%%%%%%%%%%%%%%%%
\begin{align}\label{conservation}
\frac{dp}{dr}+\frac{1}{2}\frac{d\nu}{dr}\left(\rho+p\right)=0~.
\end{align}
%%%%%%%%%%%%%%%%%%%%%%%%%%%%%%%%%%%%%%%%%%%%%%%%%%%%%%%%%
We will use \ref{gen_eq_love_01}, \ref{gen_eq_love_02} and \ref{conservation} repeatedly in the later sections of this paper. The strategy is simple, we have three differential equations involving derivatives of the unknown functions $\lambda(r)$, $\nu(r)$ and $p(r)$. Given a equation of state $p=p(\rho)$, one should be able to read off the radial profile of all these functions, which can then be used to provide the desired limit on the equilibrium configuration of stellar structure arising out of these gravity theories. 

As an aside, we would like to answer the question, what happens to the gravitational field equations presented above in an even spacetime dimension, such that $d=2N$. In such a circumstance, it turns out that all the Lovelock orders, with $N>(d/2)$ drops out of the field equations and all the Lovelock orders with $N<(d/2)$ contributes. However, the Lovelock order with $N=(d/2)$ does not contribute, since such terms appear in the field equations with coefficient $(d-2N)$. This argument crucially requires the coupling $\kappa_{N}$ to be finite. Instead, if we assume that $\kappa_{N}$ diverges as $d\rightarrow 2N$, but $\kappa_{N}(d-2N)$ is finite, then the Lovelock term with order $N=(d/2)$ will contribute. For example, if one assumes the Gauss-Bonnet coupling $\alpha$ to be finite then in four dimensions, the gravitational field equation is solely governed by the Einstein-Hilbert term. However, if we let the Gauss-Bonnet coupling to diverge, while keeping $(d-4)\alpha$ finite, then even in four dimension one can have non-trivial contribution to the gravitational field equations. This has recently been explored in \cite{Glavan:2019inb}, where vacuum solutions arising out of the gravitational field equations have been studied. As a warm-up exercise, we will first derive the limit on the stellar structure for this recently discussed Einstein-Gauss-Bonnet gravity in four spacetime dimensions and observe if some further insight can be gained through this analysis. 

Finally, we would like to provide a note of caution. Assuming the coupling constant $\kappa_{N}$ to diverge, while keeping $(d-2N)\kappa_{N}$ finite, can have serious consequences in the quantum as well as classical domain. As long as, the existence of solutions to the field equations are considered, the above procedure is possibly unambiguous and will provide new solutions. But it may not have a well-posed boundary value problem. This is because the existence of a well-behaved action principle is crucial for a theory to have a proper boundary value problem \cite{Parattu:2015gga,Chakraborty:2016yna} and choosing the coupling constant to diverge will render the action principle ill-posed (for other issues, see \cite{Ge:2020tid,Gurses:2020ofy,Ai:2020peo}). Further, in the quantum domain, the action functional takes the central stage in the path-integral formulation, which cannot be defined for the theories with diverging coupling constant. Thus even though the Lovelock theory with diverging coupling constant $\kappa_{N}$, but finite $(d-2N)\kappa_{N}$, as $d\rightarrow 2N$, can have non-trivial classical solutions, these theories will most likely fail to depict the gravitational theory at a microscopic level. 
%%%%%%%%%%%%%%%%%%%%%%%%%%%%%%%%%%%%%%%%%%%%%%%%%%%%%%%%%%%%%%%%%%%%%%%%%%%%%%%%%%%%%%%%%%%%%%%%%%%
%%%%%%%%%%%%%%%%%%%%%%%%%%%%%%%%%%%%%%%%%%%%%%%%%%%%%%%%%%%%%%%%%%%%%%%%%%%%%%%%%%%%%%%%%%%%%%%%%%%
%%%%%%%%%%%%%%%%%%%%%%%%%%%%%%%%%%%%%%%%%%%%%%%%%%%%%%%%%%%%%%%%%%%%%%%%%%%%%%%%%%%%%%%%%%%%%%%%%%%
\section{Limit on stellar structure in pure Lovelock theory with charged matter}\label{charged_pure_Section}

In this section we will discuss the limit on stellar structure for pure Lovelock theories, however in the presence of charged matter. In particular, we will first derive the corresponding limit on the stellar structure for a charged shell, which we will subsequently generalize to describe the limit on stellar structure with charged matter. The importance of this work can be grasped from its generality. All the previous results derived in the context of four dimensional general relativity are a subclass of the situation being considered here. Further, the technique adopted in this work will significantly differ from the one used to find the limit on the stellar structure without Maxwell field, so we can check how generic is the result derived in \cite{Dadhich:2016fku} in the absence of the Maxwell field.   
%%%%%%%%%%%%%%%%%%%%%%%%%%%%%%%%%%%%%%%%%%%%%%%%%%%%%%
%%%%%%%%%%%%%%%%%%%%%%%%%%%%%%%%%%%%%%%%%%%%%%%%%%%%%%
%%%%%%%%%%%%%%%%%%%%%%%%%%%%%%%%%%%%%%%%%%%%%%%%%%%%%%
\subsection{Basic gravitational field equations}

In this section we will write down the basic gravitational field equations, based on which the analysis presented in the latter half of this work will follow. We will content ourselves with the realm of static and spherically symmetric spacetime in $d$ dimensions with the gravity theory being described by pure Lovelock theory of order $N$. Thus the line element, fit for our purpose, takes the form as presented in \ref{sph_symm_ansatz}. For static and spherically symmetric case, there will not be any vector part to the four-vector potential $A_{\mu}$ of the Maxwell field and the scalar potential, which will only depend on the radial coordinate, will provide the non-zero contribution. Since $A_{t}$ is the only non-zero component of the Maxwell field $A_{\mu}$ and it depends on $r$ alone, it follows that the $F^{tr}$ component of the Maxwell field tensor will be non-zero. To determine the same, we will consider the sourced Maxwell's equations, which in the present context, becomes
%%%%%%%%%%%%%%%%%%%%%%%%%%%%%%%%%%%%%%%%%%%%%%%%%%%%%%%%%
\begin{align}
\frac{1}{\sqrt{-g}}\partial_{i}\left(\sqrt{-g}F^{ki}\right)=4\pi J^{k}~.
\end{align}
%%%%%%%%%%%%%%%%%%%%%%%%%%%%%%%%%%%%%%%%%%%%%%%%%%%%%%%%%
In the above expression, the source term $J^{k}$ has the following behaviour, $J^{k}=\rho_{\rm el}(r)u^{k}$, where $\rho_{\rm el}(r)$ is the electric charge density and $u^{k}=(e^{-\nu/2},0,0,0)$ is the four-velocity of a static observer in the static and spherically symmetric spacetime described by \ref{sph_symm_ansatz}. For this static and spherically symmetric ansatz it also follows that, $\sqrt{-g}=r^{d-2}e^{(\nu+\lambda)/2}\Omega_{d-2}$, where $\Omega_{d-2}$ provides the angular contribution and hence the above Maxwell's equations can be expressed in the following form,
%%%%%%%%%%%%%%%%%%%%%%%%%%%%%%%%%%%%%%%%%%%%%%%%%%%%%%%%%
\begin{align}
\frac{1}{r^{d-2}e^{(\nu+\lambda)/2}}\partial_{r}\left(r^{d-2}e^{(\nu+\lambda)/2}F^{tr}\right)=4\pi \rho_{\rm el}(r)e^{-\nu/2}~.
\end{align}
%%%%%%%%%%%%%%%%%%%%%%%%%%%%%%%%%%%%%%%%%%%%%%%%%%%%%%%%%
The above equation can be expressed in a suggestive form by defining the charge function $q(r)$ as, $q(r)\equiv r^{d-2}e^{(\nu+\lambda)/2}F^{tr}$, yielding,
%%%%%%%%%%%%%%%%%%%%%%%%%%%%%%%%%%%%%%%%%%%%%%%%%%%%%%%%%
\begin{align}
q(r)=4\pi \int ^{r}dr~r^{d-2}e^{\lambda/2}\rho_{\rm el}(r)~.
\end{align}
%%%%%%%%%%%%%%%%%%%%%%%%%%%%%%%%%%%%%%%%%%%%%%%%%%%%%%%%%
The above expression relates the Maxwell field tensor with the electric charge density $\rho_{\rm el}$. 

The next task is to write down the gravitational field equations. For that we need to know the source of the gravitational field. The source of gravitational field includes both the charged fluid with charge density $\rho_{\rm el}$ and matter with energy density $\rho(r)$, radial pressure $p(r)$ and transverse pressure $p_{\perp}$. Thus, the energy momentum tensor associated with the matter field correspond to, $T^{a}_{b}=\textrm{diag}(-\rho,p,p_{\perp},\ldots,p_{\perp})$. On the other hand, the components of the energy momentum tensor associated with the Maxwell field can be expressed as,
%%%%%%%%%%%%%%%%%%%%%%%%%%%%%%%%%%%%%%%%%%%%%%%%%%%%%%%%%
\begin{align}
T^{t}_{t}&=\frac{1}{8\pi}F^{tr}F_{tr}=-\frac{1}{8\pi}e^{\nu+\lambda}(F^{tr})^{2}=-\frac{1}{8\pi}e^{\nu+\lambda}\left(\frac{q(r)}{r^{d-2}e^{(\nu+\lambda)/2}}\right)^{2}
=-\frac{q^{2}(r)}{8\pi r^{2(d-2)}}~,
\\
T^{r}_{r}&=T^{t}_{t}=-\frac{q^{2}(r)}{8\pi r^{2(d-2)}}~,
\\
T^{\perp}_{\perp}&=-\frac{1}{8\pi}F^{tr}F_{tr}=\frac{q^{2}(r)}{8\pi r^{2(d-2)}}~.
\end{align}
%%%%%%%%%%%%%%%%%%%%%%%%%%%%%%%%%%%%%%%%%%%%%%%%%%%%%%%%%
Thus combining together we obtain, the total energy momentum tensor of matter and electromagnetic field, acting as the source of the gravitational field equations, to have the following form,
%%%%%%%%%%%%%%%%%%%%%%%%%%%%%%%%%%%%%%%%%%%%%%%%%%%%%%%%%
\begin{align}
T^{a}_{b}=\textrm{diag}\left(-\rho-\frac{q^{2}(r)}{8\pi r^{2(d-2)}},p-\frac{q^{2}(r)}{8\pi r^{2(d-2)}},p_{\perp}+\frac{q^{2}(r)}{8\pi r^{2(d-2)}},\ldots,p_{\perp}+\frac{q^{2}(r)}{8\pi r^{2(d-2)}}\right)~.
\end{align}
%%%%%%%%%%%%%%%%%%%%%%%%%%%%%%%%%%%%%%%%%%%%%%%%%%%%%%%%%
Therefore, one can immediately write down the temporal and radial part of the gravitational field equations associated with the $N$th order pure Lovelock gravity by taking a cue from \ref{gen_eq_love_01} and \ref{gen_eq_love_02} respectively. Taking a single term of order $N$ from the summation in these equations and choosing the coupling constant $\kappa_{N}$, such that, $2^{N-1}8\pi \kappa_{N}\{(d-2)!/(d-2N-1)!\}=1$, the temporal part of the gravitational field equations for $N$th order pure Lovelock theory can be expressed as,
%%%%%%%%%%%%%%%%%%%%%%%%%%%%%%%%%%%%%%%%%%%%%%%%%%%%%%%%%
\begin{align}\label{rho_lovelock_charge}
\frac{(1-e^{-\lambda})^{N-1}}{2^{N-1}r^{2N}}\left[rN\lambda'e^{-\lambda}+(d-2N-1)(1-e^{-\lambda})\right]=8\pi \rho+\frac{q^{2}(r)}{r^{2(d-2)}}~.
\end{align}
%%%%%%%%%%%%%%%%%%%%%%%%%%%%%%%%%%%%%%%%%%%%%%%%%%%%%%%%%
Taking the limit $d=4$ and $N=1$, one can explicitly demonstrate that the above equation identically coincides with the respective one in general relativity in geometrized units. Similarly, for pure Gauss-Bonnet gravity one arrives at the respective field equation by choosing the Gauss-Bonnet coupling constant $\alpha$, such that $(d-2)(d-3)(d-4)(\alpha/G)=1$. Proceeding in an identical manner, from \ref{gen_eq_love_02}, the radial part of the gravitational field equations can also be expressed as,
%%%%%%%%%%%%%%%%%%%%%%%%%%%%%%%%%%%%%%%%%%%%%%%%%%%%%%%%%
\begin{align}\label{radial_lovelock_charge}
\frac{(1-e^{-\lambda})^{N-1}}{2^{N-1}r^{2N}}\left[rN\nu'e^{-\lambda}-(d-2N-1)(1-e^{-\lambda})\right]=8\pi p-\frac{q^{2}(r)}{r^{2(d-2)}}~.
\end{align}
%%%%%%%%%%%%%%%%%%%%%%%%%%%%%%%%%%%%%%%%%%%%%%%%%%%%%%%%%
In this case as well the corresponding general relativistic counter part can be derived by substituting $d=4$ and $N=1$ in the above equation. We already have the equation determining $q(r)$, given $\rho_{\rm el}(r)$. The above equations determines the unknown metric functions $e^{\nu(r)}$  and $e^{\lambda(r)}$ given the matter content. Finally, the conservation of the matter energy momentum tensor yields the following equation,  
%%%%%%%%%%%%%%%%%%%%%%%%%%%%%%%%%%%%%%%%%%%%%%%%%%%%%%%%%
\begin{align}
\dfrac{d}{dr}\left(p-\frac{q^{2}(r)}{8\pi r^{2(d-2)}}\right)&+\frac{\nu'}{2}\left[\left(\rho+\frac{q^{2}(r)}{8\pi r^{2(d-2)}}\right)+\left(p-\frac{q^{2}(r)}{8\pi r^{2(d-2)}}\right)\right]
\nonumber
\\
&\hskip 2 cm +\left(\frac{d-2}{r}\right)\left[\left(p-\frac{q^{2}(r)}{8\pi r^{2(d-2)}}\right)-\left(p_{\perp}+\frac{q^{2}(r)}{8\pi r^{2(d-2)}}\right)\right]=0~,
\end{align}
%%%%%%%%%%%%%%%%%%%%%%%%%%%%%%%%%%%%%%%%%%%%%%%%%%%%%%%%%
which can be re-expressed in the following form,
%%%%%%%%%%%%%%%%%%%%%%%%%%%%%%%%%%%%%%%%%%%%%%%%%%%%%%%%%
\begin{align}\label{conservation_lovelock_charge_01}
\dfrac{d}{dr}\left(p-\frac{q^{2}(r)}{8\pi r^{2(d-2)}}\right)+\frac{\nu'}{2}\left(\rho+p\right)+\left(\frac{d-2}{r}\right)\left[\left(p-p_{\perp}\right)-\frac{q^{2}(r)}{4\pi r^{2(d-2)}}\right]=0~.
\end{align}
%%%%%%%%%%%%%%%%%%%%%%%%%%%%%%%%%%%%%%%%%%%%%%%%%%%%%%%%%
We observe that in the absence of the Maxwell field, i.e., with $q(r)=0$, the above equation reduces to the standard conservation relation for anisotropic perfect fluid in $d$ spacetime dimensions. Thus if $\rho(r)$ and $p_{\perp}(r)$ are given, the radial pressure can be derived from the above equation.

Therefore we have three independent equations, the temporal and the radial gravitational field equations depicted by \ref{rho_lovelock_charge} and \ref{radial_lovelock_charge}, as well as the conservation relation for the matter energy momentum tensor, presented in \ref{conservation_lovelock_charge_01}. One can use these three equations to derive an equation in terms of the transverse pressure alone, which will turn out to be useful in later part of the analysis. To derive that we start by taking a radial derivative of \ref{radial_lovelock_charge}, which yields,
%%%%%%%%%%%%%%%%%%%%%%%%%%%%%%%%%%%%%%%%%%%%%%%%%%%%%%%%%
\begin{align}
\dfrac{d}{dr}\left(8\pi p-\frac{q^{2}(r)}{r^{2(d-2)}}\right)&=\frac{(1-e^{-\lambda})^{N-1}}{2^{N-1}r^{2N}}\left[N\nu'e^{-\lambda}+rN\nu''e^{-\lambda}-rN\lambda'\nu'e^{-\lambda}-(d-2N-1)e^{-\lambda}\lambda'\right]
\nonumber
\\
&\hskip -2 cm +\left[rN\nu'e^{-\lambda}-(d-2N-1)(1-e^{-\lambda})\right]\left[\frac{-2N(1-e^{-\lambda})^{N-1}}{2^{N-1}r^{2N+1}}+\frac{(N-1)\lambda'e^{-\lambda}(1-e^{-\lambda})^{N-2}}{2^{N-1}r^{2N}} \right]~.
\end{align}
%%%%%%%%%%%%%%%%%%%%%%%%%%%%%%%%%%%%%%%%%%%%%%%%%%%%%%%%%
The left hand side of the above equation can be eliminated by using the conservation relation in \ref{conservation_lovelock_charge_01}, which yields,
%%%%%%%%%%%%%%%%%%%%%%%%%%%%%%%%%%%%%%%%%%%%%%%%%%%%%%%%%
\begin{align}\label{reduced_lovelock_01}
-4\pi \nu'\left(\rho+p\right)&-\left(\frac{d-2}{r}\right)\left[\left(8\pi p-\frac{q^{2}(r)}{r^{2(d-2)}}\right)-\left(8\pi p_{\perp}+\frac{q^{2}(r)}{r^{2(d-2)}}\right)\right]
\nonumber
\\
&\hskip -2 cm =\frac{(1-e^{-\lambda})^{N-1}e^{-\lambda}}{2^{N-1}r^{2N}}\left[N\nu'+rN\nu''-rN\lambda'\nu'-(d-2N-1)\lambda'\right]
\nonumber
\\
&\hskip -1 cm +\left[rN\nu'e^{-\lambda}-(d-2N-1)(1-e^{-\lambda})\right]\left(\frac{(1-e^{-\lambda})^{N-1}}{2^{N-1}r^{2N}}\right)\left[\frac{-2N}{r}+\frac{(N-1)\lambda'e^{-\lambda}}{(1-e^{-\lambda})} \right]~.
\end{align}
%%%%%%%%%%%%%%%%%%%%%%%%%%%%%%%%%%%%%%%%%%%%%%%%%%%%%%%%%
In order to determine an equation involving the transverse pressure alone, we would like to replace the quantity $(\rho+p)$ as well as the term involving pressure, appearing in the above equation, in terms of metric variables. This can be done by addition of \ref{rho_lovelock_charge} and \ref{radial_lovelock_charge}, which leads to the following expression,
%%%%%%%%%%%%%%%%%%%%%%%%%%%%%%%%%%%%%%%%%%%%%%%%%%%%%%%%%
\begin{align}\label{total_lovelock}
8\pi \left(\rho+p\right)=\frac{(1-e^{-\lambda})^{N-1}}{2^{N-1}r^{2N}}\left[rN\nu'e^{-\lambda}+rN\lambda'e^{-\lambda}\right]
=\frac{Ne^{-\lambda}(1-e^{-\lambda})^{N-1}}{2^{N-1}r^{2N-1}}\left(\lambda'+\nu'\right)~.
\end{align}
%%%%%%%%%%%%%%%%%%%%%%%%%%%%%%%%%%%%%%%%%%%%%%%%%%%%%%%%%
Substituting of \ref{total_lovelock} for $(\rho+p)$, as well as use of \ref{radial_lovelock_charge} to replace the term involving radial pressure, we obtain from \ref{reduced_lovelock_01} the following expression,
%%%%%%%%%%%%%%%%%%%%%%%%%%%%%%%%%%%%%%%%%%%%%%%%%%%%%%%%%
\begin{align}\label{reduced_lovelock_02}
\left(\frac{d-2}{r}\right)\left(8\pi p_{\perp}+\frac{q^{2}(r)}{r^{2(d-2)}}\right)
&=\frac{(1-e^{-\lambda})^{N-1}e^{-\lambda}}{2^{N-1}r^{2N}}\Bigg[N\nu'+rN\nu''-rN\lambda'\nu'-(d-2N-1)\lambda'\Bigg]
\nonumber
\\
&\hskip -3 cm +\frac{(1-e^{-\lambda})^{N-1}e^{-\lambda}}{2^{N-1}r^{2N}}\left[rN\nu'-(d-2N-1)(e^{\lambda}-1)\right]\left[-\frac{2N}{r}+\frac{(N-1)\lambda'e^{-\lambda}}{(1-e^{-\lambda})} \right]
\nonumber
\\
&\hskip -3 cm +\frac{(1-e^{-\lambda})^{N-1}}{2^{N-1}r^{2N}}\left(\frac{d-2}{r}\right)\left[rN\nu'e^{-\lambda}-(d-2N-1)(1-e^{-\lambda})\right]+\frac{Ne^{-\lambda}\nu'(1-e^{-\lambda})^{N-1}}{2^{N}r^{2N-1}}\Big(\lambda'+\nu'\Big)~.
\end{align}
%%%%%%%%%%%%%%%%%%%%%%%%%%%%%%%%%%%%%%%%%%%%%%%%%%%%%%%%%
The above expression can be further simplified and expressed in a compact form. Since there can be several possibilities, we will express the above equation in a form which will be most suitable for our later applications, yielding,
%%%%%%%%%%%%%%%%%%%%%%%%%%%%%%%%%%%%%%%%%%%%%%%%%%%%%%%%%
\begin{align}\label{reduced_lovelock_n}
\left(8\pi p_{\perp}+\frac{q^{2}(r)}{r^{2(d-2)}}\right)&=\frac{N(1-e^{-\lambda})^{N-1}e^{-\lambda}}{2^{N-1}(d-2)r^{2N-1}}\Bigg[\Bigg\{\nu'+r\nu''-\frac{r\lambda'\nu'}{2}-\left(d-2N-1\right)\lambda'+\frac{r\nu'^{2}}{2}\Bigg\}
\nonumber
\\
&+\Bigg\{\left(d-2N-2\right)\nu'+r\nu'\lambda' \frac{(N-1)}{e^{\lambda}-1}-\frac{(d-2N-2)(d-2N-1)}{Nr}\left(e^{\lambda}-1\right)\Bigg\}\Bigg]~.
\end{align}
%%%%%%%%%%%%%%%%%%%%%%%%%%%%%%%%%%%%%%%%%%%%%%%%%%%%%%%%%
This provides the sought after expression for the transverse pressure in terms of the static and spherically symmetric metric functions. Thus we can either consider \ref{rho_lovelock_charge}, \ref{radial_lovelock_charge} and \ref{conservation_lovelock_charge_01} as the three sets of independent equations to solve for, or we may consider \ref{rho_lovelock_charge}, \ref{radial_lovelock_charge} and \ref{reduced_lovelock_n} as the independent set os equations. In the second case \ref{conservation_lovelock_charge_01} follows as a consequence of these three equations and hence not independent.  

Before moving forward to the next section, let us briefly discuss a subclass of the above system, which corresponds to isotropic stellar structure. In this case $p_{\perp}=p$ and hence we can even eliminate pressure from the above equation and obtain an expression which has no reference to the matter energy momentum tensor, but depends solely on the Maxwell field. As this expression will be key to our calculation of the limit on stellar structure, we will present it below. This can be obtained by substituting $p=p_{\perp}$ in \ref{reduced_lovelock_01}, and using \ref{total_lovelock} to replace the $(\rho+p)$ term. This yields,
%%%%%%%%%%%%%%%%%%%%%%%%%%%%%%%%%%%%%%%%%%%%%%%%%%%%%%%%%
\begin{align}\label{reduced_lovelock_charge}
\frac{2(d-2)}{r}\left(\frac{q^{2}(r)}{r^{2(d-2)}}\right)&=\frac{N(1-e^{-\lambda})^{N-1}e^{-\lambda}}{2^{N-1}r^{2N}}\Bigg[\Bigg\{\nu'+r\nu''-\frac{r\lambda'\nu'}{2}-\left(d-2N-1\right)\lambda'+\frac{r\nu'^{2}}{2}\Bigg\}
\nonumber
\\
&+\Bigg\{-2N\nu'+r\nu'\lambda' \frac{(N-1)}{e^{\lambda}-1}+\frac{2(d-2N-1)}{r}\left(e^{\lambda}-1\right)\Bigg\}\Bigg]~.
\end{align}
%%%%%%%%%%%%%%%%%%%%%%%%%%%%%%%%%%%%%%%%%%%%%%%%%%%%%%%%%
This is the equation we were after, which depends solely on the Maxwell field has no dependence on the energy-momentum tensor of the stellar material. We would like to emphasize that such a relation can be derived only for isotropic fluid. The last equation will be used in the next section, after some more information about the metric elements have been obtained.

%%%%%%%%%%%%%%%%%%%%%%%%%%%%%%%%%%%%%%%%%%%%%%%%%%%%%%
%%%%%%%%%%%%%%%%%%%%%%%%%%%%%%%%%%%%%%%%%%%%%%%%%%%%%%
%%%%%%%%%%%%%%%%%%%%%%%%%%%%%%%%%%%%%%%%%%%%%%%%%%%%%%
\subsection{Metric elements exterior and interior of the stellar structure}

We have already presented all the key equations arising out of gravitational and electromagnetic field equations, in the present context of a charged fluid sphere in pure Lovelock gravity. In this section we will try to solve these equations and obtain the behaviour of the metric elements both inside and outside the charged sphere. Among all the metric coefficients, determining $e^{-\lambda}$ is the simplest and hence we will start by computation of this quantity. For this purpose, we may refer to the following identity, 
%%%%%%%%%%%%%%%%%%%%%%%%%%%%%%%%%%%%%%%%%%%%%%%%%%%%%%%%%
\begin{align}
\dfrac{d}{dr}\Bigg[r^{d-2N-1}(1-e^{-\lambda})^{N}\Bigg]=r^{d-2N-2}\Bigg[rN\lambda'e^{-\lambda}+(d-2N-1)(1-e^{-\lambda})\Bigg](1-e^{-\lambda})^{N-1}~.
\end{align}
%%%%%%%%%%%%%%%%%%%%%%%%%%%%%%%%%%%%%%%%%%%%%%%%%%%%%%%%%
Using this identity, the field equation for $e^{-\lambda}$, presented in \ref{rho_lovelock_charge}, can be expressed as,
%%%%%%%%%%%%%%%%%%%%%%%%%%%%%%%%%%%%%%%%%%%%%%%%%%%%%%%%%
\begin{align}
\dfrac{d}{dr}\Bigg[r^{d-2N-1}(1-e^{-\lambda})^{N}\Bigg]=\left(8\pi \rho+\frac{q^{2}(r)}{r^{2(d-2)}}\right)2^{N-1}r^{d-2}
\end{align}
%%%%%%%%%%%%%%%%%%%%%%%%%%%%%%%%%%%%%%%%%%%%%%%%%%%%%%%%%
It is straightforward to integrate this equation over the radial coordinate upto some radial distance $r$, starting from the origin. If the upper limit $r$ is within the stellar structure, the solution will be referred to as the interior solution. While for $r>R$, where $R$ is the radius of the star, the corresponding solution will be referred to as the exterior solution. Let us first write down the interior solution for the metric element $e^{-\lambda}$, which reads,
%%%%%%%%%%%%%%%%%%%%%%%%%%%%%%%%%%%%%%%%%%%%%%%%%%%%%%%%%
\begin{align}\label{metric_lovelock_01}
e^{-\lambda_{\rm int}}=1-\left(\frac{2^{N}m_{\rm inertial}(r)}{r^{d-2N-1}}+\frac{F(r)}{r^{d-2N-1}}\right)^{1/N}~.
\end{align}
%%%%%%%%%%%%%%%%%%%%%%%%%%%%%%%%%%%%%%%%%%%%%%%%%%%%%%%%%
Here we have defined the mass function $m_{\rm inertial}(r)$ (also referred to as the inertial mass) and the charge function $F(r)$ as,
%%%%%%%%%%%%%%%%%%%%%%%%%%%%%%%%%%%%%%%%%%%%%%%%%%%%%%%%%
\begin{align}
m_{\rm inertial}(r)\equiv 4\pi \int dr~\rho(r)r^{d-2}~;\qquad F(r)\equiv 2^{N-1}\int dr~\frac{q^{2}(r)}{r^{d-2}}~.
\end{align}
%%%%%%%%%%%%%%%%%%%%%%%%%%%%%%%%%%%%%%%%%%%%%%%%%%%%%%%%%
Note that the inertial mass accounts for the effect from stellar matter alone and has no contribution from the Maxwell field. However since all matter fields contribute to gravity, it is expected that the Maxwell field will also contribute in the `gravitational mass', which in this context is defined as,
%%%%%%%%%%%%%%%%%%%%%%%%%%%%%%%%%%%%%%%%%%%%%%%%%%%%%%%%%
\begin{align}
m_{\rm grav}(r)\equiv m_{\rm inertial}(r)+2^{-N}F(r)+2^{-N}\frac{q^{2}(r)}{r^{d-3}}~.
\end{align}
%%%%%%%%%%%%%%%%%%%%%%%%%%%%%%%%%%%%%%%%%%%%%%%%%%%%%%%%%
Using the above definition of the gravitational mass, one can indeed eliminate the inertial mass term from \ref{metric_lovelock_01} in favour of the gravitational mass and hence the radial metric component becomes,
%%%%%%%%%%%%%%%%%%%%%%%%%%%%%%%%%%%%%%%%%%%%%%%%%%%%%%%%%
\begin{align}\label{metric_lovelock_02}
e^{-\lambda_{\rm int}}=1-\left(\frac{2^{N}m_{\rm grav}(r)}{r^{d-2N-1}}-\frac{q^{2}(r)}{r^{2d-2N-4}}\right)^{1/N}~.
\end{align}
%%%%%%%%%%%%%%%%%%%%%%%%%%%%%%%%%%%%%%%%%%%%%%%%%%%%%%%%%
The metric exterior to the stellar material, which is extended upto radius $R$, can be determined by evaluating the integrals in $m_{\rm grav}$ and $q(r)$ upto radius $R$. It is natural to define $m_{\rm grav}(R)\equiv M$, the gravitational mass attributed to the stellar object by an observer at infinity and $q(R)\equiv Q$ the total electric charge of the stellar object. Thus the metric exterior to the stellar object takes the following form,
%%%%%%%%%%%%%%%%%%%%%%%%%%%%%%%%%%%%%%%%%%%%%%%%%%%%%%%%%
\begin{align}\label{charge_ext}
e^{-\lambda_{\rm ext}}=1-\left(\frac{2^{N}M}{r^{d-2N-1}}-\frac{Q^{2}}{r^{2d-2N-4}}\right)^{1/N}~.
\end{align}
%%%%%%%%%%%%%%%%%%%%%%%%%%%%%%%%%%%%%%%%%%%%%%%%%%%%%%%%%
It is possible to arrive at the above expression for $e^{-\lambda}$ by explicitly integrating the field equations in the exterior as well. Thus we have determined the metric coefficient $e^{-\lambda}$, both inside and outside the star, which will be extremely useful in the following discussion. 

The remaining bit corresponds to determination of $e^{\nu}$, which is not at all straightforward to compute in the interior of the stellar material. However in the exterior vacuum spacetime both $\rho$ and $p$ identically vanishes, thus use of \ref{total_lovelock} uniquely determines, $\nu_{\rm ext}=-\lambda_{\rm ext}$, such that, $e^{\nu_{\rm ext}}=e^{-\lambda_{\rm ext}}$, where $e^{-\lambda_{\rm ext}}$ has the expression given by \ref{charge_ext}. The determination of $e^{\nu}$ in the interior of the stellar material requires the following steps. To start with one assumes that the stellar material is isotropic, therefore \ref{reduced_lovelock_charge} becomes directly applicable in this context. Subsequently, multiplying both sides of \ref{reduced_lovelock_charge} by $\{2^{N}r^{2N}e^{\lambda}/N(1-e^{-\lambda})^{N-1}\}$ and then rearranging the terms we obtain,
%%%%%%%%%%%%%%%%%%%%%%%%%%%%%%%%%%%%%%%%%%%%%%%%%%%%%%%%%
\begin{align}
-2\nu'+2r\nu''-r\lambda'\nu'+r\nu'^{2}&-4(N-1)\nu'+2r\nu'\lambda' \frac{(N-1)e^{-\lambda}}{1-e^{-\lambda}}
\nonumber
\\
&\hskip -2 cm =\frac{2^{N-1}r^{2N}e^{\lambda}}{N(1-e^{-\lambda})^{N-1}}\left(\frac{4(d-2)}{r}\times \frac{q^{2}(r)}{r^{2(d-2)}}\right)+\frac{2\left(d-2N-1\right)}{r}\left\{r\lambda'-2(e^{\lambda}-1)\right\}~.
\end{align}
%%%%%%%%%%%%%%%%%%%%%%%%%%%%%%%%%%%%%%%%%%%%%%%%%%%%%%%%%
It is possible to re-express this equation in a very compact and useful form by using the identities depicted in \ref{AppA} (see in particular, \ref{app_identity_01} and \ref{app_identity_02} in \ref{AppA}). This yields,
%%%%%%%%%%%%%%%%%%%%%%%%%%%%%%%%%%%%%%%%%%%%%%%%%%%%%%%%%
\begin{align}\label{main_lovelock_01}
e^{-(\lambda+\nu)/2}\dfrac{d}{dr}\left[\frac{1}{r}e^{-\lambda/2}\dfrac{d}{dr}e^{\nu/2}\right]&=\dfrac{d}{dr}\left[\frac{1-e^{-\lambda}}{2r^{2}}\right]\left\{-\frac{r\nu'(N-1)e^{-\lambda}}{1-e^{-\lambda}}+\left(d-2N-1\right)\right\}
\nonumber
\\
&+\frac{2^{N-1}(d-2)}{N(1-e^{-\lambda})^{N-1}}\left(\frac{q^{2}(r)}{r^{2d-2N-1}}\right)
\end{align}
%%%%%%%%%%%%%%%%%%%%%%%%%%%%%%%%%%%%%%%%%%%%%%%%%%%%%%%%%
This is the final equation we were looking for. Note that the right hand side of the above equation depends solely on the charge term and the metric coefficient $e^{-\lambda}$ and hence using either \ref{metric_lovelock_01} or \ref{metric_lovelock_02}, it is possible to express the right hand side of this equation in terms of the matter density and charge density. Then subsequent integration should yield $e^{\nu}$. In what follows we will not attempt to integrate and solve this equation, rather we will try to derive certain restrictions on the radius of the star arising out of physically motivative requirements imposed on this equation. In this context, it will be advantageous to express the above equation, either in terms of $m_{\rm inertial}$, or in terms of $m_{\rm grav}$. Use of \ref{metric_lovelock_02} helps one to express \ref{main_lovelock_01} in terms of the gravitational mass of the stellar material, which yields,
%%%%%%%%%%%%%%%%%%%%%%%%%%%%%%%%%%%%%%%%%%%%%%%%%%%%%%%%%
\begin{align}\label{main_lovelock_02}
e^{-(\lambda+\nu)/2}&\dfrac{d}{dr}\left[\frac{1}{r}e^{-\lambda/2}\dfrac{d}{dr}e^{\nu/2}\right]
\nonumber
\\
&\hskip -1 cm =\frac{2^{N-1}}{(1-e^{-\lambda})^{N-1}}\left[\dfrac{d}{dr}\left(\frac{m_{\rm grav}(r)}{r^{d-1}}\right)-\dfrac{d}{dr}\left(\frac{2^{-N}q^{2}(r)}{r^{2d-4}}\right)\right]\left\{\frac{\left(d-2N-1\right)}{N}-\frac{r\nu'(N-1)e^{-\lambda}}{N(1-e^{-\lambda})}\right\}
\nonumber
\\
&+\frac{2^{N-1}(d-2)}{N(1-e^{-\lambda})^{N-1}}\left(\frac{q^{2}(r)}{r^{2d-2N-1}}\right)~.
\end{align}
%%%%%%%%%%%%%%%%%%%%%%%%%%%%%%%%%%%%%%%%%%%%%%%%%%%%%%%%%
On the other hand, use of \ref{metric_lovelock_01} helps one to express \ref{main_lovelock_01} in terms of the inertial mass. The corresponding differential equation for $e^{\nu}$ takes the following form,
%%%%%%%%%%%%%%%%%%%%%%%%%%%%%%%%%%%%%%%%%%%%%%%%%%%%%%%%%
\begin{align}\label{main_lovelock_03}
e^{-(\lambda+\nu)/2}&\dfrac{d}{dr}\left[\frac{1}{r}e^{-\lambda/2}\dfrac{d}{dr}e^{\nu/2}\right]
\nonumber
\\
&\hskip -1 cm =\frac{2^{N-1}}{(1-e^{-\lambda})^{N-1}}\left[\dfrac{d}{dr}\left(\frac{m_{\rm inertial}(r)}{r^{d-1}}\right)+\frac{q^{2}}{2r^{2d-3}}-(d-1)\dfrac{F(r)}{2^{N}r^{d}}\right]\left\{\frac{\left(d-2N-1\right)}{N}-\frac{r\nu'(N-1)e^{-\lambda}}{N(1-e^{-\lambda})}\right\}
\nonumber
\\
&+\frac{2^{N-1}(d-2)}{N(1-e^{-\lambda})^{N-1}}\left(\frac{q^{2}(r)}{r^{2d-2N-1}}\right)~.
\end{align}
%%%%%%%%%%%%%%%%%%%%%%%%%%%%%%%%%%%%%%%%%%%%%%%%%%%%%%%%%
It is worth empasizing that the limit to four dimensional Einstein gravity corresponds to $N\rightarrow 1$ and $d\rightarrow4$, under which the above equation reduces to,
%%%%%%%%%%%%%%%%%%%%%%%%%%%%%%%%%%%%%%%%%%%%%%%%%%%%%%%%%
\begin{align}\label{main_lovelock_gr}
e^{-(\lambda+\nu)/2}\dfrac{d}{dr}\left[\frac{1}{r}e^{-\lambda/2}\dfrac{d}{dr}e^{\nu/2}\right]=\left[\dfrac{d}{dr}\left(\frac{m_{\rm inertial}(r)}{r^{3}}\right)+\frac{5q^{2}}{2r^{5}}-\dfrac{3F(r)}{2r^{4}}\right]~.
\end{align}
%%%%%%%%%%%%%%%%%%%%%%%%%%%%%%%%%%%%%%%%%%%%%%%%%%%%%%%%%
As one can immediately verify, the above result identically matches with the one presented in \cite{Giuliani:2007zza} for a charged stellar object in four dimensional Einstein gravity. This concludes our discussion regarding the behaviour of metric elements inside and outside a stellar structure. We will now proceed to discuss possible limits on the stellar structure arising out of this analysis. 
%%%%%%%%%%%%%%%%%%%%%%%%%%%%%%%%%%%%%%%%%%%%%%%%%%%%%%
%%%%%%%%%%%%%%%%%%%%%%%%%%%%%%%%%%%%%%%%%%%%%%%%%%%%%%
%%%%%%%%%%%%%%%%%%%%%%%%%%%%%%%%%%%%%%%%%%%%%%%%%%%%%%
\subsection{The case of a charged shell}

In this section we will discuss the case of a charged shell of ignorable thickness, with total charge $Q$, surrounding a isotropic and spherical massive object of mass $M_{\rm int}$ and radius $R$. As we will see, given the mass of the spherical object and its charge the radius cannot take any arbitrary value, rather it has to satisfy certain constraint. This will provide the sought for limit on the stellar structure obeying the laws of $N$th order pure Lovelock gravity in $d$ spacetime dimensions. Since we are assuming the shell to be very thin, it does not contribute to the `inertial' mass of the system. In order to contribute to the electric field, we assume that the energy momentum tensor of the shell has the following non-zero components, $T^{\theta_{1}}_{\theta_{1}}=S\delta(r-R)=\cdots=T^{\phi}_{\phi}$. Here $\delta(r-R)$ is a $(d-1)$ dimensional delta function, peaked at the radius of the star and defined through the following integral,
%%%%%%%%%%%%%%%%%%%%%%%%%%%%%%%%%%%%%%%%%%%%%%%%%%%%%%%%%
\begin{align}
4\pi \int dr~r^{d-2}\delta (r-R)=1~.
\end{align}
%%%%%%%%%%%%%%%%%%%%%%%%%%%%%%%%%%%%%%%%%%%%%%%%%%%%%%%%%
Following which, the solution for the metric element $e^{-\lambda}$, interior to the charged shell can be determined by setting $q(r)=0$ in \ref{metric_lovelock_01} and hence it takes the following form,
%%%%%%%%%%%%%%%%%%%%%%%%%%%%%%%%%%%%%%%%%%%%%%%%%%%%%%%%%
\begin{align}
e^{-\lambda_{\rm int}}&=1-\left(\frac{2^{N}m_{\rm inertial}(r)}{r^{d-2N-1}}\right)^{1/N}~,
\end{align}
%%%%%%%%%%%%%%%%%%%%%%%%%%%%%%%%%%%%%%%%%%%%%%%%%%%%%%%%%
where $m_{\rm inertial}(R)=M_{\rm int}$. Similarly the exterior solution will be identical to the one presented in \ref{metric_lovelock_02}. From the continuity of the metric elements across the surface of the stellar object located at radius $R$, it follows that
%%%%%%%%%%%%%%%%%%%%%%%%%%%%%%%%%%%%%%%%%%%%%%%%%%%%%%%%%
\begin{align}\label{shell_mass_relation}
M=M_{\rm int}(R)+\frac{Q^{2}}{2^{N}R^{d-3}}~,
\end{align}
%%%%%%%%%%%%%%%%%%%%%%%%%%%%%%%%%%%%%%%%%%%%%%%%%%%%%%%%%
where $M$ is the gravitational mass as observed by a distant observer outside the stellar structure. Since the modifications due to the charged shell are appearing in the angular part, it follows that the temporal and the radial part of the Lovelock gravitational field equations, i.e., \ref{rho_lovelock_charge} and \ref{radial_lovelock_charge} will remain identical. However the conservation equation will be modified, yielding
%%%%%%%%%%%%%%%%%%%%%%%%%%%%%%%%%%%%%%%%%%%%%%%%%%%%%%%%%
\begin{align}\label{conservation_lovelock_charge_02}
\dfrac{d}{dr}\left(p(r)-\frac{q^{2}(r)}{8\pi r^{2(d-2)}} \right)+\frac{\nu'}{2}\left(\rho+p\right)-\frac{(d-2)S}{r}\delta (r-R)=0~.
\end{align}
%%%%%%%%%%%%%%%%%%%%%%%%%%%%%%%%%%%%%%%%%%%%%%%%%%%%%%%%%
The rest of the analysis can be performed in a straightforward manner, following the key steps depicted in the previous sections. One first uses \ref{radial_lovelock_charge} and \ref{total_lovelock} to arrive at an expression involving the charge density $S$. The resulting equation can be simplified further using the results from \ref{AppA}, see e.g., \ref{app_identity_01} and \ref{app_identity_02} in \ref{AppA}. This yields the following differential equation for the metric function $e^{\nu}$,
%%%%%%%%%%%%%%%%%%%%%%%%%%%%%%%%%%%%%%%%%%%%%%%%%%%%%%%%%
\begin{align}\label{main_lovelockn}
e^{-(\lambda+\nu)/2}&\dfrac{d}{dr}\left[\frac{1}{r}e^{-\lambda/2}\dfrac{d}{dr}e^{\nu/2}\right]
\nonumber
\\
&=\dfrac{d}{dr}\left[\frac{1-e^{-\lambda}}{2r^{2}}\right]\left\{\left(d-2N-1\right)-\frac{r\nu'(N-1)e^{-\lambda}}{1-e^{-\lambda}}\right\}+\frac{2^{N-1}(d-2)4\pi}{N(1-e^{-\lambda})^{N-1}}r^{2N-3}S\delta(r-R)~.
\end{align}
%%%%%%%%%%%%%%%%%%%%%%%%%%%%%%%%%%%%%%%%%%%%%%%%%%%%%%%%%
The above equation can be immediately integrated over the extent of the charged shell to obtain the difference between $e^{\nu/2}$ at the two ends of the charged shell in terms of the quantity $S$. In deriving this result, one needs to use the fact that $e^{-\lambda}$ is continuous across the shell. Since outside $e^{\nu_{\rm ext}}=e^{-\lambda_{\rm ext}}$, one can determine the difference $\Delta e^{\nu/2}$, explicitly in terms of $Q$ and $p_{-}$, using \ref{radial_lovelock_charge} and \ref{metric_lovelock_02} respectively. This will yield $S$ as a function of the internal pressure $p_{-}$ and the total charge of the shell $Q$. Since this relation will not be of much use in the later part, we will will present it here. 

Rather, we will integrate \ref{main_lovelockn}, from a radius $r$ within the spherical matter distribution to the lower surface of the spherical shell. Since within the lower radius of the shell, there are no charges present, the terms proportional to $S$ in \ref{main_lovelockn} will not contribute. Interestingly, the combination $r^{-2}(1-e^{-\lambda})$ is the average density of the matter distribution within the charged shell and for physically realistic material it must decreases as the radial coordinate increases. Thus the left hand side of \ref{main_lovelockn} will turn negative and hence the above integration will yield,
%%%%%%%%%%%%%%%%%%%%%%%%%%%%%%%%%%%%%%%%%%%%%%%%%%%%%%%%
\begin{align}
\frac{1}{r}e^{-\lambda_{\rm int}/2}\dfrac{d}{dr}e^{\nu_{\rm int}/2}&\geq \frac{1}{R}e^{-\lambda_{\rm int}/2}\dfrac{d}{dr}e^{\nu_{\rm int}/2}\Big\vert_{R_{-}}
\geq \frac{(d-2N-1)}{NR^{2}}\left(\frac{M_{\rm int}}{R^{d-2N-1}} \right)^{1/N}~,
\end{align}
%%%%%%%%%%%%%%%%%%%%%%%%%%%%%%%%%%%%%%%%%%%%%%%%%%%%%%%%%
where, continuity of $e^{-\lambda}$ across the shell and expression for $\nu'_{\rm int}$ from \ref{radial_lovelock_charge} has been used. Note that the above expression also requires positivity of pressure throughout the star, which is physically well-motivated. This result can be further extended by integrating the above differential equation once again and then following the steps presented in \cite{Dadhich:2016fku}. This yields the following bound on the radius of the stellar object,
%%%%%%%%%%%%%%%%%%%%%%%%%%%%%%%%%%%%%%%%%%%%%%%%%%%%%%%%
\begin{align}
M_{\rm int}^{1/N}\leq \frac{2N(d-N-1)}{(d-1)^{2}}R^{(d-2N-1)/N}~.
\end{align}
%%%%%%%%%%%%%%%%%%%%%%%%%%%%%%%%%%%%%%%%%%%%%%%%%%%%%%%%%
Using \ref{shell_mass_relation}, one can express the above inequality in terms of the compactness ratio $(\mathcal{M}/R)$, where $M=\mathcal{M}^{d-2N-1}$ and the dimensionless charge ratio $(Q/M)\mathcal{M}^{1-N}$, yielding,
%%%%%%%%%%%%%%%%%%%%%%%%%%%%%%%%%%%%%%%%%%%%%%%%%%%%%%%%
\begin{align}\label{Buchdahl_Charged_Shell}
\left[\left(\frac{\mathcal{M}}{R}\right)^{d-2N-1}-\frac{1}{2^{N}}\left(\frac{\mathcal{M}}{R}\right)^{2d-2N-4}\left(\frac{Q}{M\mathcal{M}^{N-1}}\right)^{2}\right]^{1/N}\leq \frac{2N(d-N-1)}{(d-1)^{2}}~.
\end{align}
%%%%%%%%%%%%%%%%%%%%%%%%%%%%%%%%%%%%%%%%%%%%%%%%%%%%%%%%%
The usefulness of the above formula follows from the fact that it is composed of dimensionless quantities and hence can be immediately used for comparison. In particular, for vanishing electric charge, the above inequality coincides with the one presented in \cite{Dadhich:2016fku} for pure Lovelock theories and for \gr\ it will yield the Buchdahl bound on the stellar structure. In addition, the above inequality explicitly demonstrates that the gravitational potential $\Phi$ defined as, $e^{-\lambda}\equiv 1-2\Phi$, satisfies the inequality, $\Phi \leq \{2N(d-N-1)/(d-1)^{2}\}$. This translates to the Buchdahl limit in the absence of Maxwell field in Einstein gravity, as the potential in Einstein gravity is simply $(M/R)$ in absence of the Maxwell field in four spacetime dimensions. Unfortunately, for general choices of the spacetime dimension $d$ and Lovelock order $N$, the above inequality cannot be converted into a possible constraint on the radius $R$ of the stellar object. However, for certain specific situations one can indeed arrive at the desired limit on the radius $R$ of the charged shell, some of which we will discuss below.   

%%%%%%%%%%%%%%%%%%%%%%%%%%%%%%%%%
%%%%%%%%%%%%%%%%%%%%%%%%%%%%%%%%%
%%%%%%%%%%%%%%%%%%%%%%%%%%%%%%%%%
\begin{itemize}

\item \textbf{Einstein gravity:} As a first example, consider the case of Einstein gravity (i.e., $N=1$) in $d$ spacetime dimensions. Substituting $N=1$ in \ref{Buchdahl_Charged_Shell} and manipulating the resulting expression appropriately we obtain the following constraint on the radius $R$ of the object, 
%%%%%%%%%%%%%%%%%%%%%%%%%%%%%%%%%%%%%%%%%%%%%%%%%%%%%%%%
\begin{align}
\frac{4(d-2)}{(d-1)^{2}}R^{2d-6}-2MR^{d-3}+Q^{2}\geq 0~.
\end{align}
%%%%%%%%%%%%%%%%%%%%%%%%%%%%%%%%%%%%%%%%%%%%%%%%%%%%%%%%%
The left hand side of the above inequality is an quadratic expression in $R^{d-3}$. Hence the above inequality will be satisfied if $R$ is greater than the largest root arising out of this quadratic expression, when the equality holds. This yields the following bound on the radius of the shell in terms of dimensionless quantities,
%%%%%%%%%%%%%%%%%%%%%%%%%%%%%%%%%%%%%%%%%%%%%%%%%%%%%%%%
\begin{align}\label{constraint_shell}
\frac{\mathcal{M}}{R}\leq \Bigg\{\frac{4(d-2)}{(d-1)^{2}}\left[1+\sqrt{1-\frac{4(d-2)}{(d-1)^{2}}\left(\frac{Q}{M}\right)^{2}}\right]^{-1}\Bigg\}^{1/(d-3)}~,
\end{align}
%%%%%%%%%%%%%%%%%%%%%%%%%%%%%%%%%%%%%%%%%%%%%%%%%%%%%%%%%
where the mass of the stellar mass $M$ is related to the quantity $\mathcal{M}$ as, $M=\mathcal{M}^{d-3}$. A quick check of this bound corresponds to the case of four dimensional Einstein gravity, for which $M=\mathcal{M}$. This can be obtained by setting $d=4$ in the above inequality, which yields,
%%%%%%%%%%%%%%%%%%%%%%%%%%%%%%%%%%%%%%%%%%%%%%%%%%%%%%%%
\begin{align}
\frac{M}{R}\leq \frac{8}{9}\left[1+\sqrt{1-\frac{8}{9}\left(\frac{Q}{M}\right)^{2}}\right]^{-1}~.
\end{align}
%%%%%%%%%%%%%%%%%%%%%%%%%%%%%%%%%%%%%%%%%%%%%%%%%%%%%%%%%
As one can explicitly check, this matches with the result derived in \cite{Giuliani:2007zza} and for $Q=0$, it reduces to the Buchdahl limit for general relativity. It is instructive to plot the parameter space allowed by the above inequality in the space of compactness ratio and charge ratio, which have been presented in \ref{Fig_Einstein_Shell}. As both the plots, for four and seven spacetime dimensions, in \ref{Fig_Einstein_Shell} demonstrate, the compactness ratio increases with increases of the charge ratio. This is evident, since presence of electric charge will act against gravity and hence the radius of the stable stellar structure will decrease to compensate for the same, thus increasing the compactness ratio. In addition, increase of spacetime dimension increases the allowed parameter space, which a quick comparison between the two plots in \ref{Fig_Einstein_Shell} demonstrates. This is because, increase of spacetime dimension, makes gravity stronger and hence the radius of the stellar structure decreases. It should be emphasized that, the compactness ratio depends on the spacetime dimension $d$ and the order of the Lovelock polynomial $N$. In general this ratio reads, $(M^{1/(d-2N-1)}/R)$, which becomes $(M/R)$ for four dimensional Einstein gravity and $(M^{1/4}/R)$ for seven dimensional Einstein gravity. On the other hand for pure Gauss-Bonnet gravity, the compactness ratio becomes, $(M/R)$ in six dimension and $(M^{1/2}/R)$ in seven spacetime dimension. This explains the change of compactness ratio in the labels of \ref{Fig_Einstein_Shell} and \ref{Fig_GB_Shell}, respectively.

%%%%%%%%%%%%%%%%%%%%%%%
%%%%%%%%%%%%%%%%%%%%%%%
%%%%%%%%%%%%%%%%%%%%%%%
%%%%%%%%%%%%%%%%%%%%%%%
\begin{figure}
\includegraphics[scale=0.3]{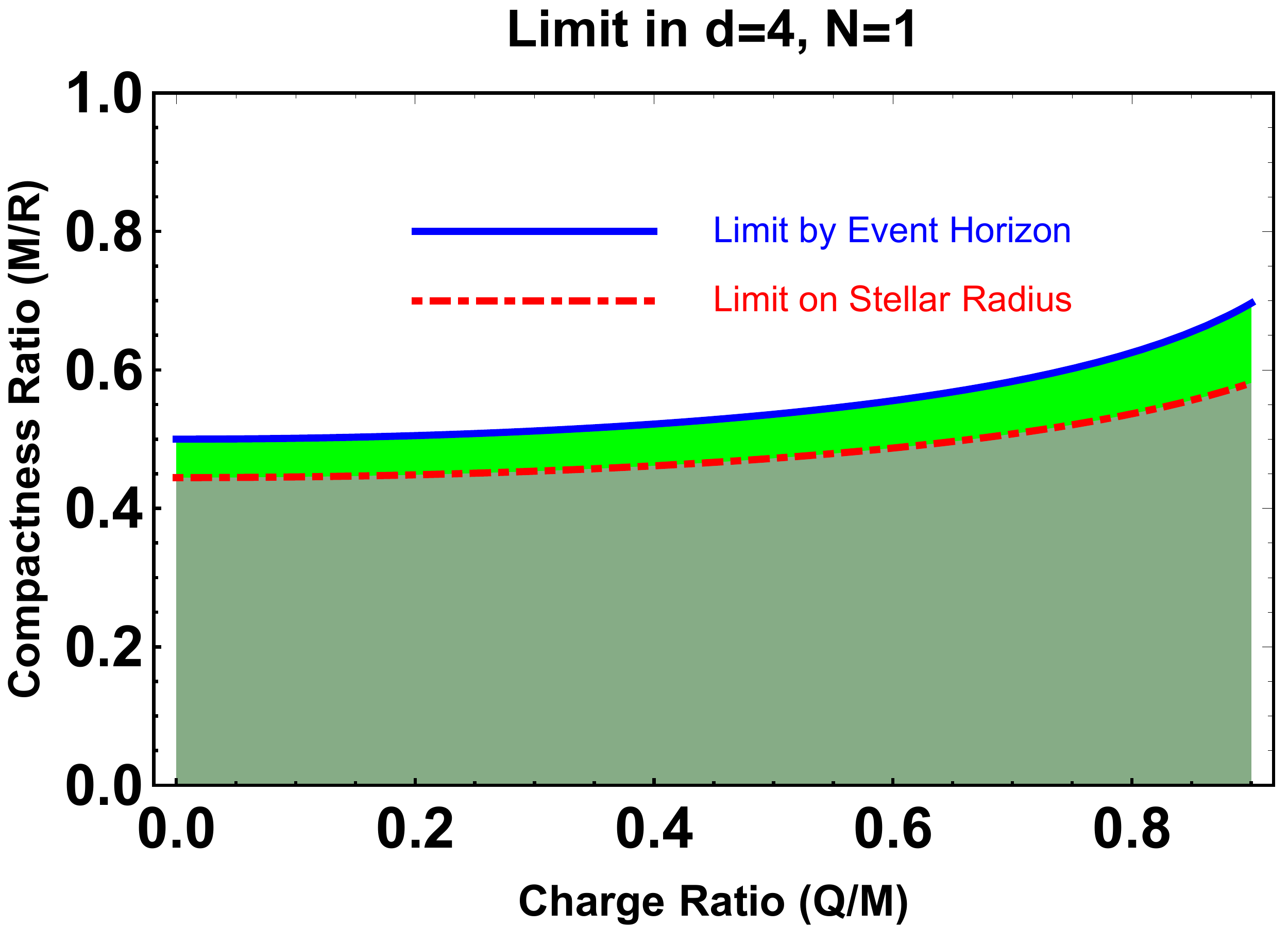}~~
\includegraphics[scale=0.3]{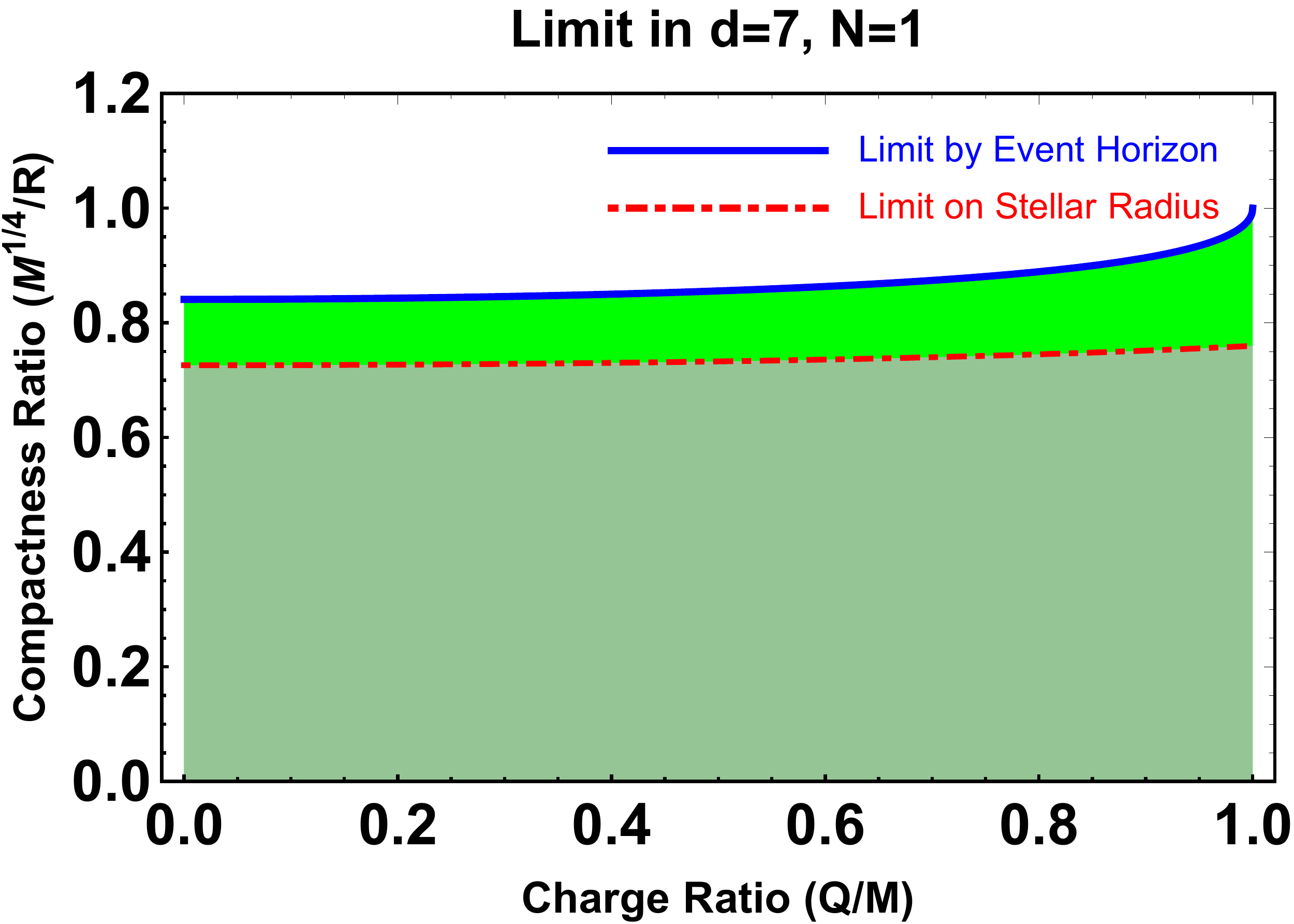}
\caption{The above figures depict the allowed parameter space in the compactness ratio $(\mathcal{M}/R)$ and charge ratio $(Q/M)$ plane for the existence of a stable stellar structure in Einstein gravity (here $\mathcal{M}^{d-3}=M$). As evident, the compactness ratio is always smaller than the event horizon. The left figure is for $d=4$ and the right figure is for $d=7$. The green zone, whose boundary is the blue line, depicts the event horizon, while the existence of stable stellar structure is allowed below the red, dot-dashed line.}
\label{Fig_Einstein_Shell}
\end{figure}
%%%%%%%%%%%%%%%%%%%%%%%
%%%%%%%%%%%%%%%%%%%%%%%
%%%%%%%%%%%%%%%%%%%%%%%
%%%%%%%%%%%%%%%%%%%%%%%

%%%%%%%%%%%%%%%%%%%%%%%
%%%%%%%%%%%%%%%%%%%%%%%
%%%%%%%%%%%%%%%%%%%%%%%
%%%%%%%%%%%%%%%%%%%%%%%
\begin{figure}
\includegraphics[scale=0.3]{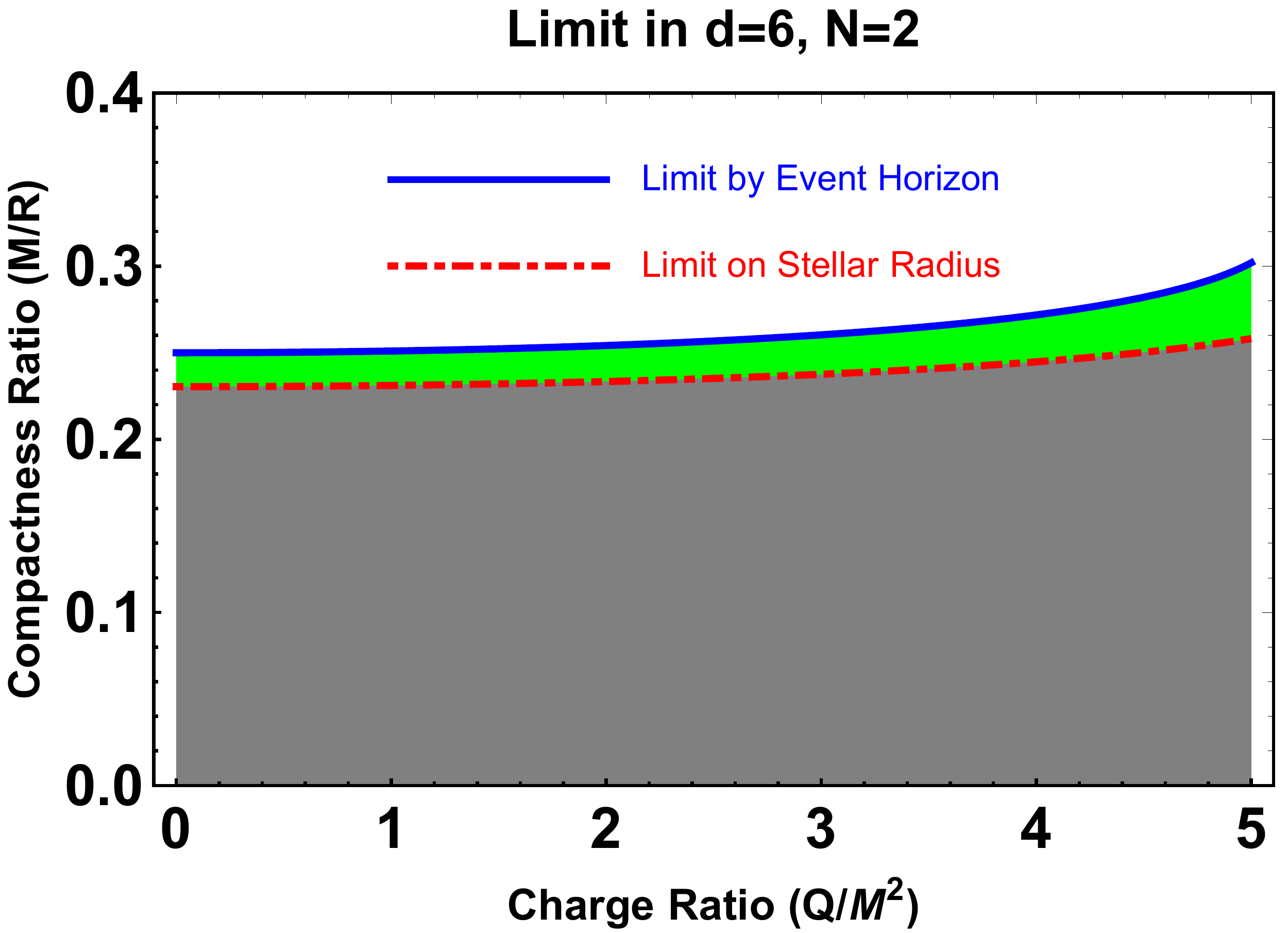}~~
\includegraphics[scale=0.3]{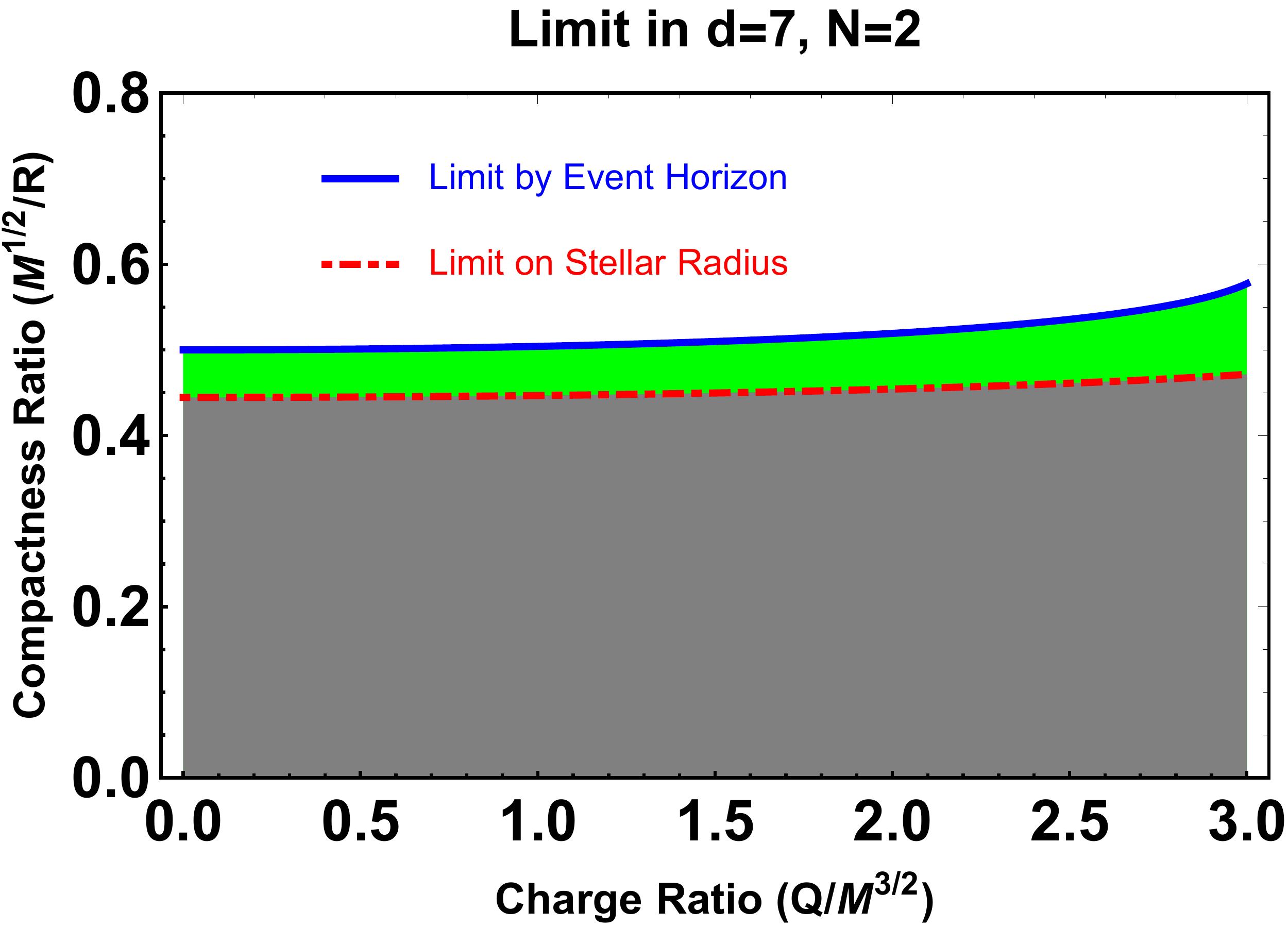}
\caption{The above figures depict the allowed parameter space in the compactness ratio $(\mathcal{M}/R)$ and charge ratio $(Q/\mathcal{M}M)$ plane for the existence of a stable stellar structure in pure Gauss-Bonnet gravity. The left figure is for $d=6$ and the right figure is for $d=7$ (here $M=\mathcal{M}^{d-5}$). The green zone, whose boundary is the blue line, depicts the event horizon, while the existence of stable stellar structure is allowed below the red dot-dashed line. As evident, the compactness ratio of the event horizon is always larger than any stable stellar structure.}
\label{Fig_GB_Shell}
\end{figure}
%%%%%%%%%%%%%%%%%%%%%%%
%%%%%%%%%%%%%%%%%%%%%%%
%%%%%%%%%%%%%%%%%%%%%%%
%%%%%%%%%%%%%%%%%%%%%%%

%%%%%%%%%%%%%%%%%%%%%%%
%%%%%%%%%%%%%%%%%%%%%%%
%%%%%%%%%%%%%%%%%%%%%%%
%%%%%%%%%%%%%%%%%%%%%%%
\begin{figure}
\center
\includegraphics[scale=0.4]{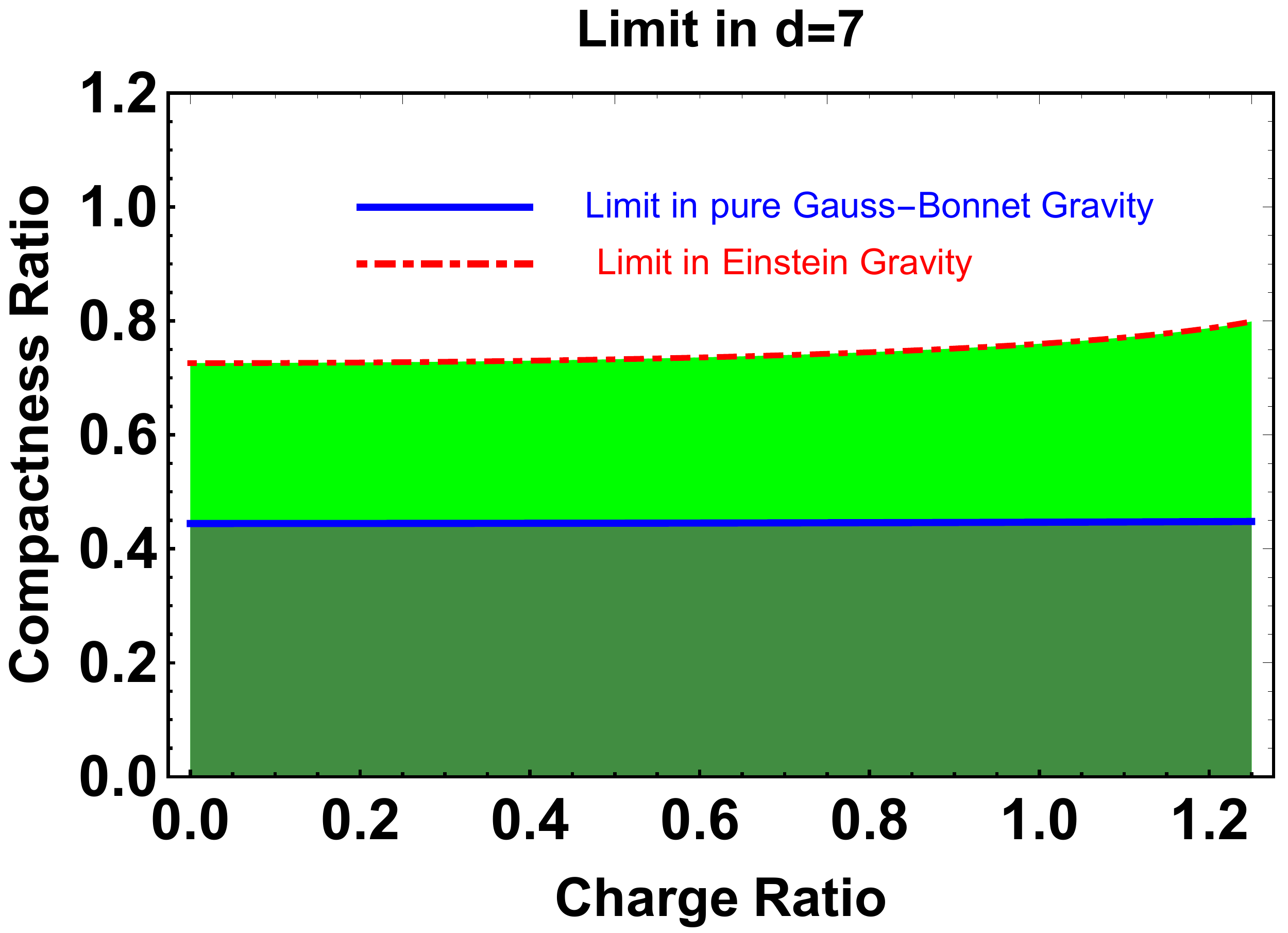}
\caption{The above figure compares the allowed region in the compactness ratio and charge ratio plane for the existence of a stable stellar structure in Einstein and in pure Gauss-Bonnet gravity in $d=7$. As evident, the allowed region for pure Gauss-Bonnet gravity for stable stellar structure is smaller compared to the Einstein gravity.}
\label{Fig_Comparison_Shell}
\end{figure}
%%%%%%%%%%%%%%%%%%%%%%%
%%%%%%%%%%%%%%%%%%%%%%%
%%%%%%%%%%%%%%%%%%%%%%%
%%%%%%%%%%%%%%%%%%%%%%%

\item \textbf{Seven dimensional pure Gauss Bonnet gravity:} As another example, consider the case of seven dimensional pure Gauss-Bonnet gravity. In this case we may substitute $d=7$ and $N=2$ in the inequality presented in \ref{Buchdahl_Charged_Shell}, from which we obtain,
%%%%%%%%%%%%%%%%%%%%%%%%%%%%%%%%%%%%%%%%%%%%%%%%%%%%%%%%
\begin{align}
\sqrt{\left(\frac{M}{R^{2}}-\frac{Q^{2}}{4R^{6}}\right)}\leq \frac{16}{36}=\frac{4}{9}~.
\end{align}
%%%%%%%%%%%%%%%%%%%%%%%%%%%%%%%%%%%%%%%%%%%%%%%%%%%%%%%%%
Note that for $Q=0$, the above inequality provides the Buchdahl limit for a massive star in four dimensional general relativity. This is because in vacuum spacetimes, pure Lovelock gravity in $d=3N+1$ dimension is indistinguishable from general relativity in four dimension, which is in exact agreement with the results presented in \cite{Chakraborty:2016qbw}. By squaring the above expression and manipulating the relevant terms we immediately obtain the following inequality,
%%%%%%%%%%%%%%%%%%%%%%%%%%%%%%%%%%%%%%%%%%%%%%%%%%%%%%%%
\begin{align}
\frac{64}{81}R^{6}-4MR^{4}+Q^{2}\geq 0~,
\end{align}
%%%%%%%%%%%%%%%%%%%%%%%%%%%%%%%%%%%%%%%%%%%%%%%%%%%%%%%%%
on the radius of a stable stellar structure with mass $M$ and electric charge $Q$. It turns out that the above algebraic equation, when the equality holds, has a single positive real root and the above inequality demands that the radius of the star should be larger than this root. This yields, the following bound on the stellar radius, 
%%%%%%%%%%%%%%%%%%%%%%%%%%%%%%%%%%%%%%%%%%%%%%%%%%%%%%%%
\begin{align}
R^{2}&\geq \frac{3}{16}\Bigg[9M+\frac{27\times 3^{2/3}M^{2}}{16\left(243 M^{3}-32Q^{2}+8Q\sqrt{-243M^{3}+16Q^{2}}\right)^{1/3}}
\nonumber
\\
&\hskip 2 cm +3^{1/3}\left(243 M^{3}-32Q^{2}+8Q\sqrt{-243M^{3}+16Q^{2}}\right)^{1/3}\Bigg]~.
\end{align}
%%%%%%%%%%%%%%%%%%%%%%%%%%%%%%%%%%%%%%%%%%%%%%%%%%%%%%%%%
It may appear that the above solution is complex for larger values of the mass, owing to the factor $(-243M^{3}+16Q^{2})$, which may turn negative and is present under a square root. However further manipulations reveal that the above solution is actually real. To see this, one first notices that the terms within the cube root in the above expression actually forms a perfect square yielding $(\sqrt{243M^{3}-16Q^{2}}\pm4iQ)^{2/3}$. Subsequent manipulation involving such quantity and its complex conjugate yields the following simplified and manifestly real inequality for the radius of the stellar structure,
%%%%%%%%%%%%%%%%%%%%%%%%%%%%%%%%%%%%%%%%%%%%%%%%%%%%%%%%
\begin{align}
\frac{M}{R^{2}}\leq \frac{16}{27}M\Bigg[1+2\cos\left(\frac{2\theta}{3}\right)\Bigg]^{-1}~;\qquad \theta=\tan^{-1}\left(\frac{4(Q/M^{3/2})}{\sqrt{243-16(Q/M^{3/2})^{2}}}\right)~.
\end{align}
%%%%%%%%%%%%%%%%%%%%%%%%%%%%%%%%%%%%%%%%%%%%%%%%%%%%%%%%%
As emphasized earlier, the inequality, in the above expression is manifestly real and more importantly is constructed out of dimensionless quantities. For a quick check of the above inequality, note that for $Q=0$, the above inequality becomes consistent with the Buchdahl's limit for four dimensional general relativity, as expected from a pure Lovelock theory in $d=3N+1$ dimensions \cite{Chakraborty:2016qbw}. To bring out the picture in a clearer fashion we have plotted the bound on the compactness ratio $(M/R^{2})$, of a stable stellar structure for pure Gauss-Bonnet gravity in six and seven dimensions in \ref{Fig_GB_Shell}. Following the trend in Einstein gravity, in pure Gauss-Bonnet theory as well, the compactness ratio allowed for stable stellar structure increases as the spacetime dimension increases and as the charge ratio increases. Moreover, a comparison between seven dimensional Einstein gravity and pure Gauss-Bonnet gravity reveals that for higher order Lovelock theories the allowed parameter space for the compactness ratio decreases (see \ref{Fig_Comparison_Shell}), this is because gravitational interaction is weak in higher order Lovelock theories in comparison with general relativity.

\end{itemize}
%%%%%%%%%%%%%%%%%%%%%%%%%%%%%%%%%
%%%%%%%%%%%%%%%%%%%%%%%%%%%%%%%%%
%%%%%%%%%%%%%%%%%%%%%%%%%%%%%%%%%

The above examples explicitly demonstrate the implications of the bound on the compactness ratio presented in \ref{Buchdahl_Charged_Shell} for both Einstein gravity and pure Gauss-Bonnet gravity. It should be noted that we have not considered the $d=5$ case, since the pure Gauss-Bonnet gravity has no non-trivial asymptotically flat solution in this spacetime dimension \cite{Dadhich:2015lra}, alike the Einstein gravity in three spacetime dimensions. As we have mentioned earlier, even though the bound on the stellar structure is generic enough to be applied to any pure Lovelock gravity theory in arbitrary spacetime dimensions, it is difficult to obtain an analytic expression for the bound in generic contexts. In which case one must determine the allowed compactness ratios by solving the inequality numerically. The above discussion was for a charged shell surrounding a matter distribution and in the next section we will take up the general case of a charged sphere and shall derive the bound arising thereof.

%%%%%%%%%%%%%%%%%%%%%%%%%%%%%%%%%%%%%%%%%%%%%%%%%%%%%%
%%%%%%%%%%%%%%%%%%%%%%%%%%%%%%%%%%%%%%%%%%%%%%%%%%%%%%
%%%%%%%%%%%%%%%%%%%%%%%%%%%%%%%%%%%%%%%%%%%%%%%%%%%%%%
\subsection{General analysis of the bound on stellar structure}

In the previous section we have discussed the case of a charged shell as a warm up exercise. In this section we will consider the general analysis of the corresponding bound on the stellar structure. For which we will closely follow the analysis presented in \cite{Boehmer:2007gq}. The analysis effectively depends on certain manipulations of \ref{main_lovelock_01} and its subsequent integration. We start, by defining the following quantites,
%%%%%%%%%%%%%%%%%%%%%%%%%%%%%%%%%%%%%%%%%%%%%%%%%%%%%%%%%
\begin{align}
\eta(r)&\equiv \int^{r}dr'~r'e^{\lambda/2}\int^{r'}dr''~e^{\frac{\lambda+\nu}{2}}\left[\frac{2^{N-1}(d-2)}{N(1-e^{-\lambda})^{N-1}}\left(\frac{q^{2}(r)}{r^{2d-2N-1}}\right)\right]~,
\\
\psi&\equiv e^{\nu/2}-\eta~;\qquad \xi \equiv \int^{r}dr'~r'e^{\lambda/2}~.
\end{align}
%%%%%%%%%%%%%%%%%%%%%%%%%%%%%%%%%%%%%%%%%%%%%%%%%%%%%%%%%
Thus one can compute both $(d\psi/d\xi)$ as well as $(d^{2}\psi/d\xi^{2})$ and substitute those expressions in \ref{main_lovelock_01}. This leads to the following form for the central inequality, presented in \ref{main_lovelock_01} as,
%%%%%%%%%%%%%%%%%%%%%%%%%%%%%%%%%%%%%%%%%%%%%%%%%%%%%%%%%
\begin{align}
re^{-\nu/2}\frac{d^{2}\psi}{d\xi^{2}}=\frac{2r^{3}}{e^{-\lambda}}\dfrac{d}{dr}\left[\frac{1-e^{-\lambda}}{2r^{2}}\right]\left\{-\frac{2\nu'(N-1)e^{-\lambda}}{1-e^{-\lambda}}+\frac{2\left(d-2N-1\right)}{r}\right\}\leq 0~.
\end{align}
%%%%%%%%%%%%%%%%%%%%%%%%%%%%%%%%%%%%%%%%%%%%%%%%%%%%%%%%%
In order to arrive at the final inequality, we have assumed that, $r^{-2}(1-e^{-\lambda})$ decreases as the radial distance increases. For the case of vanishing electric charge, the above condition corresponds to decreasing mass density as one approaches the radius of the star. In the case of charged sphere as well it seems reasonable to assume that the above condition holds. The above inequality can be trivially integrated, yielding,
%%%%%%%%%%%%%%%%%%%%%%%%%%%%%%%%%%%%%%%%%%%%%%%%%%%%%%%%%
\begin{align}\label{inequality_gen}
\dfrac{d\psi}{d\xi}\leq \frac{\psi(\xi)-\psi(0)}{\xi}~.
\end{align}
%%%%%%%%%%%%%%%%%%%%%%%%%%%%%%%%%%%%%%%%%%%%%%%%%%%%%%%%%
Note that $\xi=0$ corresponds to the central region of the star. Since at the center there is no electric charge, it follows that $\eta(0)=0$. Also there should be no interchange of time and space at the center, which suggests $e^{\nu}(0)>0$ and hence $\psi(0)>0$. Thus from \ref{inequality_gen} one arrives at the following condition,
%%%%%%%%%%%%%%%%%%%%%%%%%%%%%%%%%%%%%%%%%%%%%%%%%%%%%%%%%
\begin{align}
\dfrac{d\psi}{d\xi}\leq \frac{\psi(\xi)}{\xi}
\end{align}
%%%%%%%%%%%%%%%%%%%%%%%%%%%%%%%%%%%%%%%%%%%%%%%%%%%%%%%%%
Substituting for $\psi$, $\xi$ and $(d\psi/d\xi)$, it follows from the above inequality, that
%%%%%%%%%%%%%%%%%%%%%%%%%%%%%%%%%%%%%%%%%%%%%%%%%%%%%%%%%
\begin{align}\label{gen_limit_01}
\left[ \int^{r}dr'~r'e^{\lambda/2}\right]\Bigg\{\frac{e^{-\lambda/2}}{r}&\frac{d}{dr}e^{\nu/2}- \int^{r}dr'~e^{\frac{\lambda+\nu}{2}}\left[\frac{2^{N-1}(d-2)}{N(1-e^{-\lambda})^{N-1}}\left(\frac{q^{2}(r)}{r^{2d-2N-1}}\right)\right]\Bigg\}
\nonumber
\\
&\leq e^{\nu/2}-\int^{r}dr'~r'e^{\lambda/2}\int^{r'}dr''~e^{\frac{\lambda+\nu}{2}}\left[\frac{2^{N-1}(d-2)}{N(1-e^{-\lambda})^{N-1}}\left(\frac{q^{2}(r)}{r^{2d-2N-1}}\right)\right]~.
\end{align}
%%%%%%%%%%%%%%%%%%%%%%%%%%%%%%%%%%%%%%%%%%%%%%%%%%%%%%%%%
The above inequality can be reduced further, provided certain realistic assumptions are used. The first such assumption corresponds to $\{(1-e^{-\lambda})/r^{2}\}$ to be decreasing with an increase of $r$. This condition, as emphasized earlier, relates to the fact that density of the stellar material decreases outward. It yields, 
%%%%%%%%%%%%%%%%%%%%%%%%%%%%%%%%%%%%%%%%%%%%%%%%%%%%%%%%%
\begin{align}\label{assumption_01}
\alpha(r')\left(\frac{m_{\rm grav}(r')}{r'^{d-2N-1}}\right)^{1/N}\geq \alpha(r)\left(\frac{m_{\rm grav}(r)}{r^{d-2N-1}}\right)^{1/N}\left(\frac{r'}{r}\right)^{2}~,
\end{align}
%%%%%%%%%%%%%%%%%%%%%%%%%%%%%%%%%%%%%%%%%%%%%%%%%%%%%%%%%
where, $r'\leq r$ and we have defined the quantity $\alpha(r)$, through the following relation,
%%%%%%%%%%%%%%%%%%%%%%%%%%%%%%%%%%%%%%%%%%%%%%%%%%%%%%%%%
\begin{align}\label{gen_def_01}
e^{-\lambda}=1-2\alpha(r)\left(\frac{m_{\rm grav}(r)}{r^{d-2N-1}}\right)^{1/N}~;\qquad \alpha(r)^{N}=1-\frac{q^{2}}{2^{N}m_{\rm grav}r^{d-3}}~.
\end{align}
%%%%%%%%%%%%%%%%%%%%%%%%%%%%%%%%%%%%%%%%%%%%%%%%%%%%%%%%%
Substituting for $\alpha$ from \ref{gen_def_01} in \ref{assumption_01} one can re-express the above inequality in terms of mass and charge of the stellar object within some radius $r$. Therefore, using \ref{assumption_01} and integrating the relevant expression, one can demonstrate that the quantity $\xi$, defined earlier, satisfies the following inequality,
%%%%%%%%%%%%%%%%%%%%%%%%%%%%%%%%%%%%%%%%%%%%%%%%%%%%%%%%%
\begin{align}\label{condition_01}
\xi^{-1}&\leq 2\alpha(r)\left(\frac{m_{\rm grav}(r)}{r^{d-1}}\right)^{1/N}\left[1-\sqrt{1-2\alpha(r)\left(\frac{m_{\rm grav}(r)}{r^{d-2N-1}}\right)^{1/N}}\right]^{-1}~.
\end{align}
%%%%%%%%%%%%%%%%%%%%%%%%%%%%%%%%%%%%%%%%%%%%%%%%%%%%%%%%%
We also need to invoke certain additional assumptions regarding the behaviour of the charge $q(r)$ in order to proceed further with the integrals in \ref{gen_limit_01} involving charge. It turns out that in the present context, the following assumption will suffice for our purpose,
%%%%%%%%%%%%%%%%%%%%%%%%%%%%%%%%%%%%%%%%%%%%%%%%%%%%%%%%%
\begin{align}\label{assumption_02}
\frac{1}{(1-e^{-\lambda(r')})^{N-1}}\left(\frac{q^{2}(r')}{r'^{2d-2N-1}}\right)e^{\frac{\nu(r')}{2}}\geq \frac{1}{(1-e^{-\lambda(r)})^{N-1}}\left(\frac{q^{2}(r)}{r^{2d-2N-1}}\right)e^{\frac{\nu(r)}{2}}~,
\end{align}
%%%%%%%%%%%%%%%%%%%%%%%%%%%%%%%%%%%%%%%%%%%%%%%%%%%%%%%%%
for $r'\leq r$. Using both the assumptions, namely, \ref{assumption_01} and \ref{assumption_02} we arrive at the following inequality, which reads,
%%%%%%%%%%%%%%%%%%%%%%%%%%%%%%%%%%%%%%%%%%%%%%%%%%%%%%%%%
\begin{align}\label{condition_02}
\int^{r}dr'&~e^{\frac{\lambda(r')+\nu(r')}{2}}\left[\frac{2^{N-1}(d-2)}{N(1-e^{-\lambda(r')})^{N-1}}\left(\frac{q^{2}(r')}{r'^{2d-2N-1}}\right)\right]
\nonumber
\\
&\geq \left[\frac{2^{N-1}(d-2)e^{\frac{\nu(r)}{2}}}{N(1-e^{-\lambda(r)})^{N-1}}\left(\frac{q^{2}(r)}{r^{2d-2N-1}}\right)\right]\left(2\alpha(r)\left(\frac{m_{\rm grav}(r)}{r^{d-1}}\right)^{1/N}\right)^{-1/2}
\nonumber
\\
&\hskip 4 cm \times \sin^{-1}\left(\sqrt{2\alpha(r)\left(\frac{m_{\rm grav}(r)}{r^{d-2N-1}}\right)^{1/N}}\right)~.
\end{align}
%%%%%%%%%%%%%%%%%%%%%%%%%%%%%%%%%%%%%%%%%%%%%%%%%%%%%%%%%
Finally, using \ref{condition_02} and the assumptions introduced above, we obtain the following inequality for double integration of the charge with appropriate coefficients appearing on the right hand side of \ref{gen_limit_01} as,
%%%%%%%%%%%%%%%%%%%%%%%%%%%%%%%%%%%%%%%%%%%%%%%%%%%%%%%%%
\begin{align}\label{condition_03}
\int^{r}dr'~&r'e^{\lambda/2}\int^{r'}dr''~e^{\frac{\lambda+\nu}{2}}\left[\frac{2^{N-1}(d-2)}{N(1-e^{-\lambda})^{N-1}}\left(\frac{q^{2}(r'')}{r''^{2d-2N-1}}\right)\right]
\nonumber
\\
&\hskip -1 cm \geq \left[\frac{2^{N-1}(d-2)e^{\frac{\nu(r)}{2}}}{N(1-e^{-\lambda(r)})^{N-1}}\left(\frac{q^{2}(r)}{r^{2d-2N-1}}\right)\right]\left(2\alpha(r)\left(\frac{m_{\rm grav}(r)}{r^{d-1}}\right)^{1/N}\right)^{-3/2}\Bigg[r\left(2\alpha(r)\left(\frac{m_{\rm grav}(r)}{r^{d-1}}\right)^{1/N}\right)^{1/2}
\nonumber
\\
&-\sqrt{1-2\alpha(r)\left(\frac{m_{\rm grav}(r)}{r^{d-2N-1}}\right)^{1/N}} \sin^{-1}\left(\sqrt{2\alpha(r)\left(\frac{m_{\rm grav}(r)}{r^{d-2N-1}}\right)^{1/N}}\right)\Bigg]~.
\end{align}
%%%%%%%%%%%%%%%%%%%%%%%%%%%%%%%%%%%%%%%%%%%%%%%%%%%%%%%%%
In order to arrive at the above inequality, we have used the following integration,
%%%%%%%%%%%%%%%%%%%%%%%%%%%%%%%%%%%%%%%%%%%%%%%%%%%%%%%%%
\begin{align}
\int ^{r}dr'~\frac{r'}{\sqrt{1-a^{2}r'^{2}}}\sin^{-1}(ar')=\frac{1}{a^{2}}\left[ar-\sqrt{1-a^{2}r^{2}}\sin^{-1}(ar)\right]
\end{align}
%%%%%%%%%%%%%%%%%%%%%%%%%%%%%%%%%%%%%%%%%%%%%%%%%%%%%%%%%
with the identification, $a^{2}=2\alpha(r)\{m_{\rm grav}(r)/r^{d-1}\}^{1/N}$. We now have all the necessary ingredients to re-express the \ref{gen_limit_01} in a more manageable form. We want to revert back to the metric coefficients, rather than explicitly writing down mass and charge terms as well as further simplifications based on the assumptions described above can be made. Use of the inequalities derived in \ref{condition_01}, \ref{condition_02} and \ref{condition_03} in \ref{gen_limit_01}, yields the following modified inequality,
%%%%%%%%%%%%%%%%%%%%%%%%%%%%%%%%%%%%%%%%%%%%%%%%%%%%%%%%%
\begin{align}\label{gen_limit_04}
\left(1-e^{-\lambda/2}\right)&\frac{e^{-(\lambda+\nu)/2}}{r}\frac{d}{dr}e^{\nu/2}
\leq 
\frac{(1-e^{-\lambda})}{r^{2}}+\left[\frac{2^{N-1}(d-2)r}{N(1-e^{-\lambda})^{N-1}}\left(\frac{q^{2}(r)}{r^{2d-2N-1}}\right)\right]\Bigg\{\frac{\sin^{-1}\left(\sqrt{1-e^{-\lambda}}\right)}{\sqrt{1-e^{-\lambda}}}-1\Bigg\}
\end{align}
%%%%%%%%%%%%%%%%%%%%%%%%%%%%%%%%%%%%%%%%%%%%%%%%%%%%%%%%%
The above inequality depends explicitly on $\nu'$, which can be determined in terms of $e^{-\lambda}$ by using the gravitational field equation presented in \ref{radial_lovelock_charge}. This yields,
%%%%%%%%%%%%%%%%%%%%%%%%%%%%%%%%%%%%%%%%%%%%%%%%%%%%%%%%%
\begin{align}
e^{-\nu/2}\frac{d}{dr}e^{\nu/2}&=\frac{e^{\lambda}}{rN}\left[\frac{d-2N-1}{2}\left(1-e^{-\lambda}\right)+\frac{r^{2N}2^{N-2}}{(1-e^{-\lambda})^{N-1}}\left\{8\pi p-\frac{q^{2}}{r^{2(d-2)}} \right\}\right]
\nonumber
\\
&=\frac{2^{N-1}(d-2N-1)e^{\lambda}}{N(1-e^{-\lambda})^{N-1}r^{d-2N}}\left[m_{\rm grav}(r)+4\pi (d-2N-1)p r^{d-1}-\frac{q^{2}}{r^{d-3}}\frac{1+2^{N-1}}{2^{N}(d-2N-1)}\right]~.
\end{align}
%%%%%%%%%%%%%%%%%%%%%%%%%%%%%%%%%%%%%%%%%%%%%%%%%%%%%%%%%
Substituting the above expression in \ref{gen_limit_04} and making certain manipulations we finally obtain the inequality we were after,
%%%%%%%%%%%%%%%%%%%%%%%%%%%%%%%%%%%%%%%%%%%%%%%%%%%%%%%%%
\begin{align}\label{gen_limit_06}
\frac{\left(1-e^{-\lambda/2}\right)}{e^{-\lambda/2}}&\frac{(d-2N-1)}{r^{d-2N+1}}\left[m_{\rm grav}(r)+4\pi (d-2N-1)p r^{d-1}-\frac{q^{2}}{r^{d-3}}\frac{1+2^{N-1}}{2^{N}(d-2N-1)}\right]
\nonumber
\\
&\leq 
\frac{N(1-e^{-\lambda})^{N}}{2^{N-1}r^{2}}+\left[(d-2)r\left(\frac{q^{2}(r)}{r^{2d-2N-1}}\right)\right]\Bigg\{\frac{\sin^{-1}\left(\sqrt{1-e^{-\lambda}}\right)}{\sqrt{1-e^{-\lambda}}}-1\Bigg\}~.
\end{align}
%%%%%%%%%%%%%%%%%%%%%%%%%%%%%%%%%%%%%%%%%%%%%%%%%%%%%%%%%
Even though this result appears complicated, further manipulations along with appropriate assumptions can lead to the desired limit on the stellar radius in pure Lovelock gravity with a charge term. However, first we will confirm that in the absence of a charge term the above inequality indeed reproduces the Buchdahl limit presented in \cite{Chakraborty:2016qbw}. 
%%%%%%%%%%%%%%%%%%%%%%%%%%%%
%%%%%%%%%%%%%%%%%%%%%%%%%%%%
%%%%%%%%%%%%%%%%%%%%%%%%%%%%
\subsubsection{The case of zero electric charge}

In the case of zero electric charge, i.e., with $q=0$, the above inequality presented in \ref{gen_limit_06} will yield, keeping in mind that radial pressure is always positive ($p>0$), 
%%%%%%%%%%%%%%%%%%%%%%%%%%%%%%%%%%%%%%%%%%%%%%%%%%%%%%%%%
\begin{align}\label{gen_limit_zeroq}
\frac{\left(1-e^{-\lambda/2}\right)}{e^{-\lambda/2}}\frac{(d-2N-1)}{r^{d-2N+1}}m(r)&\leq \frac{N(1-e^{-\lambda})^{N}}{2^{N-1}r^{2}}~.
\end{align}
%%%%%%%%%%%%%%%%%%%%%%%%%%%%%%%%%%%%%%%%%%%%%%%%%%%%%%%%%
In the above inequality if we replace $e^{-\lambda}$ with the corresponding solution arising out of gravitational field equations presented in \ref{metric_lovelock_02} with zero electric charge, then we obtain,
%%%%%%%%%%%%%%%%%%%%%%%%%%%%%%%%%%%%%%%%%%%%%%%%%%%%%%%%%
\begin{align}\label{finalmass}
\left(\frac{2^{N}m_{\rm grav}(r)}{r^{d-2N-1}}\right)^{1/N}&\leq 1-\frac{1}{\left(1+\frac{2N}{(d-2N-1)}\right)^{2}}=\frac{4N(d-N-1)}{(d-1)^{2}}~.
\end{align}
%%%%%%%%%%%%%%%%%%%%%%%%%%%%%%%%%%%%%%%%%%%%%%%%%%%%%%%%%
As evident from the above expression, this inequality is nothing but the analogue of the Buchdahl limit for pure Lovelock gravity without the Maxwell term in the action. This explicitly demonstrates that the general inequality presented in \ref{gen_limit_06} indeed matches with the results of \cite{Chakraborty:2016qbw} for uncharged massive stellar structures within the realm of pure Lovelock theories. We will now take up the case of the above inequality with electric charge and shall simplify it further.

%%%%%%%%%%%%%%%%%%%%%%%%%%%%
%%%%%%%%%%%%%%%%%%%%%%%%%%%%
%%%%%%%%%%%%%%%%%%%%%%%%%%%%
\subsubsection{Final inequality with non-zero electric charge}

\begin{figure}
\includegraphics[scale=0.3]{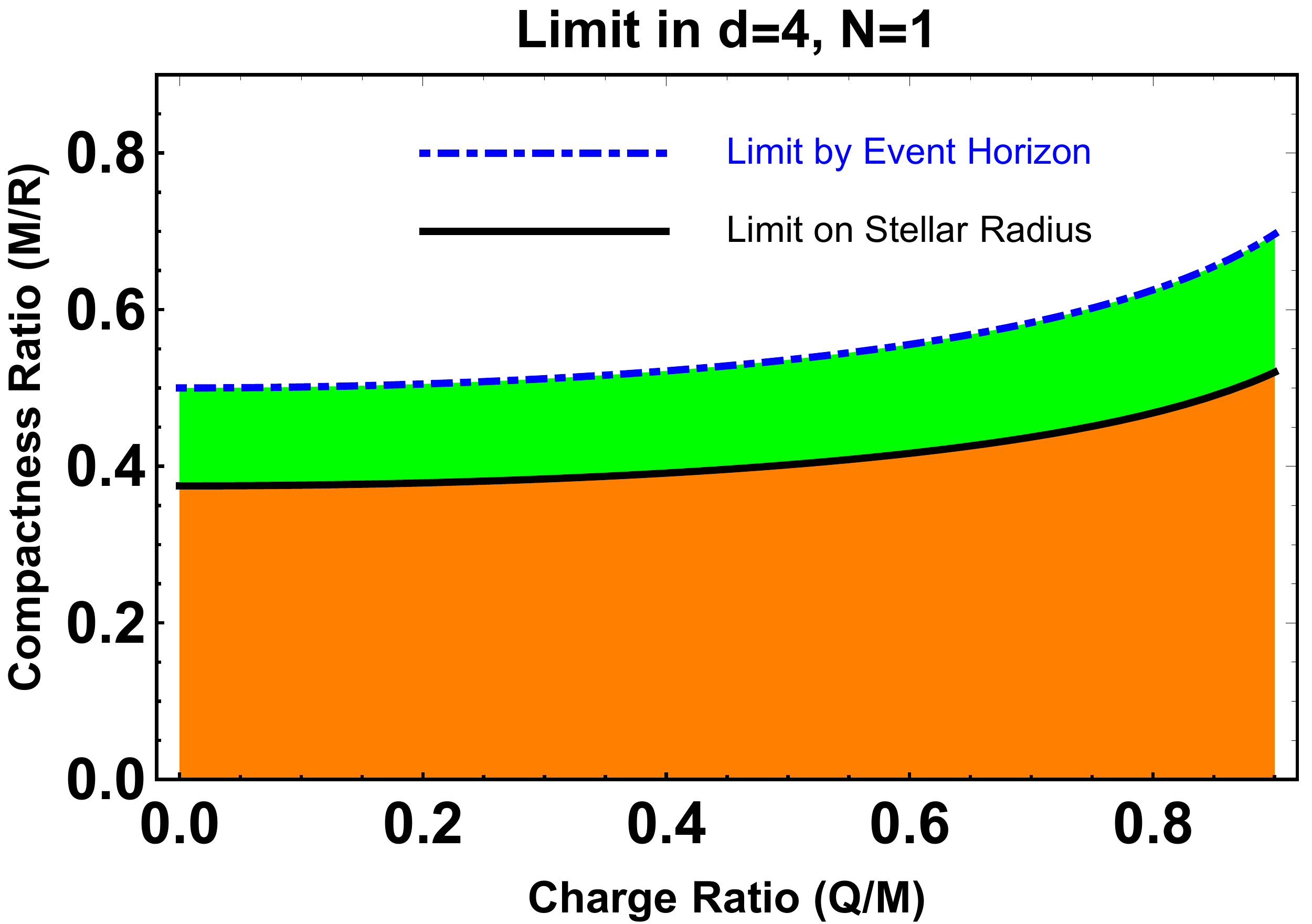}~~
\includegraphics[scale=0.3]{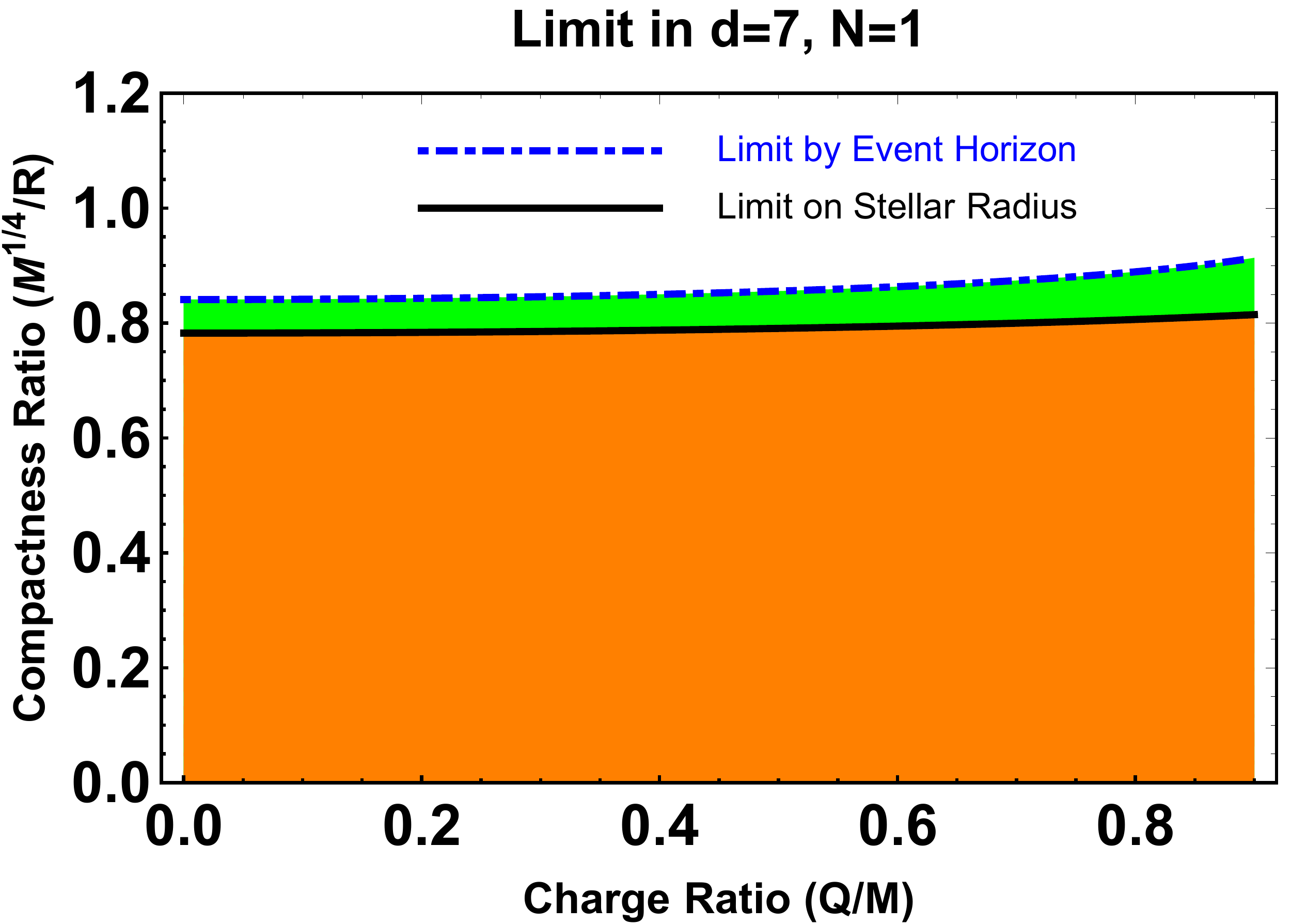}
\caption{The above figures present the allowed parameter space for the existence of a stable charged stellar object in Einstein gravity and shows that the compactness ratio $(\mathcal{M}/R)$ of a stellar object is always smaller than the event horizon (here, $M=\mathcal{M}^{d-3}$). Also as the charge ratio $(Q/M)$ increases the compactness ratio increases, since the radius of the star has to decrease to balance the repulsive electromagnetic force. Among the above plots, the left figure is for $d=4$ and the right figure is for $d=7$. The green zone, whose boundary is the blue dot-dashed line, depicts the region outside the event horizon, while the existence of stable stellar structure is allowed below the black thick line.}
\label{Fig_Sphere_Einstein}
\end{figure}

The inequality presented in \ref{gen_limit_06} can be further simplified with certain additional assumptions. The first of such assumption, corresponds to $(1-e^{-\lambda})$ being small. In which case it follows that 
%%%%%%%%%%%%%%%%%%%%%%%%%%%%%%%%%%%%%%%%%%%%%%%%%%%%%%%%%
\begin{align}
\frac{\sin^{-1}\left(\sqrt{1-e^{-\lambda}}\right)}{\sqrt{1-e^{-\lambda}}}-1=\frac{1-e^{-\lambda}}{6}~. 
\end{align}
%%%%%%%%%%%%%%%%%%%%%%%%%%%%%%%%%%%%%%%%%%%%%%%%%%%%%%%%%
Substituting this result in \ref{gen_limit_06} and using the fact that at the surface of the star we have $q(R)=Q$ and $m_{\rm grav}(R)=M$, after certain manipulations we obtain the following modified bound on the metric elements,
%%%%%%%%%%%%%%%%%%%%%%%%%%%%%%%%%%%%%%%%%%%%%%%%%%%%%%%%%
\begin{align}\label{gen_limit_09}
e^{-\lambda/2}\geq\frac{(d-2N-1)\left[M-\frac{Q^{2}}{R^{d-3}}\frac{1+2^{N-1}}{2^{N}(d-2N-1)}\right]}{(d-1)M-\frac{Q^{2}}{2^{N}R^{d-3}}\left(2N+1+2^{N-1}\right)+\frac{(d-2)}{6}\left(\frac{Q^{2}}{R^{d-3}}\right)\left(\frac{M}{R^{d-2N-1}}-\frac{Q^{2}}{2^{N}R^{2d-2N-4}}\right)^{1/N}}~.
\end{align}
%%%%%%%%%%%%%%%%%%%%%%%%%%%%%%%%%%%%%%%%%%%%%%%%%%%%%%%%%
To proceed further we may express $e^{-\lambda}$ in terms of its constituents, i.e., the mass of the stellar object and the electric charge it carries. In addition it is instructive to express all the relevant quantities in terms of the dimensionless ratios, namely the compactness ratio $(\mathcal{M}/R)$ and the charge ratio $(Q/M)\mathcal{M}^{1-N}$, where $M=\mathcal{M}^{d-2N-1}$. Therefore one may introduce the following definitions,
%%%%%%%%%%%%%%%%%%%%%%%%%%%%%%%%%%%%%%%%%%%%%%%%%%%%%%%%%
\begin{align}
u&\equiv\left(\frac{\mathcal{M}}{R}\right)^{d-2N-1}-\frac{1}{2^{N}}\left(\frac{\mathcal{M}}{R}\right)^{2d-2N-4}\left(\frac{Q}{M\mathcal{M}^{N-1}}\right)^{2}~;
\nonumber
\\
A&=\frac{B}{2^{N}}\left(\frac{\mathcal{M}}{R}\right)^{2d-2N-4}\left(\frac{Q}{M\mathcal{M}^{N-1}}\right)^{2}~;\qquad B=d-2N-2-2^{N-1}~,
\end{align}
%%%%%%%%%%%%%%%%%%%%%%%%%%%%%%%%%%%%%%%%%%%%%%%%%%%%%%%%%
in terms of which \ref{gen_limit_09} can be rewritten after appropriate manipulations as,
%%%%%%%%%%%%%%%%%%%%%%%%%%%%%%%%%%%%%%%%%%%%%%%%%%%%%%%%%
\begin{align}\label{gen_limit_12}
\left(1-2u^{1/N}\right)\left[(d-1)u+A+Bu^{1/N}\right]^{2}\geq\Big[(d-2N-1)u+A\Big]^{2}~.
\end{align}
%%%%%%%%%%%%%%%%%%%%%%%%%%%%%%%%%%%%%%%%%%%%%%%%%%%%%%%%%
This is the expression we were after. Note that for vanishing electric charge and in four spacetime dimensions, we have $A=0=B$, such that the above inequality boils down to the Buchdahl's limit for fluid sphere in general relativity. To see its effect on Einstein gravity, we have substituted $N=1$ in \ref{gen_limit_12} and hence plotted the parameter space of the compactness ratio allowed by the above inequality in \ref{Fig_Sphere_Einstein} for four and seven dimensions. As the figures demonstrate, the compactness ratio of the stellar structure is always smaller than the compactness ratio of the event horizon. This is because, the radius of the event horizon is always smaller than the radius of any stable stellar object.  Furthermore, with increase of spacetime dimensions the allowed parameter space becomes larger, as gravity becomes stronger, suggested by analysis of the previous section as well. 

%%%%%%%%%%%%%%%%%%%%%%%
%%%%%%%%%%%%%%%%%%%%%%%
%%%%%%%%%%%%%%%%%%%%%%%
%%%%%%%%%%%%%%%%%%%%%%%
%%%%%%%%%%%%%%%%%%%%%%%
%%%%%%%%%%%%%%%%%%%%%%%
%%%%%%%%%%%%%%%%%%%%%%%
%%%%%%%%%%%%%%%%%%%%%%%

\begin{figure}
\includegraphics[scale=0.3]{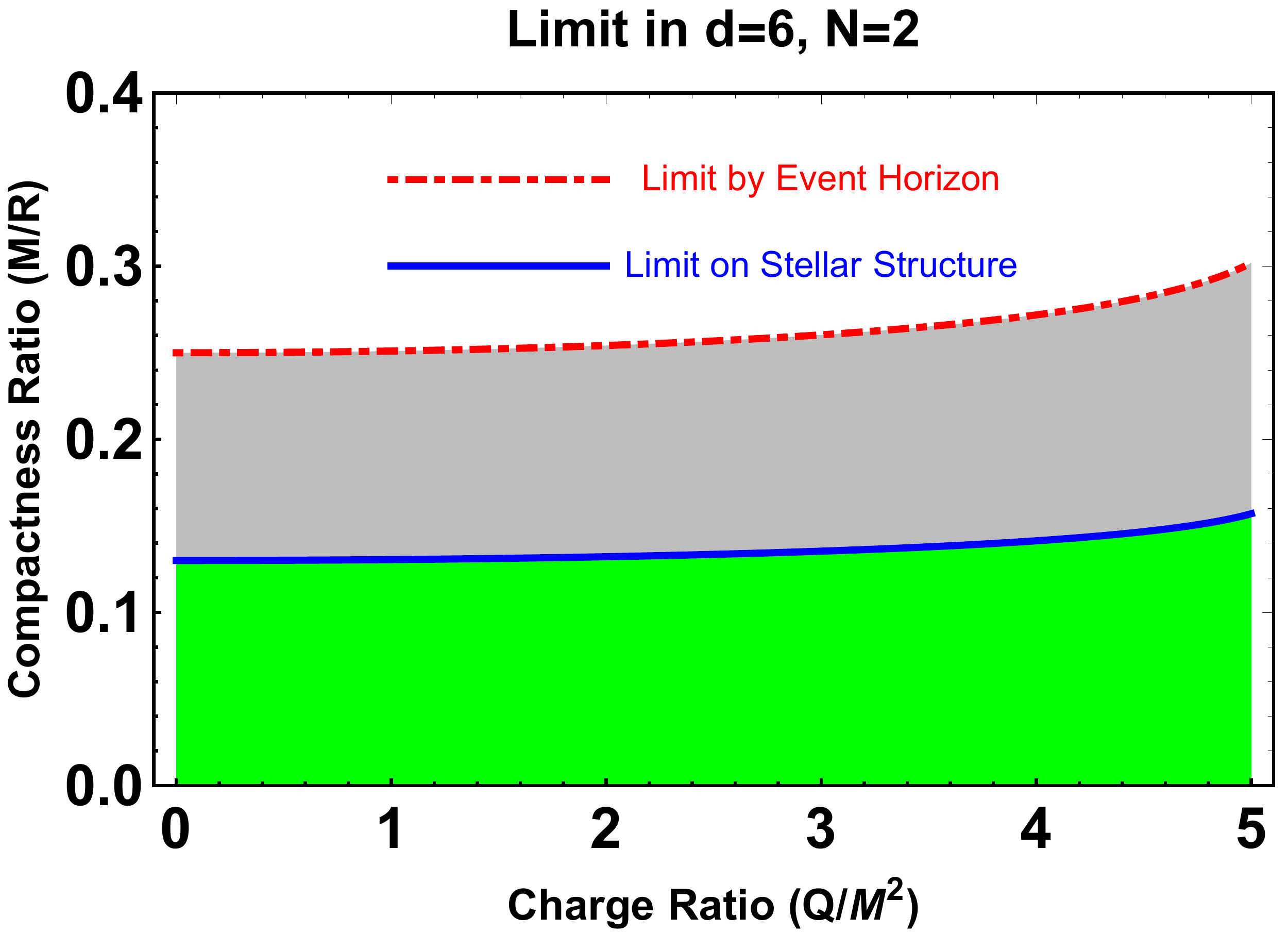}~~
\includegraphics[scale=0.3]{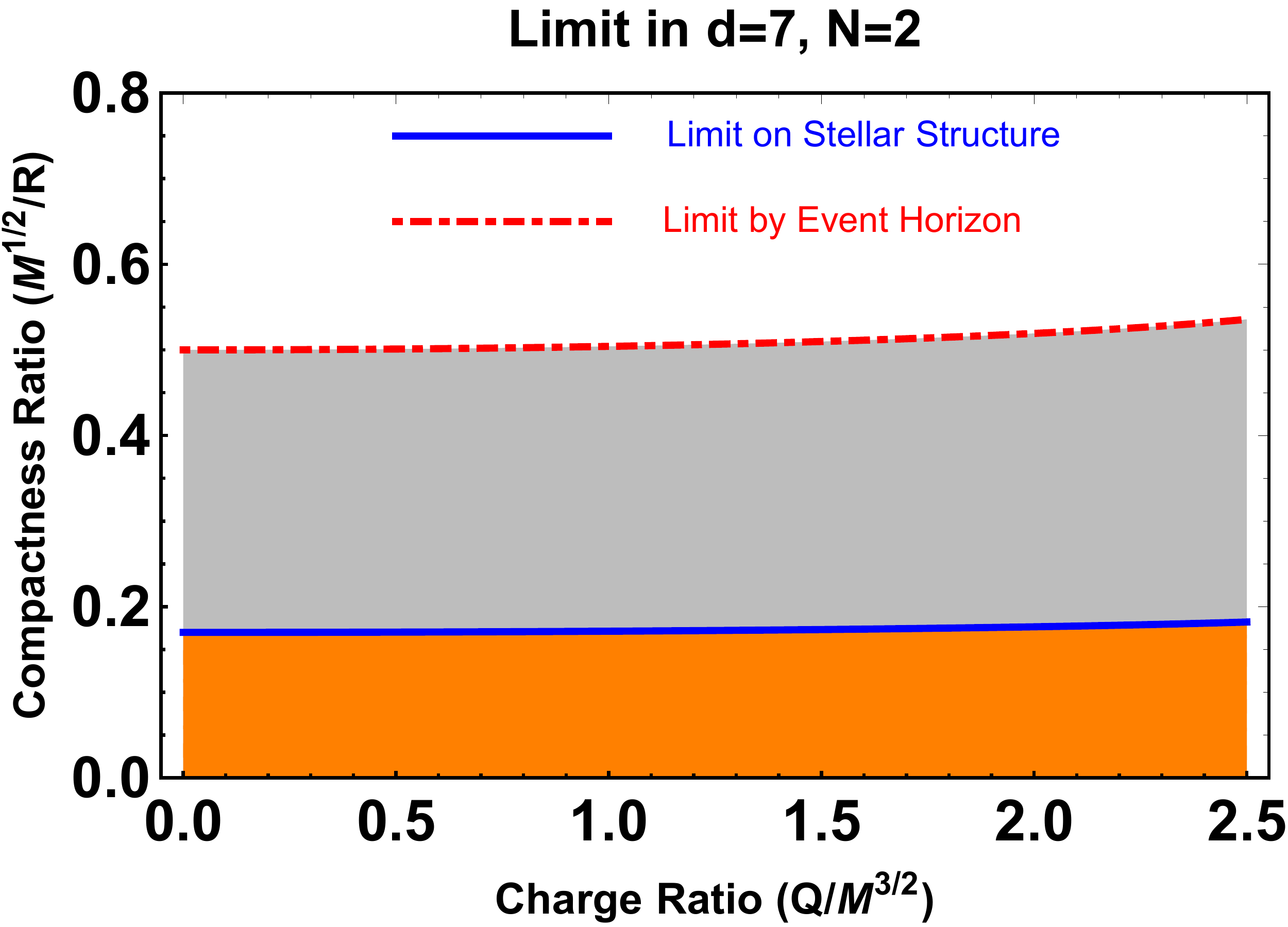}
\caption{The above figures demonstrate the allowed parameter space for the existence of stable charged objects in pure Gauss-Bonnet gravity and shows that the minimum radius is always larger than the event horizon. The figure on the left is for six dimensions and the figure on the right is for seven spacetime dimensions.  The grey zone, whose boundary is the red dot-dashed line, depicts the region outside the event horizon, while the existence of stable stellar structure is allowed below the blue thick line. Here the compactness ratio is $(\mathcal{M}/R)$ and the charge ratio being $(Q/M\mathcal{M})$, where $M=\mathcal{M}^{d-5}$.}
\label{Fig_Sphere_GB}
\end{figure}

\begin{figure}
\center
\includegraphics[scale=0.37]{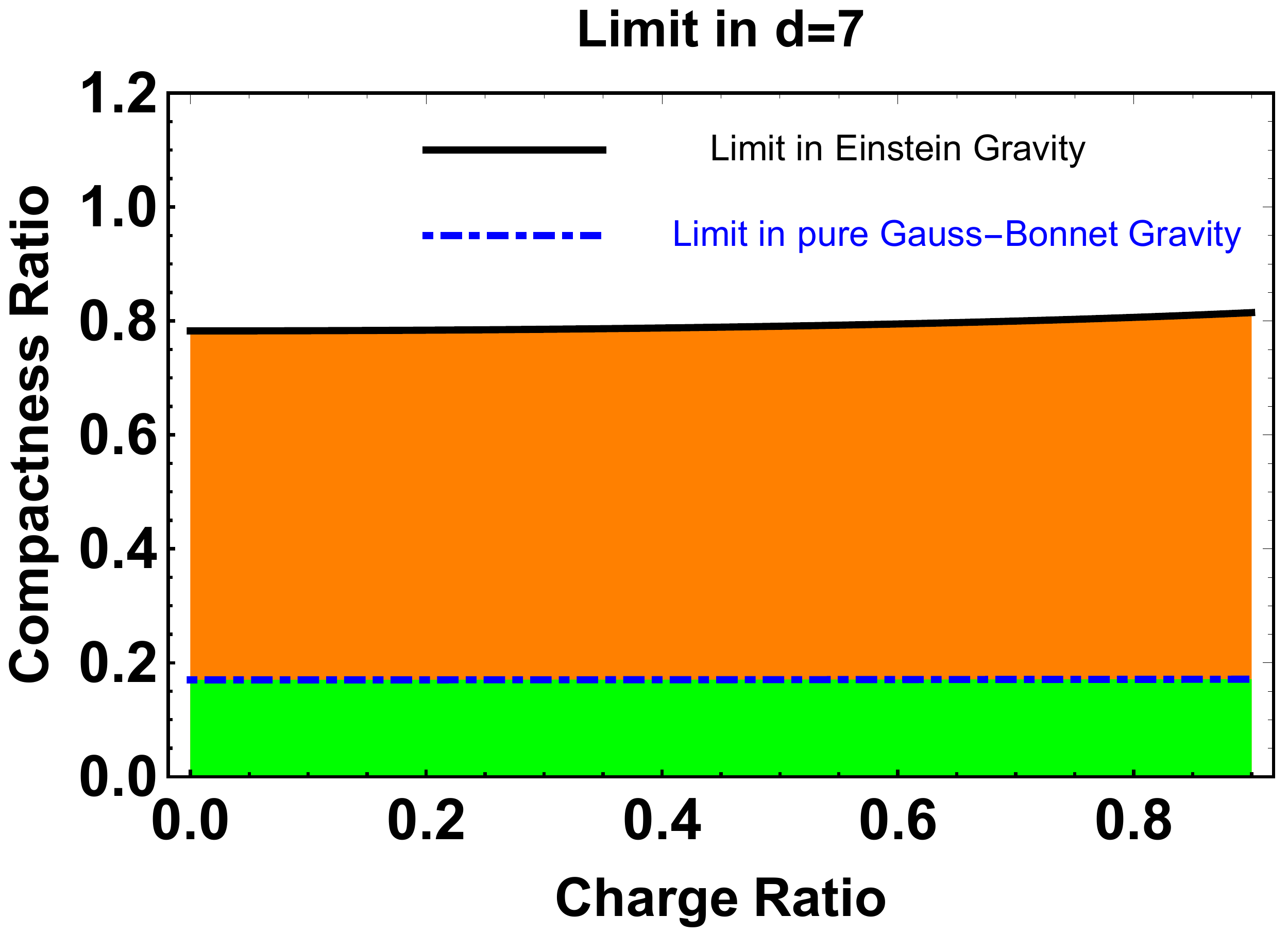}
\caption{The above figure provides a comparison between the allowed parameter space of compactness ratio for Einstein and pure Gauss-Bonnet gravity in $d=7$ for stable charged object. As evident, the allowed region for pure Gauss-Bonnet gravity for the existence of stable charged object is much smaller compared to the Einstein gravity. This is simply because in a given spacetime dimension the gravitational force is much stronger in Einstein gravity than in higher order pure Lovelock theories.}
\label{Fig_Sphere_Comparison}
\end{figure}

For pure Gauss-Bonnet gravity, the parameter space in the compactness ratio allowed by the inequality presented in \ref{gen_limit_12} has been depicted in \ref{Fig_Sphere_GB}. We have presented the limit in both six and seven spacetime dimensions. As the plots in \ref{gen_limit_12} shows, the radius of the stable stellar structure always remains larger than the event horizon and the allowed parameter space for compactness ratio significantly decreases for pure Lovelock theories for charged sphere in comparison with the case of a charged shell discussed earlier. Here also with an increase of the spacetime dimension the allowed parameter space for compactness ratio becomes larger. Finally, a comparison between seven dimension Einstein and pure Gauss-Bonnet theory depicts that the radius of a stable charged sphere will be much larger than the corresponding one in Einstein gravity (see \ref{Fig_Sphere_Comparison}). This is consistent with the corresponding analysis of the case of a charged shell presented in the previous section.   

%%%%%%%%%%%%%%%%%%%%%%%
%%%%%%%%%%%%%%%%%%%%%%%
%%%%%%%%%%%%%%%%%%%%%%%
%%%%%%%%%%%%%%%%%%%%%%%
%%%%%%%%%%%%%%%%%%%%%%%
%%%%%%%%%%%%%%%%%%%%%%%
%%%%%%%%%%%%%%%%%%%%%%%
%%%%%%%%%%%%%%%%%%%%%%%

%%%%%%%%%%%%%%%%%%%%%%%
%%%%%%%%%%%%%%%%%%%%%%%
%%%%%%%%%%%%%%%%%%%%%%%
%%%%%%%%%%%%%%%%%%%%%%%
%%%%%%%%%%%%%%%%%%%%%%%
%%%%%%%%%%%%%%%%%%%%%%%
%%%%%%%%%%%%%%%%%%%%%%%
%%%%%%%%%%%%%%%%%%%%%%%

To summarize, in $d$ spacetime dimensions, for pure Lovelock gravity of order $N$, the compactness ratio is defined as, $(M^{1/(d-2N-1)}/R)$, which for Einstein gravity ($N=1$) and in four spacetime dimensions, become identical to $(M/R)$, the familiar expression for compactness. The bounds on the compactness depends on the dimension as well as the order of pure Lovelock term. This is best seen in the case of vanishing charge, where the compactness limit becomes, $(M/R^{d-2N-1})<2N\{(d-N-1)/(d-1)^{2}\}$, reducing to $(M/R)<(4/9)$ in the case of four dimensional general relativity. Thus the definition of compactness ratio as well as the bound on it depends crucially on the spacetime dimension and order of the pure Lovelock theory. Similarly, for four dimensional Einstein gravity, the electric charge was bounded by $Q\leq M$, to prohibit formation of naked singularity. In higher spacetime dimensions and in pure Lovelock theory this bound on electric charge also gets modified. Due to complicated nature of the equation, it is not possible to provide a closed form expression for arbitrary spacetime dimension $d$ and for arbitrary Lovelock order $N$. However, some examples should drive this point home. If we consider pure Gauss-Bonnet gravity in six spacetime dimensions, the bound becomes, $Q^{2}\leq 27M^{4}$ and for pure Gauss-Bonnet gravity in seven spacetime dimensions, we obtain, $27Q^{2}\leq 256 M^{3}$. Thus the critical value of electric charge depends crucially on spacetime dimensions and order of Lovelock polynomial and must be determined on a case by case basis.

%%%%%%%%%%%%%%%%%%%%%%%%%%%%%%%%%%%%%%%%%%%%%%%%%%%%%%%%%%%%%%%%%%%%%%%%%%%%%%%%%%%%%%%%%%%%%%%%%%%
%%%%%%%%%%%%%%%%%%%%%%%%%%%%%%%%%%%%%%%%%%%%%%%%%%%%%%%%%%%%%%%%%%%%%%%%%%%%%%%%%%%%%%%%%%%%%%%%%%%
%%%%%%%%%%%%%%%%%%%%%%%%%%%%%%%%%%%%%%%%%%%%%%%%%%%%%%%%%%%%%%%%%%%%%%%%%%%%%%%%%%%%%%%%%%%%%%%%%%%
\section{Bound on stellar structure in four dimensional Einstein-Gauss-Bonnet gravity}\label{section_4degb}

In this section, we will discuss the stellar structure and limits to the same in the context of four dimensional Einstein-Gauss-Bonnet gravity \cite{Glavan:2019inb}. We would like to emphasize that this is to illustrate the techniques developed in the earlier sections for another example, which appeared recently in the literature. This requires understanding both the exterior as well as the interior solution along with appropriate matching conditions at the surface of the star. We will first review the exterior solution, which will also involve some ingredients necessary for the analysis of the interior geometry.  

%%%%%%%%%%%%%%%%%%%%%%%%%%%%%%%%%%%%%%%%%%%%%%%%%%
%%%%%%%%%%%%%%%%%%%%%%%%%%%%%%%%%%%%%%%%%%%%%%%%%%
%%%%%%%%%%%%%%%%%%%%%%%%%%%%%%%%%%%%%%%%%%%%%%%%%%
\subsection{Exterior vacuum solution}

In what follows we will briefly review the exterior vacuum solution of a static and spherically symmetric star in four-dimensional Einstein-Gauss-Bonnet theory \cite{Glavan:2019inb}. The effective coupling constants are the Newton's gravitational constant $G$, such that $\kappa_{1}=(1/16\pi G)$ and the redefined Gauss-Bonnet coupling $(d-4)\alpha \equiv \beta$, such that when $d\rightarrow 4$, $\beta$ is finite. It must be emphasized that the validity of this limit is far from obvious and may not even be valid. This is because, the action principle becomes infinite and hence the path integral becomes ill defined, even in the Euclidean domain. Further, this requires the Gauss-Bonnet coupling parameter $\alpha$ to be large (actually infinite) and hence will have serious implications for string theory, since $\alpha$ can be expressed in terms of the string length scale, which must remain finite at all cost. Thus as emphasized earlier, we will consider this scenario as merely a case where we wish to illustrate the techniques developed in the previous sections and see if some further insight can be gained through an understanding of the stellar structure. We start by writing down the temporal part of the four dimensional gravitational field equations within the spherical star, presented in \ref{gen_eq_love_01}, which in the present context takes the following form,
%%%%%%%%%%%%%%%%%%%%%%%%%%%%%%%%%%%%%%%%%%%%%%%%%%%%%%%%%
\begin{align}\label{density_eq_01}
\left[r\lambda'e^{-\lambda}+\left(1-e^{-\lambda}\right)\right]
+16\pi G\beta\frac{(1-e^{-\lambda})}{r^{2}}\left[2r\lambda'e^{-\lambda}-\left(1-e^{-\lambda}\right)\right]=8\pi G \rho(r) r^{2}~.
\end{align}
%%%%%%%%%%%%%%%%%%%%%%%%%%%%%%%%%%%%%%%%%%%%%%%%%%%%%%%%%
The terms on the left most side are coming from the Einstein-Hilbert term in the Lovelock Lagrangian, while the terms with $\beta$ have their origin in the Gauss-Bonnet theory. It is instructive to re-express this equation in the following form, 
%%%%%%%%%%%%%%%%%%%%%%%%%%%%%%%%%%%%%%%%%%%%%%%%%%%%%%%%%
\begin{align}\label{density_neq_01}
\frac{d}{dr}\left[r\left(1-e^{-\lambda}\right)\right]+16\pi G\beta \frac{d}{dr}\left[\frac{(1-e^{-\lambda})^{2}}{r}\right]=8\pi G \rho(r) r^{2}~,
\end{align}
%%%%%%%%%%%%%%%%%%%%%%%%%%%%%%%%%%%%%%%%%%%%%%%%%%%%%%%%%
which can be immediately integrated upto a radius $r$, within the extent of the spherical star, yielding the following algebraic equation for $(1-e^{-\lambda})$,
%%%%%%%%%%%%%%%%%%%%%%%%%%%%%%%%%%%%%%%%%%%%%%%%%%%%%%%%%
\begin{align}
r\left(1-e^{-\lambda}\right)+16\pi G\beta\frac{(1-e^{-\lambda})^{2}}{r}=2Gm(r)~;\qquad m(r)\equiv 4\pi \int dr~r^{2}\rho(r)~.
\end{align}
%%%%%%%%%%%%%%%%%%%%%%%%%%%%%%%%%%%%%%%%%%%%%%%%%%%%%%%%%
The above expression defines the mass function $m(r)$, which denotes the total gravitational mass within a radial distance $r$ from the origin. Given the above quadratic equation in the quantity $(1-e^{-\lambda})$, one can solve for the same and hence determine the metric element $e^{-\lambda}$. This leads to the following expression for the $g^{rr}$ component of the spacetime metric presented in \ref{sph_symm_ansatz},  
%%%%%%%%%%%%%%%%%%%%%%%%%%%%%%%%%%%%%%%%%%%%%%%%%%%%%%%%%
\begin{align}\label{solution_genn}
e^{-\lambda}=1+\frac{r^{2}}{\Upsilon}\left[1-\sqrt{1+\frac{4\Upsilon Gm(r)}{r^{3}}}\right]~,
\end{align}
%%%%%%%%%%%%%%%%%%%%%%%%%%%%%%%%%%%%%%%%%%%%%%%%%%%%%%%%%
where, we have introduced a new parameter $\Upsilon \equiv 32\pi G \beta$ ($[\Upsilon]=M^{-2}=L^{2}$). From the above solution the metric outside the spherical star can be immediately computed by considering the radial coordinate to take values $r\geq R$, where $R$ is the radius of the star. Then we have $m(r)=M$, the total mass of the star. Thus the metric outside the static and spherically symmetric star becomes \cite{Glavan:2019inb}, 
%%%%%%%%%%%%%%%%%%%%%%%%%%%%%%%%%%%%%%%%%%%%%%%%%%%%%%%%%
\begin{align}\label{solution_ext}
e^{-\lambda_{\rm ext}}=1+\frac{r^{2}}{\Upsilon}\left[1-\sqrt{1+\frac{4\Upsilon GM}{r^{3}}}\right]~,
\end{align}
%%%%%%%%%%%%%%%%%%%%%%%%%%%%%%%%%%%%%%%%%%%%%%%%%%%%%%%%%
where the subscript `ext' reminds us that it is the external metric. In principle the solution for $e^{-\lambda}$ (and also for $e^{-\lambda_{\rm ext}}$) has two branches and we have taken the `-' ve branch. This is because, we are interested in the asymptotically flat spacetime and we demand the solution to have a proper general relativity limit, i.e., the above solution must go smoothly to the Schwarschild solution in the limit $\Upsilon \rightarrow 0$. 

Another point must be stressed here. The above solution, for vacuum spacetime, depicts a black hole. Interestingly, the horizon(s) of the black hole will satisfy the following algebraic equation, $e^{-\lambda_{\rm ext}}=0$, which in this case translates into, the following equation for the horizon radius $r_{\rm h}$, $2\Upsilon r_{\rm h}^{2}-4\Upsilon GM r_{\rm h}+\Upsilon^{2}=0$. This has the following solutions,
%%%%%%%%%%%%%%%%%%%%%%%%%%%%%%%%%%%%%%%%%%%%%%%%%%%%%%%%%
\begin{align}
r_{\rm h}=GM\left[1\pm\sqrt{1-\frac{\Upsilon}{2G^{2}M^{2}}}\right]
\end{align}
%%%%%%%%%%%%%%%%%%%%%%%%%%%%%%%%%%%%%%%%%%%%%%%%%%%%%%%%%
Therefore, the causal structure is radically altered, instead of a single horizon, the spacetime now has two horizons. This is in complete contrast to the higher dimensional solutions in Einstein-Gauss-Bonnet theory where there is only one horizon even after the introduction of the Gauss-Bonnet coupling. Therefore, the $D\rightarrow 4$ limit is in some sense spurious as it is generating two horizons out of one. Also the singularity becomes timelike, rather than being spacelike, as in the case of a charged black hole. Therefore, in this scheme of dimensional reduction, the spacetime structure itself is radically altered, however small $\Upsilon$ may be. This points towards a potential failure of the dimensional reduction scheme mentioned above. 

The existence of a second horizon also brings in additional complications in the theory, which are absent in higher dimensional Einstein-Gauss-Bonnet theory. This has to do with strong cosmic censorship conjecture. Generically any spacetime inheriting a Cauchy horizon is plagued with the violation of strong cosmic censorship conjecture, where the metric can be extended across the Cauchy horizon as a weak solution of the gravitational field equations \cite{Mishra:2019ged,Rahman:2018oso,Dias:2018ynt,Cardoso:2017soq}. It turns out that the same is true is true in this scenario as well \cite{Mishra:2020gce} and hence is a potential problem, which is absent in Einstein-Gauss-Bonnet theories in higher spacetime dimensions. 

Returning back to our main theme, note that the other metric element, i.e., $e^{\nu(r)}$ must also be solved for. This requires solving the radial part of the gravitational field equations, which in four spacetime dimensions can be expressed as, 
%%%%%%%%%%%%%%%%%%%%%%%%%%%%%%%%%%%%%%%%%%%%%%%%%%%%%%%%%
\begin{align}\label{pressure_eq_01}
\left[r\nu'e^{-\lambda}-\left(1-e^{-\lambda}\right)\right]
+16\pi G\beta\frac{(1-e^{-\lambda})}{r^{2}}\left[2r\nu'e^{-\lambda}+\left(1-e^{-\lambda}\right)\right]=8\pi Gpr^{2}~.
\end{align}
%%%%%%%%%%%%%%%%%%%%%%%%%%%%%%%%%%%%%%%%%%%%%%%%%%%%%%%%%
This follows immediately from \ref{gen_eq_love_02}, after defining the new coupling constant $\beta$ and specializing to four spacetime dimensions. Subsequent addition of \ref{density_eq_01} and \ref{pressure_eq_01} yields, with the redefinition of $32\pi G\beta=\Upsilon$, as in the earlier discusion, the following equation
%%%%%%%%%%%%%%%%%%%%%%%%%%%%%%%%%%%%%%%%%%%%%%%%%%%%%%%%%
\begin{align}\label{sum_eq_01}
8\pi G\left(\rho+p\right)r^{2}=re^{-\lambda}\left(\lambda'+\nu'\right)+\frac{\Upsilon}{r} e^{-\lambda}(1-e^{-\lambda})\left(\lambda'+\nu'\right)~.
\end{align}
%%%%%%%%%%%%%%%%%%%%%%%%%%%%%%%%%%%%%%%%%%%%%%%%%%%%%%%%%
In the exterior region, there is no matter source and hence $p_{\rm ext}(r)=0=\rho_{\rm ext}(r)$. Setting the same condition in \ref{sum_eq_01}, we obtain $\lambda'_{\rm ext}+\nu'_{\rm ext}=0$, such that, $e^{\nu_{\rm ext}}=e^{-\lambda_{\rm ext}}$. This completes the derivation of the four dimensional static spherically symmetric solution in the exterior of a compact stellar object in Einstein-Gauss-Bonnet gravity and matches with the result presented in \cite{Glavan:2019inb}.

%%%%%%%%%%%%%%%%%%%%%%%%%%%%%%%%%%%%%%%%%%%%%%%%%%
%%%%%%%%%%%%%%%%%%%%%%%%%%%%%%%%%%%%%%%%%%%%%%%%%%
%%%%%%%%%%%%%%%%%%%%%%%%%%%%%%%%%%%%%%%%%%%%%%%%%%
\subsection{Interior solution with uniform density}

We have worked out the exterior solution of a stellar object in the context of a four dimensional Einstein-Gauss-Bonnet gravity, which matches with the solution presented in \cite{Glavan:2019inb}. Let us now work out the interior solution of the stellar structure. For simplicity, we will first work out the interior solution of a uniform density spherical star, i.e., with $\rho=\textrm{constant}=\rho_{\rm c}$. In the next section we will generalize our result for non-uniform density star as well. Even though in this section we assume uniform density, but we will not make any assumption about the pressure, which will be taken to be isotropic and function of the radial coordinate alone. If the spherical star has a radius $R$, then the total mass of the spherical star will be given by, $M=(4\pi/3)\rho_{\rm c}R^{3}$. In this scenario, integration of \ref{density_neq_01} is straightforward, yielding the following algebraic equation for $(1-e^{-\lambda_{\rm int}})$,
%%%%%%%%%%%%%%%%%%%%%%%%%%%%%%%%%%%%%%%%%%%%%%%%%%%%%%%%%
\begin{align}
r(1-e^{-\lambda_{\rm int}})+\frac{16\pi G \beta}{r}(1-e^{-\lambda_{\rm int}})^{2}=\frac{8\pi G\rho_{c}r^{3}}{3}=2GM\left(\frac{r}{R}\right)^{3}~,
\end{align}
%%%%%%%%%%%%%%%%%%%%%%%%%%%%%%%%%%%%%%%%%%%%%%%%%%%%%%%%%
where the subscript `int' implies that the quantities involved are associated with the interior solution. The above quadratic equation can be immediately solved, yielding,
%%%%%%%%%%%%%%%%%%%%%%%%%%%%%%%%%%%%%%%%%%%%%%%%%%%%%%%%%
\begin{align}\label{solution_int_01}
e^{-\lambda_{\rm int}}=1-\mu r^{2}~;\qquad \mu \equiv \frac{1}{\Upsilon}\left[\sqrt{1+\frac{4GM \Upsilon}{R^{3}}}-1 \right]~,
\end{align}
%%%%%%%%%%%%%%%%%%%%%%%%%%%%%%%%%%%%%%%%%%%%%%%%%%%%%%%%%
where we have introduced, $32\pi G\beta \equiv \Upsilon$ and have choosen the `+' ve branch of the solution so that correct general relativity limit can be obtained. Note that, from \ref{solution_genn} it follows that, $e^{-\lambda_{\rm int}(R)}=e^{-\lambda_{\rm ext}(R)}$, i.e., the $g^{rr}$ component of the metric is continuous across the stellar surface located at $r=R$. However, the derivatives of $e^{-\lambda}$ does not match on the stellar surface at $r=R$. This is because, the density has a discontinuity at the surface of the star, i.e., $\rho(r\rightarrow R^{-})=\rho_{\rm c}$, while $\rho(r\rightarrow R^{+})=0$. This finishes one part of the interior solution and it explicitly demonstrates that the $g^{rr}$ component of the interior metric has the same universal structure as other Lovelock classes for uniform density sphere \cite{Dadhich:2010qh}. 

Let us now concentrate on the solution for the other metric component, $e^{\nu_{\rm int}(r)}$, for which the gravitational field equation along with the conservation equation can be appropriately manipulated, yielding (for a derivation see \ref{AppA}),
%%%%%%%%%%%%%%%%%%%%%%%%%%%%%%%%%%%%%%%%%%%%%%%%%%%%%%%%%
\begin{align}\label{reduced_eq_02}
\left(1- \frac{\Upsilon}{r^{2}}\left[r\nu'_{\rm int}e^{-\lambda_{\rm int}}+(1-e^{-\lambda_{\rm int}})\right]\right)&\frac{d}{dr}\left[\frac{1-e^{-\lambda_{\rm int}}}{2r^{2}}\right]
\nonumber
\\
&\hskip -1 cm =\left(1+ \frac{\Upsilon (1-e^{-\lambda_{\rm int}})}{r^{2}}\right)e^{-(\nu_{\rm int}+\lambda_{\rm int})/2}\frac{d}{dr}\left[\frac{1}{r}e^{-\lambda_{\rm int}/2}\frac{d}{dr}e^{\nu_{\rm int}/2}\right]~.
\end{align}
%%%%%%%%%%%%%%%%%%%%%%%%%%%%%%%%%%%%%%%%%%%%%%%%%%%%%%%%%
Since $r^{-2}(1-e^{-\lambda_{\rm int}})$ is constant for a uniform density sphere, it follows that the left hand side of \ref{reduced_eq_02} identically vanishes. Thus  $e^{\nu_{\rm int}}$ satisfies the following differential equation,
%%%%%%%%%%%%%%%%%%%%%%%%%%%%%%%%%%%%%%%%%%%%%%%%%%%%%%%%%
\begin{align}
\frac{d}{dr}e^{\nu_{\rm int}/2}=Are^{\lambda_{\rm int}/2}=\frac{Ar}{\sqrt{1-\mu r^{2}}}~,
\end{align}
%%%%%%%%%%%%%%%%%%%%%%%%%%%%%%%%%%%%%%%%%%%%%%%%%%%%%%%%%
which can be immediately integrated, yielding,
%%%%%%%%%%%%%%%%%%%%%%%%%%%%%%%%%%%%%%%%%%%%%%%%%%%%%%%%%
\begin{align}\label{solution_int_02}
e^{\nu_{\rm int}/2}=A+B\sqrt{1-\mu r^{2}}=A+Be^{-\lambda_{\rm int}/2}~.
\end{align}
%%%%%%%%%%%%%%%%%%%%%%%%%%%%%%%%%%%%%%%%%%%%%%%%%%%%%%%%%
where $A$ and $B$ are to be obtained by matching $e^{\nu_{\rm int}}$ and its derivative with the exterior solution. The matching will exist for both the metric component and its derivative, since the pressure at the surface of the star identically vanishes, matching with the vacuum exterior. These conditions yield the following two equations,
%%%%%%%%%%%%%%%%%%%%%%%%%%%%%%%%%%%%%%%%%%%%%%%%%%%%%%%%%
\begin{align}
A+Be^{-\lambda_{\rm int}(R)/2}&=e^{\nu_{\rm ext}(R)/2}~,
\\
B\left(\frac{d}{dr}e^{-\lambda_{\rm int}/2}\right)_{R}&=\left(\frac{d}{dr}e^{\nu_{\rm ext}/2}\right)_{R}~.
\end{align}
%%%%%%%%%%%%%%%%%%%%%%%%%%%%%%%%%%%%%%%%%%%%%%%%%%%%%%%%%
Using the result that $e^{\nu_{\rm ext}/2}=e^{-\lambda_{\rm ext}/2}$ and on the surface of the star, $e^{-\lambda_{\rm ext}(R)/2}=e^{-\lambda_{\rm int}(R)/2}$, the matching equations can be immediately solved to determine the unknown constants $A$ and $B$ as,
%%%%%%%%%%%%%%%%%%%%%%%%%%%%%%%%%%%%%%%%%%%%%%%%%%%%%%%%%
\begin{align}
B=\frac{1}{\mu\Upsilon}\left[\frac{1+\frac{GM\Upsilon}{R^{3}}}{\sqrt{1+\frac{4GM \Upsilon}{R^{3}}}}-1\right]~;\qquad
A=\sqrt{1-\mu R^{2}}\left\{1- \frac{1}{\mu\Upsilon}\left[\frac{1+\frac{GM\Upsilon}{R^{3}}}{\sqrt{1+\frac{4GM \Upsilon}{R^{3}}}}-1\right]\right\}~.
\end{align}
%%%%%%%%%%%%%%%%%%%%%%%%%%%%%%%%%%%%%%%%%%%%%%%%%%%%%%%%%
It is worth emphasizing that alike $e^{-\lambda_{\rm int}}$, $e^{\nu_{\rm int}}$ also shows universal behaviour, in the sense the form of the solution is identical to that of pure Lovelock theory of any order \cite{Dadhich:2010qh}. This further reinforces the belief that all Lovelock theories behave in an identical manner as long as interior solution to the uniform density stellar structure is considered. Further, one can immediately verify that the above expressions for $A$ and $B$ matches with those in general relativity in the $\Upsilon \rightarrow 0$ limit. 

Having derived the full interior solution for the metric elements, let us work out the limit on the stellar structure, also known as the Buchdahl limit. The derivation crucially hinges on the physical argument that pressure of the matter forming the star should be positive and finite everywhere, including the centre. It also follows that $(1-e^{-\lambda_{\rm int}})>0$ for all points within the stellar structure, which along with positivity of pressure demands from \ref{pressure_eq_01},
%%%%%%%%%%%%%%%%%%%%%%%%%%%%%%%%%%%%%%%%%%%%%%%%%%%%%%%%%
\begin{align}\label{ineq_01}
r\nu_{\rm i}'e^{-\lambda_{\rm int}}>1-e^{-\lambda_{\rm int}}~.
\end{align}
%%%%%%%%%%%%%%%%%%%%%%%%%%%%%%%%%%%%%%%%%%%%%%%%%%%%%%%%%
There is also the additional but, well motivated assumption that the Gauss-Bonnet coupling parameter $\alpha$ (and hence $\Upsilon$) is positive. From \ref{solution_int_01} and \ref{solution_int_02}, we can read off the quantity $\nu'_{\rm int}$,
%%%%%%%%%%%%%%%%%%%%%%%%%%%%%%%%%%%%%%%%%%%%%%%%%%%%%%%%%
\begin{align}
\nu'_{\rm int}=\frac{B}{A+B\sqrt{1-\mu r^{2}}}\left(\frac{-2\mu r}{\sqrt{1-\mu r^{2}}}\right)~,
\end{align}
%%%%%%%%%%%%%%%%%%%%%%%%%%%%%%%%%%%%%%%%%%%%%%%%%%%%%%%%%
which when substituted in the inequality presented by \ref{ineq_01} yields,
%%%%%%%%%%%%%%%%%%%%%%%%%%%%%%%%%%%%%%%%%%%%%%%%%%%%%%%%%
\begin{align}
\frac{-2B\sqrt{1-\mu r^{2}}}{A+B\sqrt{1-\mu r^{2}}}>1~.
\end{align}
%%%%%%%%%%%%%%%%%%%%%%%%%%%%%%%%%%%%%%%%%%%%%%%%%%%%%%%%%
This inequality must hold true even at the center of the star, since the pressure is everywhere positive and finite, which implies $(A+B)>0$, since $B$ is intrinsically negative. This provides the following inequality among the radius of the star and various parameters of the star, which takes the following form, 
%%%%%%%%%%%%%%%%%%%%%%%%%%%%%%%%%%%%%%%%%%%%%%%%%%%%%%%%%
\begin{align}
A+B=\sqrt{1-\mu R^{2}}\left\{1- \frac{1}{\mu\Upsilon}\left[\frac{1+\frac{GM\Upsilon}{R^{3}}}{\sqrt{1+\frac{4GM \Upsilon}{R^{3}}}}-1\right]\right\}
+\frac{1}{\mu\Upsilon}\left[\frac{1+\frac{GM\Upsilon}{R^{3}}}{\sqrt{1+\frac{4GM \Upsilon}{R^{3}}}}-1\right]>0~.
\end{align}
%%%%%%%%%%%%%%%%%%%%%%%%%%%%%%%%%%%%%%%%%%%%%%%%%%%%%%%%%
This relation can be further simplified by invoking \ref{solution_int_01}, from which one can derive the following identities,
%%%%%%%%%%%%%%%%%%%%%%%%%%%%%%%%%%%%%%%%%%%%%%%%%%%%%%%%%
\begin{align}
1+\Upsilon \mu=\sqrt{1+\frac{4GM \Upsilon}{R^{3}}}~;\qquad 1+\frac{GM \Upsilon}{R^{3}}=\frac{\left(1+\Upsilon \mu\right)^{2}-1}{4}+1=\frac{4+2\Upsilon \mu+\Upsilon^{2} \mu^{2}}{4}~,
\end{align}
%%%%%%%%%%%%%%%%%%%%%%%%%%%%%%%%%%%%%%%%%%%%%%%%%%%%%%%%%
which helps to write down the condition $A+B>0$ in the following form,
%%%%%%%%%%%%%%%%%%%%%%%%%%%%%%%%%%%%%%%%%%%%%%%%%%%%%%%%%
\begin{align}\label{limit_01}
\sqrt{1-\mu R^{2}}\left(1+\frac{\Upsilon \mu}{2}\right)>\frac{1}{3}\left(1-\frac{\Upsilon \mu}{2}\right)~.
\end{align}
%%%%%%%%%%%%%%%%%%%%%%%%%%%%%%%%%%%%%%%%%%%%%%%%%%%%%%%%%
We will now manipulate the above equation and hence determine a possible bound on $\mu$, which will subsequently yield the desired bound on the compactness ratio $(M/R)$. As a first step in this direction, we square \ref{limit_01}, yielding
%%%%%%%%%%%%%%%%%%%%%%%%%%%%%%%%%%%%%%%%%%%%%%%%%%%%%%%%%
\begin{align}\label{final_eq_08}
\left(1-\mu R^{2}\right)\geq \frac{1}{9}\left(\frac{2-\mu \Upsilon}{2+\mu \Upsilon}\right)^{2}~.
\end{align}
%%%%%%%%%%%%%%%%%%%%%%%%%%%%%%%%%%%%%%%%%%%%%%%%%%%%%%%%%
Till this point the results were most general and no assumption about $\Upsilon$ was taken. However, to proceed further, we ignore terms $\mathcal{O}(\Upsilon^{2})$ and obtain the following algebraic condition on $\mu$,
%%%%%%%%%%%%%%%%%%%%%%%%%%%%%%%%%%%%%%%%%%%%%%%%%%%%%%%%%
\begin{align}\label{final_eq_09}
0\leq - 9\Upsilon R^{2}\left(\mu-\mu_{+}\right)\left(\mu-\mu_{-}\right)~,
\end{align}
%%%%%%%%%%%%%%%%%%%%%%%%%%%%%%%%%%%%%%%%%%%%%%%%%%%%%%%%%
where, the quantities $\mu_{\pm}$ are given by,
%%%%%%%%%%%%%%%%%%%%%%%%%%%%%%%%%%%%%%%%%%%%%%%%%%%%%%%%%
\begin{align}
\mu_{\pm}=\frac{\left(-10\Upsilon+9R^{2}\right)\pm \left\{\left(9R^{2}\right)+6\Upsilon -\frac{16\Upsilon^{2}}{8R^{2}}+\frac{(10\Upsilon)^{2}}{18R^{2}}\right\}}{-18\Upsilon R^{2}}~.
\end{align}
%%%%%%%%%%%%%%%%%%%%%%%%%%%%%%%%%%%%%%%%%%%%%%%%%%%%%%%%%
Thus for the inequality presented in \ref{final_eq_09} to hold true, the parameter $\mu$ must satisfy the condition $\mu<\mu_{-}$, which after simplification provides the following bound on $\mu$,
%%%%%%%%%%%%%%%%%%%%%%%%%%%%%%%%%%%%%%%%%%%%%%%%%%%%%%%%%
\begin{align}
\mu<\frac{8}{9R^{2}}+\frac{16\Upsilon}{81R^{4}}~.
\end{align}
%%%%%%%%%%%%%%%%%%%%%%%%%%%%%%%%%%%%%%%%%%%%%%%%%%%%%%%%%
Since $\mu$ is a function of the ratio between mass of the stellar object $M$ and its radius $R$, one can convert the above limit on $\mu$ to a limit on $(M/R)$. This is most conveniently done by using the expression for $\mu$ from \ref{solution_int_01} and expanding the same till first order in $\Upsilon$. Such an analysis leaves us with the following inequality,
%%%%%%%%%%%%%%%%%%%%%%%%%%%%%%%%%%%%%%%%%%%%%%%%%%%%%%%%%
\begin{align}\label{final_limit}
\frac{GM}{R}\leq \frac{4}{9}+\frac{8}{27}\left(\frac{\Upsilon}{R^{2}}\right)~.
\end{align}
%%%%%%%%%%%%%%%%%%%%%%%%%%%%%%%%%%%%%%%%%%%%%%%%%%%%%%%%%
It is possible to state the above as the desired inequality that the mass and radius of the star must satisfy. However, the primary aim of the Buchdahl limit is to provide a constraint on the minimum radius given the mass of the star. Thus it is more legitimate to re-express the above inequality with all the radius dependence shifted to the left hand side of the above inequality, this yields,
%%%%%%%%%%%%%%%%%%%%%%%%%%%%%%%%%%%%%%%%%%%%%%%%%%%%%%%%%
\begin{align}\label{final_limit_mod}
\frac{GM}{R}\leq \frac{4}{9}+\frac{2}{3}\left(\frac{4}{9}\right)^{3}\left(\frac{\Upsilon}{G^{2}M^{2}}\right)~.
\end{align}
%%%%%%%%%%%%%%%%%%%%%%%%%%%%%%%%%%%%%%%%%%%%%%%%%%%%%%%%%
The above inequality provides the limit on the minimum radius a stellar structure of a given mass $M$ can have in four dimensional Einstein-Gauss-Bonnet gravity. 
For $\Upsilon=0$, we immediately get back the Buchdahl limit, $(GM/R)<(4/9)$, while for $\Upsilon\neq 0$, the bound on the ratio of the mass and radius of the star is larger than the limit on general relativity. Also note that we have here $\Upsilon=(d-4)32\pi G\alpha$, thus if we take the $d\rightarrow 4$ limit, while keeping $\alpha$ finite, it follows that the above limit will reduce to Buchdahl limit for four dimensional general relativity. 

\begin{figure}
\includegraphics[scale=0.3]{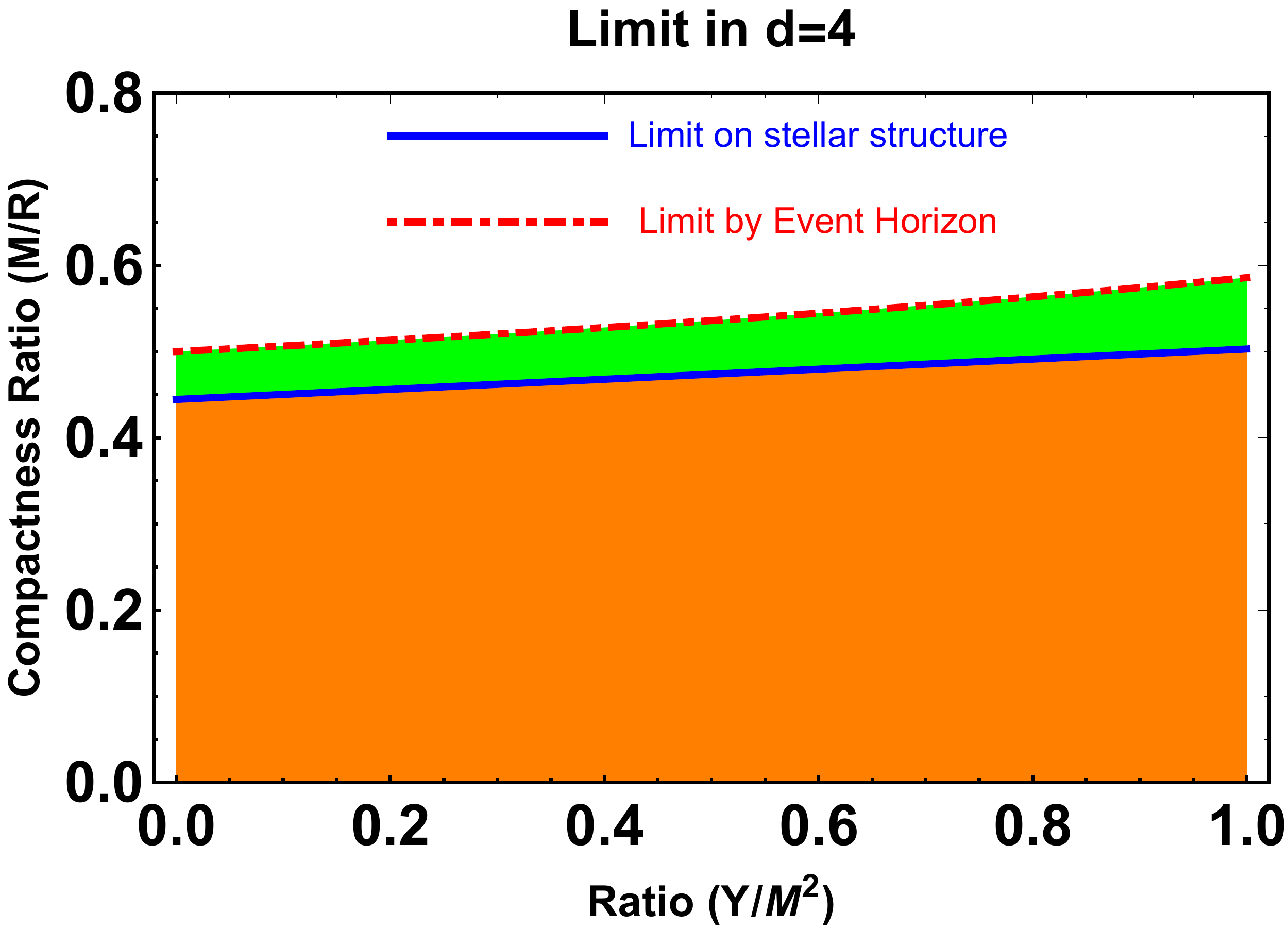}~~
\includegraphics[scale=0.3]{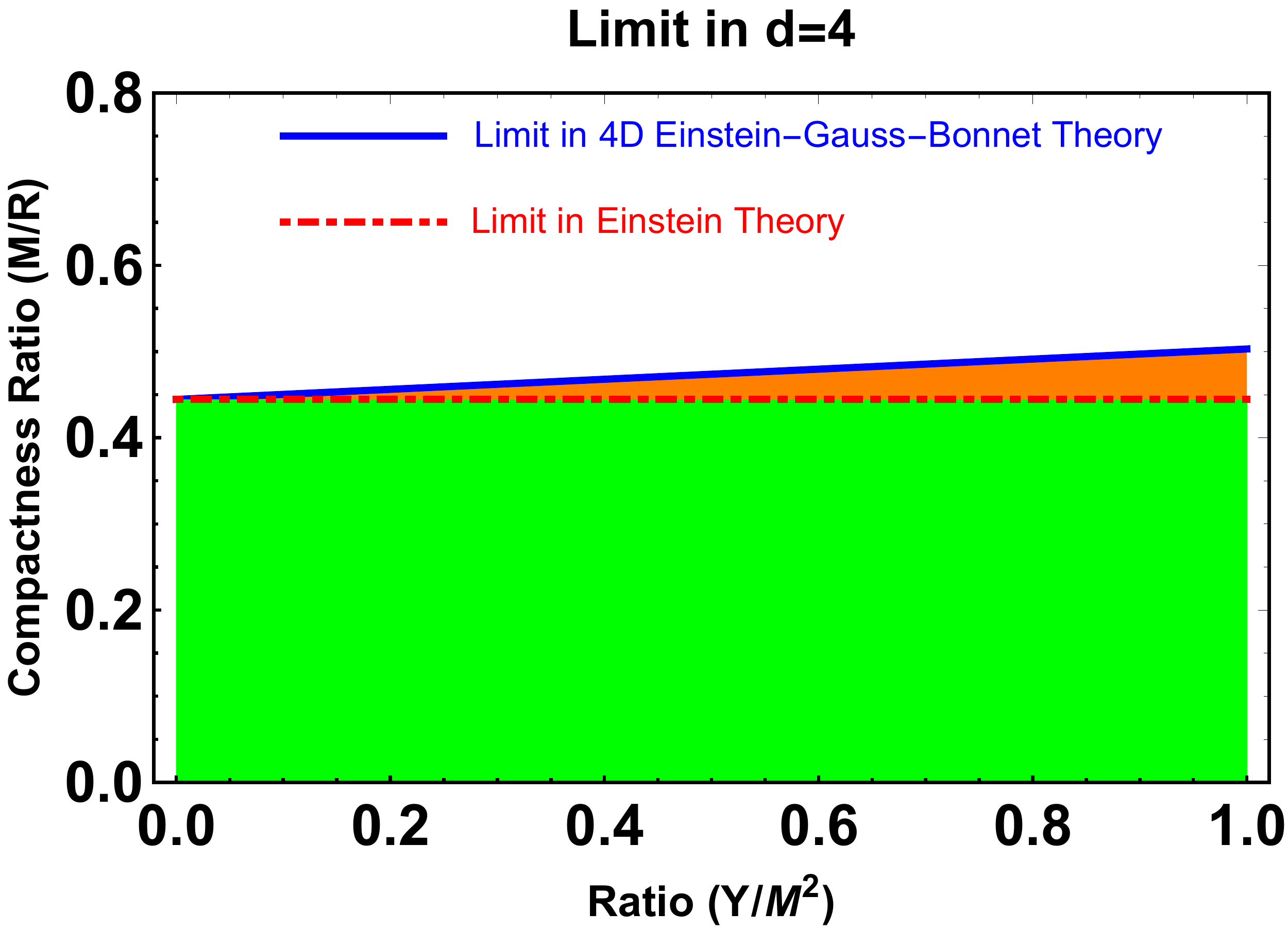}
\caption{The above figures present the allowed parameter space for the existence of a stable stellar structure in four dimensional Einstein-Gauss-Bonnet gravity. The left figure, which is for four dimensional Einstein-Gauss-Bonnet gravity, clearly demonstrates that the minimum stable radius for a stellar structure is always larger than the event horizon. On the other hand, the right figure is for comparison with the corresponding situation in Einstein gravity. The orange zone, at the top of the figure at the right hand side, between blue and red, dot dashed line, depicts the region where discrepancy with general relativity may arise. Here the compactness ratio stands for $(M/R)$.}
\label{Fig_EGB}
\end{figure}

For clarity we have presented the allowed parameter space for the compactness ratio (GM/R), so that a stable stellar object can exist in the four dimensional Einstein-Gauss-Bonnet gravity, in \ref{Fig_EGB}. As the plots suggest, radius of the stellar object remains larger than the event horizon, or equivalently the compactness ratio of the stellar object remains smaller than the event horizon. Further, for larger mass, i.e., smaller $(\Upsilon/M^{2})$ ratio, the difference of the four dimensional Einstein-Gauss-Bonnet gravity with general relativity is smaller. Only for small mass stellar objects (or, larger $(\Upsilon/M^{2})$ ratio) there can be some measurable departure from general relativity (see \ref{Fig_EGB}).

%%%%%%%%%%%%%%%%%%%%%%%
%%%%%%%%%%%%%%%%%%%%%%%
%%%%%%%%%%%%%%%%%%%%%%%
%%%%%%%%%%%%%%%%%%%%%%%
%%%%%%%%%%%%%%%%%%%%%%%
%%%%%%%%%%%%%%%%%%%%%%%
%%%%%%%%%%%%%%%%%%%%%%%
%%%%%%%%%%%%%%%%%%%%%%%

We would like to emphasize that the above inequality was derived assuming the interior of the star to have uniform density, which is an idealized situation. In general,  the density should be a function of the radial coordinate as well, i.e., the density should be non-uniform within the stellar structure. In the next section we show that even with non-uniform density the same limit on the ratio of mass and radius of the star can be obtained. 

%%%%%%%%%%%%%%%%%%%%%%%%%%%%%%%%%%%%%%%%%%%%%%%%%%
%%%%%%%%%%%%%%%%%%%%%%%%%%%%%%%%%%%%%%%%%%%%%%%%%%
%%%%%%%%%%%%%%%%%%%%%%%%%%%%%%%%%%%%%%%%%%%%%%%%%%
\subsection{Interior solution with non-uniform density}

Let us generalize the result we have derived in the previous section for uniform density to the case of non-uniform density as well. The temporal part of the differential equation can be solved for the radial component of the interior metric, $e^{-\lambda_{\rm int}}$ in closed form, which is given by \ref{solution_genn}.  Arising out of which we have the following result,
%%%%%%%%%%%%%%%%%%%%%%%%%%%%%%%%%%%%%%%%%%%%%%%%%%%%%%%%%
\begin{align}
\frac{d}{dr}\left[\frac{\left(1-e^{-\lambda_{\rm int}}\right)}{2r^{2}}\right]=\frac{1}{\sqrt{1+\frac{4\Upsilon Gm(r)}{r^{3}}}}\frac{d}{dr}\left(\frac{Gm(r)}{r^{3}}\right)~.
\end{align}
%%%%%%%%%%%%%%%%%%%%%%%%%%%%%%%%%%%%%%%%%%%%%%%%%%%%%%%%%
The quantity $m(r)/r^{3}$ can be interpreted as the average density of the fluid material forming the interior of the stellar structure, which in normal circumstances will decrease in the outward direction. Thus one can argue that $m(r)/r^{3}$ will also decrease with increase in $r$, therefore the left hand side of \ref{reduced_eq_02} will be negative (since the term multiplying it is necessarily positive, as $\Upsilon>0$). Thus the left hand side of \ref{reduced_eq_02} is negative, which in turn imposes the following inequality 
%%%%%%%%%%%%%%%%%%%%%%%%%%%%%%%%%%%%%%%%%%%%%%%%%%%%%%%%%
\begin{align}\label{final_eq_02}
e^{-(\nu_{\rm int}+\lambda_{\rm int})/2}&\frac{d}{dr}\left[\frac{1}{r}e^{-\lambda_{\rm int}/2}\frac{d}{dr}e^{\nu_{\rm int}/2}\right]\leq 0~,
\end{align}
%%%%%%%%%%%%%%%%%%%%%%%%%%%%%%%%%%%%%%%%%%%%%%%%%%%%%%%%%
provided the following two conditions hold,
%%%%%%%%%%%%%%%%%%%%%%%%%%%%%%%%%%%%%%%%%%%%%%%%%%%%%%%%%
\begin{align}
\left(1- \frac{\Upsilon}{r^{2}}\left[r\nu_{\rm int}'e^{-\lambda_{\rm int}}+(1-e^{-\lambda_{\rm int}})\right]\right)>0~;
\quad \textrm{as well as} \quad 
\left(1+ \frac{\Upsilon (1-e^{-\lambda_{\rm int}})}{r^{2}}\right)>0~.
\end{align}
%%%%%%%%%%%%%%%%%%%%%%%%%%%%%%%%%%%%%%%%%%%%%%%%%%%%%%%%%
Since we are interested with positive $\Upsilon$ and ultimately we will consider the case of small $\Upsilon$, the above conditions are expected to hold. Thus for our purpose, the inequality presented by \ref{final_eq_02} will certainly hold true. One can now integrate \ref{final_eq_02} from the surface of the star at $r=R$ to a radius $r$ in the interior of the star, yielding,
%%%%%%%%%%%%%%%%%%%%%%%%%%%%%%%%%%%%%%%%%%%%%%%%%%%%%%%%%
\begin{align}\label{final_eq_03}
\frac{1}{r}e^{-\lambda_{\rm int}/2}\frac{d}{dr}e^{\nu_{\rm int}/2}\geq \frac{1}{R}e^{-\lambda_{\rm int}(R)/2}\left(\frac{d}{dr}e^{\nu_{\rm int}/2}\right)_{r=R}~,
\end{align}
%%%%%%%%%%%%%%%%%%%%%%%%%%%%%%%%%%%%%%%%%%%%%%%%%%%%%%%%%
where the terms on the right hand side have been evaluated at the surface of the star. Since $e^{\nu}$ and its first derivative must be continuous at the surface of the star, one can compute the derivative of the interior metric element $e^{\nu_{\rm int}}$ at the surface of the star by using the exterior metric element, which is known. From a straightforward calculation, it follows that,
%%%%%%%%%%%%%%%%%%%%%%%%%%%%%%%%%%%%%%%%%%%%%%%%%%%%%%%%%
\begin{align}
\left(\frac{d}{dr}e^{\nu_{\rm int}/2}\right)_{r=R}=\frac{R}{\Upsilon e^{-\lambda_{\rm s}/2}}\left[1-\frac{1+\frac{\Upsilon GM}{R^{3}}}{\sqrt{1+\frac{4\Upsilon GM}{R^{3}}}}\right]~.
\end{align}
%%%%%%%%%%%%%%%%%%%%%%%%%%%%%%%%%%%%%%%%%%%%%%%%%%%%%%%%%
Thus the inequality presented in \ref{final_eq_03} takes the following form,
%%%%%%%%%%%%%%%%%%%%%%%%%%%%%%%%%%%%%%%%%%%%%%%%%%%%%%%%%
\begin{align}\label{final_eq_04}
\frac{1}{r}e^{-\lambda_{\rm int}/2}\frac{d}{dr}e^{\nu_{\rm int}/2}\geq \frac{1}{\Upsilon}\left[1-\frac{1+\frac{\Upsilon GM}{R^{3}}}{\sqrt{1+\frac{4\Upsilon GM}{R^{3}}}}\right]~.
\end{align}
%%%%%%%%%%%%%%%%%%%%%%%%%%%%%%%%%%%%%%%%%%%%%%%%%%%%%%%%%
A simple calculation will assure one about the correct general relativity limit of the above equation by taking the limit $\Upsilon \rightarrow 0$ in a careful manner. Integrating \ref{final_eq_04} from the surface of the star to the centre of the star, we obtain, 
%%%%%%%%%%%%%%%%%%%%%%%%%%%%%%%%%%%%%%%%%%%%%%%%%%%%%%%%%
\begin{align}
e^{\nu_{\rm int}/2}(r=R)-e^{\nu_{\rm int}/2}(r=0)\geq \int _{0}^{R}dr ~re^{\lambda/2}~\frac{1}{\Upsilon}\left[1-\frac{1+\frac{\Upsilon GM}{R^{3}}}{\sqrt{1+\frac{4\Upsilon GM}{R^{3}}}}\right]~.
\end{align}
%%%%%%%%%%%%%%%%%%%%%%%%%%%%%%%%%%%%%%%%%%%%%%%%%%%%%%%%%
Using the continuity of the metric elements across the surface of the star, the above inequality can also be rewritten in the following form,
%%%%%%%%%%%%%%%%%%%%%%%%%%%%%%%%%%%%%%%%%%%%%%%%%%%%%%%%%
\begin{align}
e^{\nu_{\rm int}/2}(r=0)&\leq \sqrt{1+\frac{R^{2}}{\Upsilon}\left[1-\sqrt{1+\frac{4\Upsilon GM}{R^{3}}}\right]}
\nonumber
\\
&-\frac{1}{\Upsilon}\left[1-\frac{1+\frac{\Upsilon GM}{R^{3}}}{\sqrt{1+\frac{4\Upsilon GM}{R^{3}}}}\right]\int _{0}^{R}dr ~\frac{r}{\sqrt{1+\frac{r^{2}}{\Upsilon}\left[1-\sqrt{1+\frac{4\Upsilon Gm(r)}{r^{3}}}\right]}}~.
\end{align}
%%%%%%%%%%%%%%%%%%%%%%%%%%%%%%%%%%%%%%%%%%%%%%%%%%%%%%%%%
Since the average density decreases outward, then it follows that the following inequality $(m(r)/r^{3})>(M/R^{3})$ must hold, where $M$ is the total mass and $R$ is the radius of the sphere. Use of the above condition, makes the inequality presented in the previous equation stronger and hence we have, 
%%%%%%%%%%%%%%%%%%%%%%%%%%%%%%%%%%%%%%%%%%%%%%%%%%%%%%%%%
\begin{align}\label{final_eq_05}
e^{\nu_{\rm int}/2}(r=0)&\leq \sqrt{1+\frac{R^{2}}{\Upsilon}\left[1-\sqrt{1+\frac{4\Upsilon GM}{R^{3}}}\right]}
-\frac{1}{\Upsilon}\left[1-\frac{1+\frac{\Upsilon GM}{R^{3}}}{\sqrt{1+\frac{4\Upsilon GM}{R^{3}}}}\right]\int _{0}^{R}dr ~\frac{r}{\sqrt{1-\mu r^{2}}}~,
\end{align}
%%%%%%%%%%%%%%%%%%%%%%%%%%%%%%%%%%%%%%%%%%%%%%%%%%%%%%%%%
where, $\mu$ has been defined in \ref{solution_int_01}. Integrating \ref{final_eq_05} and manipulating the resulting equation appropriately, we obtain, 
%%%%%%%%%%%%%%%%%%%%%%%%%%%%%%%%%%%%%%%%%%%%%%%%%%%%%%%%%
\begin{align}\label{final_eq_06}
e^{\nu_{\rm int}/2}(r=0)\leq \frac{3}{2}\sqrt{1-\mu R^{2}}\left(\frac{1+(\mu \Upsilon/2)}{1+\mu \Upsilon}\right)-\frac{1}{2}\left[\frac{1-(\mu \Upsilon/2)}{1+\mu \Upsilon}\right]~.
\end{align}
%%%%%%%%%%%%%%%%%%%%%%%%%%%%%%%%%%%%%%%%%%%%%%%%%%%%%%%%%
Since the metric elements should be finite everywhere and the signature convention for the metric must be respected, it follows that $e^{\nu_{\rm int}/2}(r=0)>0$. Thus applying this condition to \ref{final_eq_06}, we obtain, 
%%%%%%%%%%%%%%%%%%%%%%%%%%%%%%%%%%%%%%%%%%%%%%%%%%%%%%%%%
\begin{align}\label{final_eq_07}
\sqrt{1-\mu R^{2}}\left(1+\frac{\mu \Upsilon}{2}\right)\geq \frac{1}{3}\left(1-\frac{\mu \Upsilon}{2}\right)
\end{align}
%%%%%%%%%%%%%%%%%%%%%%%%%%%%%%%%%%%%%%%%%%%%%%%%%%%%%%%%%
This is same as \ref{limit_01}, which we had derived in the context of uniform density star. The rest of the analysis will exactly parallel the one presented in the case of uniform density star and thus ultimately we will arrive at an identical limit on the compactness ratio $(M/R)$ as in \ref{final_limit}. Hence the limit on the stellar structure, derived in \ref{final_limit}, is indeed valid, irrespective of the constitutes of the star, as long as spherical symmetry and isotropy is respected.  

%%%%%%%%%%%%%%%%%%%%%%%%%%%%%%%%%%%%%%%%%%%%%%%%%%
%%%%%%%%%%%%%%%%%%%%%%%%%%%%%%%%%%%%%%%%%%%%%%%%%%
%%%%%%%%%%%%%%%%%%%%%%%%%%%%%%%%%%%%%%%%%%%%%%%%%%
\subsection{Can a similar situation exist within pure Lovelock gravity?}

The above discussion about four dimensional Einstein-Gauss-Bonnet gravity prompts one to search for similar situation in pure Lovelock theories as well. We know that a pure Lovelock theory of order $N$ does not contribute to the field equations in spacetime dimension $d=2N$, thanks to the factor $(d-2N)$ appearing in the gravitational field equations, see \ref{gen_eq_love_01} and \ref{gen_eq_love_02} respectively. Thus one may ask, can there be any non-trivial static and spherically symmetric solution in pure Lovelock theories as we consider the limit $d\rightarrow 2N$, such that $(d-2N)\kappa_{N}=\textrm{constant}$. To see the consequence, let us express \ref{gen_eq_love_01} for $N$th order pure Lovelock gravity with a perfect fluid source in the following form,
%%%%%%%%%%%%%%%%%%%%%%%%%%%%%%%%%%%%%%%%%%%%%%%%%%%%%%%%%
\begin{align}
\left[(d-2)(d-3)\cdots (d-2N)\right]\kappa_{N}\frac{(1-e^{-\lambda})^{N-1}}{2r^{2N}}\left[Nr\lambda'e^{-\lambda}+\left(d-2N-1\right)\left(1-e^{-\lambda}\right)\right]=\frac{1}{2}\rho~.
\end{align}
%%%%%%%%%%%%%%%%%%%%%%%%%%%%%%%%%%%%%%%%%%%%%%%%%%%%%%%%%
Thus the contribution from $N$th order pure Lovelock gravity vanishes in spacetime dimension $d=2N$, if we keep $\kappa_{N}$ finite, as expected. However, following the limiting procedure pointed out in \cite{Glavan:2019inb}, if consider the limit $d\rightarrow 2N$, while keeping the quantity $(d-2N)\kappa_{N}\equiv \beta_{N}$ finite, then the above equation becomes,
%%%%%%%%%%%%%%%%%%%%%%%%%%%%%%%%%%%%%%%%%%%%%%%%%%%%%%%%%
\begin{align}
\left[\frac{N}{r}\lambda'e^{-\lambda}\left(1-e^{-\lambda}\right)^{N-1}-\frac{\left(1-e^{-\lambda}\right)^{N}}{r^{2}}\right]
&=\frac{2^{N}}{\gamma_{N}}\times\left(4\pi \rho~r^{2N-2}\right)~,
\end{align}
%%%%%%%%%%%%%%%%%%%%%%%%%%%%%%%%%%%%%%%%%%%%%%%%%%%%%%%%%
where, the coupling constant $\gamma_{N}$ is defined in terms of the original coupling $\beta_{N}$ as, 
%%%%%%%%%%%%%%%%%%%%%%%%%%%%%%%%%%%%%%%%%%%%%%%%%%%%%%%%%
\begin{align}
\frac{\gamma_{N}}{16\pi}=\{(d-2)(d-3)\cdots (d-2N+1)\}2^{N-2}\beta_{N}~.
\end{align}
%%%%%%%%%%%%%%%%%%%%%%%%%%%%%%%%%%%%%%%%%%%%%%%%%%%%%%%%%
The left hand side of the temporal part of the gravitational field equation, presented above, can be expressed as a total derivative term, which can be integrated, leading to the following expression for the metric element,
%%%%%%%%%%%%%%%%%%%%%%%%%%%%%%%%%%%%%%%%%%%%%%%%%%%%%%%%%
\begin{align}
e^{-\lambda}=1-\left(\frac{2^{N}M}{\gamma_{N}}\right)^{1/N}r^{1/N}~.
\end{align}
%%%%%%%%%%%%%%%%%%%%%%%%%%%%%%%%%%%%%%%%%%%%%%%%%%%%%%%%%
Thus for $N$th order pure Lovelock theory one cannot obtain a static and spherically symmetric solution in $d=2N$ dimension, which is asymptotically flat. This shows that somehow pure Lovelock theories do not such a limiting procedure to obtain any non-trivial solution in spacetime dimensions $d=2N$. This acts as another discerning feature of $N$th order pure Lovelock theories. To reiterate, just as pure Einstein gravity in two dimensions is vacuous, the pure Lovelock theory in $d=2N$ dimension is also not of any physical significance, even when appropriate limiting procedure to keep $(d-2N)\kappa_{N}$ as a constant has been used. Thus we can safely conclude, unlike the case of Einstein-Lovelock theories, where the limiting procedure, $d\rightarrow 2N$, while keeping $(d-2N)\kappa_{N}$ finite, may lead to non-trivial asymptotically flat solution, such a procedure does not work for pure lovelock theories. 

%%%%%%%%%%%%%%%%%%%%%%%%%%%%%%%%%%%%%%%%%%%%%%%%%%%%%%%%%%%%%%%%%%%%%%%%%%%%%%%%%%%%%%%%%%%%%%%%%%%
%%%%%%%%%%%%%%%%%%%%%%%%%%%%%%%%%%%%%%%%%%%%%%%%%%%%%%%%%%%%%%%%%%%%%%%%%%%%%%%%%%%%%%%%%%%%%%%%%%%
%%%%%%%%%%%%%%%%%%%%%%%%%%%%%%%%%%%%%%%%%%%%%%%%%%%%%%%%%%%%%%%%%%%%%%%%%%%%%%%%%%%%%%%%%%%%%%%%%%%
\section{Concluding Remarks}

Stability of stellar structure is one of the most important requirement for feasibility of a gravitational theory to be realized in nature. Even though in general contexts it is difficult to determine such criterion for stability of a stellar structure, surprisingly with reasonable assumptions one can indeed make significant progress and determine a bound on the compactness ratio of the stellar object. The bound on the compactness ratio tells us that given a certain mass for the stellar object, its radius cannot be smaller than certain value predicted by the upper limit of the above mentioned bound. For example, in the context of general relativity, the above bound on the stellar structure corresponds to $(M/R)<(4/9)$, assuming the matter distribution to be isotropic and matter density inside the stellar structure to be decreasing as one progresses to the outside region of the stellar structure. It turns out that the bound on the stellar structure gets modified if it inherits electric charge.

Motivated by the interesting and desirable properties of pure Lovelock theories in higher spacetime dimensions we have studied the above bound for stellar structures in pure Lovelock theories with or without the Maxwell field. To keep the analysis general we have considered all possible means of arriving at the bound on charged stellar structures. To begin with, we have expressed the gravitational field equations in pure Lovelock theories of arbitrary order $N$, in arbitrary spacetime dimension $d$, in presence of the Maxwell field. It turns out that in the context of static and spherical symmetry, the temporal part of the field equation can be explicitly solved, yielding the radial metric component $e^{\lambda}$, both inside and outside the stellar object. Due to the presence of additional energy from the Maxwell field, the gravitational mass as experienced by an observer outside the stellar structure differs from the inertial mass. Since the external observer experiences the gravitational mass, rather than the inertial mass, the bound we obtain is also in terms of the gravitational mass of the stellar object. By manipulating these gravitational field equations, along with the conservation relation for the matter energy momentum tensor, we have been able to arrive at the desired bound, both for a charged shell and a charged stellar object. Interestingly, both the limits reduce to the general relativistic counterparts in four dimensions derived in earlier literatures. Through these bounds we have been able to provide the limit on the stellar structures inheriting the Maxwell field for arbitrary orders of pure Lovelock theories and in arbitrary spacetime dimensions. For example, we have explicitly demonstrated the case for a $d$ dimensional Einstein gravity and seven dimensional pure Gauss-Bonnet theory. As the trends indicate, with increase of the dimensionless charge ratio $(Q/M)\mathcal{M}^{1-N}$, where $M=\mathcal{M}^{d-2N-1}$, the radius of the star decreases and hence the compactness ratio $(\mathcal{M}/R)$ increases. Also with the increase of spacetime dimensions, the compactness ratio increases, this is because gravity becomes stronger in higher spacetime dimensions. This behaviour of the compactness ratio holds true for a charged shell and also for a charged sphere in pure Lovelock theories. Interestingly, the degeneracy present between pure Lovelock theories in $d=3N+1$ dimensions can be broken using Maxwell field and thus the bound on compactness ratio becomes different for different pure Lovelock theories in $d=3N+1$ dimensions. 

As an aside, we have also discussed the case of four-dimensional Einstein-Gauss-Bonnet theories (with $\alpha$ being the Gauss-Bonnet coupling parameter) which are obtained by taking $d\rightarrow 4$ limit, while keeping $(d-4)\alpha$ to be a constant. Even though this limiting procedure can have several disadvantages, e.g., it renders the action formulation ill-posed, we have explored whether stable structures can exist within this gravity theory by using the techniques developed earlier in this paper. The compatibility of the results with the charged case follows from the fact that the four dimensional static and spherically symmetric black hole in Einstein-Gauss-Bonnet gravity changes the spacetime structure radically and introduces Cauchy horizons. This is akin to the introduction of an electric charge. As our results explicitly demonstrate, the effect of the redefined Gauss-Bonnet coupling parameter has an effect on the stellar structure, identical to that of an electric charge. Here also with an increase of the dimensionless coupling parameter $(\Upsilon/M^{2})$, the compactness ratio $(M/R)$ increases, i.e., the radius of the stellar object decreases. Finally, we have shown that in pure Lovelock theories such a limiting procedure does not lead to any meaningful static and spherically symmetric solution and hence is free from any ambiguities and complications arising out of this dimensional reduction. 

%%%%%%%%%%%%%%%%%%%%%%%%%%%%%%%%%%%%%%%%%%%%%%%%%%%%%%%%%%%%%%%%%%%%%%%%%%%%%%%%%%%%%%%%%%%%%%%%%%%
%%%%%%%%%%%%%%%%%%%%%%%%%%%%%%%%%%%%%%%%%%%%%%%%%%%%%%%%%%%%%%%%%%%%%%%%%%%%%%%%%%%%%%%%%%%%%%%%%%%
%%%%%%%%%%%%%%%%%%%%%%%%%%%%%%%%%%%%%%%%%%%%%%%%%%%%%%%%%%%%%%%%%%%%%%%%%%%%%%%%%%%%%%%%%%%%%%%%%%%
\section*{Acknowledgements}

Research of S.C. is funded by a INSPIRE Faculty fellowship from Department of Science and Technology, Government of India (Reg. No. DST/INSPIRE/04/2018/000893). SC thanks AEI Potsdam for hospitality towards the initial stages of this project. 
%%%%%%%%%%%%%%%%%%%%%%%%%%%%%%%%%%%%%%%%%%%%%%%%%%%%%%%%%%%%%%%%%%%%%%%%%%%%%%%%%%%%%%%%%%%%%%%
%%%%%%%%%%%%%%%%%%%%%%%%%%%%%%%%%%%%%%%%%%%%%%%%%%%%%%%%%%%%%%%%%%%%%%%%%%%%%%%%%%%%%%%%%%%%%%%
%%%%%%%%%%%%%%%%%%%%%%%%%%%%%%%%%%%%%%%%%%%%%%%%%%%%%%%%%%%%%%%%%%%%%%%%%%%%%%%%%%%%%%%%%%%%%%%
\appendix
\labelformat{section}{Appendix #1} 
\labelformat{subsection}{Appendix #1}
\numberwithin{equation}{section}
%%%%%%%%%%%%%%%%%%%%%%%%%%%%%%%%%%%%%%%%%%%%%%%%%%%%%%%%%%%%%%%%%%%%%%%%%%%%%%%%%%%%%%%%%%%%%%%%%%%
%%%%%%%%%%%%%%%%%%%%%%%%%%%%%%%%%%%%%%%%%%%%%%%%%%%%%%%%%%%%%%%%%%%%%%%%%%%%%%%%%%%%%%%%%%%%%%%%%%%
%%%%%%%%%%%%%%%%%%%%%%%%%%%%%%%%%%%%%%%%%%%%%%%%%%%%%%%%%%%%%%%%%%%%%%%%%%%%%%%%%%%%%%%%%%%%%%%%%%%
\section{Derivation of the interior solution for four dimensional Einstein-Gauss-Bonnet gravity}\label{AppA}

In this section we will derive \ref{reduced_eq_02} in the context of four dimensional Einstein-Gauss-Bonnet gravity, which will be central to our analysis of the stability of stellar structures in this model. For this purpose, we start by differentiating the pressure equation, presented in \ref{pressure_eq_01}, which reads,
%%%%%%%%%%%%%%%%%%%%%%%%%%%%%%%%%%%%%%%%%%%%%%%%%%%%%%%%%
\begin{align}
8\pi G \frac{dp}{dr}&=\frac{d}{dr}\left[\frac{1}{r^{2}}\left\{r\nu'e^{-\lambda}-\left(1-e^{-\lambda}\right)\right\}
+16\pi G\beta\frac{(1-e^{-\lambda})}{r^{4}}\left\{2r\nu'e^{-\lambda}+\left(1-e^{-\lambda}\right)\right\} \right]
\nonumber
\\
&=-\frac{2}{r^{3}}\left\{r\nu'e^{-\lambda}-\left(1-e^{-\lambda}\right)\right\}+\frac{1}{r^{2}}\left\{r\nu''e^{-\lambda}+\nu'e^{-\lambda}-r\nu'\lambda'e^{-\lambda}
-\lambda'e^{-\lambda}\right\}
\nonumber
\\
&+16\pi G\beta\Big[-\frac{4(1-e^{-\lambda})}{r^{5}}\left\{2r\nu'e^{-\lambda}+\left(1-e^{-\lambda}\right)\right\}
\nonumber
\\
&+\frac{(1-e^{-\lambda})}{r^{4}}\left\{2r\nu''e^{-\lambda}+2\nu'e^{-\lambda}-2r\nu'\lambda'e^{-\lambda}+\lambda'e^{-\lambda}\right\}
\nonumber
\\
&\hskip 2cm+\frac{\lambda'e^{-\lambda}}{r^{4}}\left\{2r\nu'e^{-\lambda}+\left(1-e^{-\lambda}\right)\right\}\Big]~.
\end{align}
%%%%%%%%%%%%%%%%%%%%%%%%%%%%%%%%%%%%%%%%%%%%%%%%%%%%%%%%%
Rearranging terms, the above equation can also be expressed as, 
%%%%%%%%%%%%%%%%%%%%%%%%%%%%%%%%%%%%%%%%%%%%%%%%%%%%%%%%%
\begin{align}
16\pi G r^{2}\frac{dp}{dr}&=-\frac{4}{r}\left\{r\nu'e^{-\lambda}-\left(1-e^{-\lambda}\right)\right\}+\left\{2r\nu''e^{-\lambda}+2\nu'e^{-\lambda}-2r\nu'\lambda'e^{-\lambda}
-2\lambda'e^{-\lambda}\right\}
\nonumber
\\
&+16\pi G\beta\Big[-\frac{8(1-e^{-\lambda})}{r^{3}}\left\{2r\nu'e^{-\lambda}+\left(1-e^{-\lambda}\right)\right\}
\nonumber
\\
&+\frac{2(1-e^{-\lambda})}{r^{2}}\left\{2r\nu''e^{-\lambda}+2\nu'e^{-\lambda}-2r\nu'\lambda'e^{-\lambda}+\lambda'e^{-\lambda}\right\}
\nonumber
\\
&\hskip 2cm+\frac{2\lambda'e^{-\lambda}}{r^{2}}\left\{2r\nu'e^{-\lambda}+\left(1-e^{-\lambda}\right)\right\}\Big]~.
\end{align}
%%%%%%%%%%%%%%%%%%%%%%%%%%%%%%%%%%%%%%%%%%%%%%%%%%%%%%%%%
Using the conservation equation, i.e., \ref{conservation}, we obtain, $16\pi Gr^{2}(dp/dr)=-8\pi G\nu'r^{2}(\rho+p)$. Thus replacing the left hand side of the above equation with $-8\pi G\nu'r^{2}(\rho+p)$ and using \ref{sum_eq_01}, arising out of addition of radial and temporal part of the gravitational field equations, we obtain,
%%%%%%%%%%%%%%%%%%%%%%%%%%%%%%%%%%%%%%%%%%%%%%%%%%%%%%%%%
\begin{align}
&-\frac{4}{r}\left\{r\nu'e^{-\lambda}-\left(1-e^{-\lambda}\right)\right\}+\left\{2r\nu''e^{-\lambda}+2\nu'e^{-\lambda}-2r\nu'\lambda'e^{-\lambda}
-2\lambda'e^{-\lambda}\right\}
\nonumber
\\
&+16\pi G\beta\Big[-\frac{8(1-e^{-\lambda})}{r^{3}}\left\{2r\nu'e^{-\lambda}+\left(1-e^{-\lambda}\right)\right\}
+\frac{2(1-e^{-\lambda})}{r^{2}}\left\{2r\nu''e^{-\lambda}+2\nu'e^{-\lambda}-2r\nu'\lambda'e^{-\lambda}
+\lambda'e^{-\lambda}\right\}
\nonumber
\\
&+\frac{2\lambda'e^{-\lambda}}{r^{2}}\left\{2r\nu'e^{-\lambda}+\left(1-e^{-\lambda}\right)\right\}\Big]
+r\nu'e^{-\lambda}\left(\lambda'+\nu'\right)+\frac{32\pi G\beta}{r} \nu'e^{-\lambda}(1-e^{-\lambda})\left(\lambda'+\nu'\right)=0~.
\end{align}
%%%%%%%%%%%%%%%%%%%%%%%%%%%%%%%%%%%%%%%%%%%%%%%%%%%%%%%%%
This equation can be rearranged in the following manner,
%%%%%%%%%%%%%%%%%%%%%%%%%%%%%%%%%%%%%%%%%%%%%%%%%%%%%%%%%
\begin{align}\label{reduced_eq_01n}
&-\frac{2}{r}e^{-\lambda}\left[-2\left(e^{\lambda}-1\right)+r\lambda'\right]+e^{-\lambda}\left\{2r\nu''+r\nu'^{2}-2\nu'-r\nu'\lambda'\right\}
\nonumber
\\
&+\frac{32\pi G\beta (1-e^{-\lambda})}{r^{2}}\Bigg[\frac{2}{r}e^{-\lambda}\left\{r\lambda'-2\left(e^{\lambda}-1\right)\right\}
+e^{-\lambda}\left\{2r\nu''+r\nu'^{2}-2\nu'-r\nu'\lambda'
\right\}-4\nu'e^{-\lambda}
\nonumber
\\
&\hskip 3cm+\frac{2r\nu'\lambda'e^{-2\lambda}}{(1-e^{-\lambda})}\Bigg]=0~.
\end{align}
%%%%%%%%%%%%%%%%%%%%%%%%%%%%%%%%%%%%%%%%%%%%%%%%%%%%%%%%%
Using the following two identities,
%%%%%%%%%%%%%%%%%%%%%%%%%%%%%%%%%%%%%%%%%%%%%%%%%%%%%%%%%
\begin{align}
\frac{d}{dr}\left[\frac{1}{r}e^{-\lambda/2}\frac{d}{dr}e^{\nu/2}\right]&=\frac{e^{(\nu-\lambda)/2}}{4r^{2}}\left\{2r\nu''+r\nu'^{2}-2\nu'-r\nu'\lambda'\right\}~,
\label{app_identity_01}
\\
\frac{d}{dr}\left[\frac{1-e^{-\lambda}}{2r^{2}}\right]&=\frac{e^{-\lambda}}{2r^{3}}\left[-2\left(e^{\lambda}-1\right)+r\lambda'\right]~,
\label{app_identity_02}
\end{align}
%%%%%%%%%%%%%%%%%%%%%%%%%%%%%%%%%%%%%%%%%%%%%%%%%%%%%%%%%
we can re-express \ref{reduced_eq_01n}, with $32\pi G\beta=\Upsilon$ as,
%%%%%%%%%%%%%%%%%%%%%%%%%%%%%%%%%%%%%%%%%%%%%%%%%%%%%%%%%
\begin{align}\label{reduced_eq_01}
&-4r^{2}\left(1- \frac{\Upsilon (1-e^{-\lambda})}{r^{2}}\right)\frac{d}{dr}\left[\frac{1-e^{-\lambda}}{2r^{2}}\right]
+\left(1+ \frac{\Upsilon (1-e^{-\lambda})}{r^{2}}\right)e^{-\lambda}\frac{4r^{2}}{e^{(\nu-\lambda)/2}}\frac{d}{dr}\left[\frac{1}{r}e^{-\lambda/2}\frac{d}{dr}e^{\nu/2}\right]
\nonumber
\\
&\hskip 4 cm+\frac{\Upsilon (1-e^{-\lambda})}{r^{2}}\Bigg[-4\nu'e^{-\lambda}+\frac{2r\nu'\lambda'e^{-2\lambda}}{(1-e^{-\lambda})}\Bigg]=0~.
\end{align}
%%%%%%%%%%%%%%%%%%%%%%%%%%%%%%%%%%%%%%%%%%%%%%%%%%%%%%%%%
The last term in the above expression can be expressed as,
%%%%%%%%%%%%%%%%%%%%%%%%%%%%%%%%%%%%%%%%%%%%%%%%%%%%%%%%%
\begin{align}
\frac{\Upsilon (1-e^{-\lambda})}{r^{2}}\Bigg[-4\nu'e^{-\lambda}+\frac{2r\nu'\lambda'e^{-2\lambda}}{(1-e^{-\lambda})}\Bigg]
&=\frac{2\Upsilon \nu' e^{-2\lambda}}{r^{2}}\Bigg[-2(e^{\lambda}-1)+r\lambda'\Bigg]
\nonumber
\\
&=4\Upsilon \nu' e^{-\lambda}r\frac{d}{dr}\left[\frac{1-e^{-\lambda}}{2r^{2}}\right]~.
\end{align}
%%%%%%%%%%%%%%%%%%%%%%%%%%%%%%%%%%%%%%%%%%%%%%%%%%%%%%%%%
Using the above expression in \ref{reduced_eq_01}, we arrive at the desired expression presented in \ref{reduced_eq_02} in the main text.

%%%%%%%%%%%%%%%%%%%%%%%%%%%%%%%%%%%%%%%%%%%%%%%%%%%%%%%%%%%%%%%%%%%%%%%%%%%%%%%%%%%%%%%%%%%%%%%%%%%
%%%%%%%%%%%%%%%%%%%%%%%%%%%%%%%%%%%%%%%%%%%%%%%%%%%%%%%%%%%%%%%%%%%%%%%%%%%%%%%%%%%%%%%%%%%%%%%%%%%
%%%%%%%%%%%%%%%%%%%%%%%%%%%%%%%%%%%%%%%%%%%%%%%%%%%%%%%%%%%%%%%%%%%%%%%%%%%%%%%%%%%%%%%%%%%%%%%%%%%
%Bibliography
\bibliography{References}

\providecommand{\href}[2]{#2}\begingroup\raggedright\begin{thebibliography}{10}

\bibitem{Buchdahl:1959zz}
H.~A. Buchdahl, ``{General Relativistic Fluid Spheres},''
  \href{http://dx.doi.org/10.1103/PhysRev.116.1027}{{\em Phys. Rev.} {\bfseries
  116} (1959) 1027}.

\bibitem{Mak:2001gg}
M.~Mak, J.~Dobson, Peter~N., and T.~Harko, ``{Maximum mass radius ratio for
  compact general relativistic objects in Schwarzschild-de Sitter geometry},''
  \href{http://dx.doi.org/10.1142/S0217732300002723}{{\em Mod. Phys. Lett. A}
  {\bfseries 15} (2000) 2153--2158},
  \href{http://arxiv.org/abs/gr-qc/0104031}{{\ttfamily arXiv:gr-qc/0104031}}.

\bibitem{Stuchlik:2008xe}
Z.~Stuchlik, ``{Spherically Symmetric Static Configurations of Uniform Density
  in Spacetimes with a Non-Zero Cosmological Constant},'' {\em Acta Phys.
  Slov.} {\bfseries 50} (2000) 219--228,
  \href{http://arxiv.org/abs/0803.2530}{{\ttfamily arXiv:0803.2530 [gr-qc]}}.

\bibitem{Andreasson:2007ck}
H.~Andreasson, ``{Sharp bounds on 2m/r of general spherically symmetric static
  objects},'' \href{http://dx.doi.org/10.1016/j.jde.2008.05.010}{{\em J. Diff.
  Eq.} {\bfseries 245} (2008) 2243--2266},
  \href{http://arxiv.org/abs/gr-qc/0702137}{{\ttfamily arXiv:gr-qc/0702137}}.

\bibitem{Karageorgis:2007cy}
P.~Karageorgis and J.~G. Stalker, ``{Sharp bounds on 2m/r for static spherical
  objects},'' \href{http://dx.doi.org/10.1088/0264-9381/25/19/195021}{{\em
  Class. Quant. Grav.} {\bfseries 25} (2008) 195021},
  \href{http://arxiv.org/abs/0707.3632}{{\ttfamily arXiv:0707.3632 [gr-qc]}}.

\bibitem{Goswami:2015dma}
R.~Goswami, S.~D. Maharaj, and A.~M. Nzioki, ``{Buchdahl-Bondi limit in
  modified gravity: Packing extra effective mass in relativistic compact
  stars},'' \href{http://dx.doi.org/10.1103/PhysRevD.92.064002}{{\em Phys. Rev.
  D} {\bfseries 92} (2015) 064002},
  \href{http://arxiv.org/abs/1506.04043}{{\ttfamily arXiv:1506.04043 [gr-qc]}}.

\bibitem{Zarro:2009gd}
C.~A. Zarro, ``{Buchdahl limit for d-dimensional spherical solutions with a
  cosmological constant},''
  \href{http://dx.doi.org/10.1007/s10714-008-0675-8}{{\em Gen. Rel. Grav.}
  {\bfseries 41} (2009) 453--468}.

\bibitem{Barraco:2002ds}
D.~Barraco and V.~H. Hamity, ``{Maximum mass of a spherically symmetric
  isotropic star},'' \href{http://dx.doi.org/10.1103/PhysRevD.65.124028}{{\em
  Phys. Rev. D} {\bfseries 65} (2002) 124028}.

\bibitem{2000JMP....41.4752H}
T.~{Harko} and M.~K. {Mak}, ``{Anisotropic charged fluid spheres in D
  space-time dimensions},'' \href{http://dx.doi.org/10.1063/1.533375}{{\em
  Journal of Mathematical Physics} {\bfseries 41} no.~7, (July, 2000)
  4752--4764}.

\bibitem{Dadhich:2019jyf}
N.~Dadhich, ``{Buchdahl compactness limit and gravitational field energy},''
  \href{http://dx.doi.org/10.1088/1475-7516/2020/04/035}{{\em JCAP} {\bfseries
  04} (2020) 035}, \href{http://arxiv.org/abs/1903.03436}{{\ttfamily
  arXiv:1903.03436 [gr-qc]}}.

\bibitem{Dadhich:2010qh}
N.~Dadhich, A.~Molina, and A.~Khugaev, ``{Uniform density static fluid sphere
  in Einstein-Gauss-Bonnet gravity and its universality},''
  \href{http://dx.doi.org/10.1103/PhysRevD.81.104026}{{\em Phys. Rev. D}
  {\bfseries 81} (2010) 104026},
  \href{http://arxiv.org/abs/1001.3922}{{\ttfamily arXiv:1001.3922 [gr-qc]}}.

\bibitem{Giuliani:2007zza}
A.~Giuliani and T.~Rothman, ``{Absolute stability limit for relativistic
  charged spheres},'' \href{http://dx.doi.org/10.1007/s10714-007-0539-7}{{\em
  Gen. Rel. Grav.} {\bfseries 40} (2008) 1427--1447},
  \href{http://arxiv.org/abs/0705.4452}{{\ttfamily arXiv:0705.4452 [gr-qc]}}.

\bibitem{Boehmer:2007gq}
C.~Boehmer and T.~Harko, ``{Minimum mass-radius ratio for charged gravitational
  objects},'' \href{http://dx.doi.org/10.1007/s10714-007-0417-3}{{\em Gen. Rel.
  Grav.} {\bfseries 39} (2007) 757--775},
  \href{http://arxiv.org/abs/gr-qc/0702078}{{\ttfamily arXiv:gr-qc/0702078}}.

\bibitem{Andreasson:2012dj}
H.~Andreasson, C.~G. Boehmer, and A.~Mussa, ``{Bounds on M/R for Charged
  Objects with positive Cosmological constant},''
  \href{http://dx.doi.org/10.1088/0264-9381/29/9/095012}{{\em Class. Quant.
  Grav.} {\bfseries 29} (2012) 095012},
  \href{http://arxiv.org/abs/1201.5725}{{\ttfamily arXiv:1201.5725 [gr-qc]}}.

\bibitem{Wright:2015yda}
M.~Wright, ``{Buchdahl's inequality in five dimensional Gauss--Bonnet
  gravity},'' \href{http://dx.doi.org/10.1007/s10714-016-2091-9}{{\em Gen. Rel.
  Grav.} {\bfseries 48} no.~7, (2016) 93},
  \href{http://arxiv.org/abs/1507.05560}{{\ttfamily arXiv:1507.05560 [gr-qc]}}.

\bibitem{Dadhich:2015lra}
N.~Dadhich, ``{A distinguishing gravitational property for gravitational
  equation in higher dimensions},''
  \href{http://dx.doi.org/10.1140/epjc/s10052-016-3933-z}{{\em Eur. Phys. J. C}
  {\bfseries 76} no.~3, (2016) 104},
  \href{http://arxiv.org/abs/1506.08764}{{\ttfamily arXiv:1506.08764 [gr-qc]}}.

\bibitem{Dadhich:2012cv}
N.~Dadhich, S.~G. Ghosh, and S.~Jhingan, ``{The Lovelock gravity in the
  critical spacetime dimension},''
  \href{http://dx.doi.org/10.1016/j.physletb.2012.03.084}{{\em Phys. Lett. B}
  {\bfseries 711} (2012) 196--198},
  \href{http://arxiv.org/abs/1202.4575}{{\ttfamily arXiv:1202.4575 [gr-qc]}}.

\bibitem{Camanho:2015hea}
X.~O. Camanho and N.~Dadhich, ``{On Lovelock analogs of the Riemann tensor},''
  \href{http://dx.doi.org/10.1140/epjc/s10052-016-3891-5}{{\em Eur. Phys. J. C}
  {\bfseries 76} no.~3, (2016) 149},
  \href{http://arxiv.org/abs/1503.02889}{{\ttfamily arXiv:1503.02889 [gr-qc]}}.

\bibitem{Chakraborty:2015wma}
S.~Chakraborty, ``{Lanczos-Lovelock gravity from a thermodynamic
  perspective},'' \href{http://dx.doi.org/10.1007/JHEP08(2015)029}{{\em JHEP}
  {\bfseries 08} (2015) 029}, \href{http://arxiv.org/abs/1505.07272}{{\ttfamily
  arXiv:1505.07272 [gr-qc]}}.

\bibitem{Chakraborty:2014rga}
S.~Chakraborty and T.~Padmanabhan, ``{Evolution of Spacetime arises due to the
  departure from Holographic Equipartition in all Lanczos-Lovelock Theories of
  Gravity},'' \href{http://dx.doi.org/10.1103/PhysRevD.90.124017}{{\em Phys.
  Rev. D} {\bfseries 90} no.~12, (2014) 124017},
  \href{http://arxiv.org/abs/1408.4679}{{\ttfamily arXiv:1408.4679 [gr-qc]}}.

\bibitem{Chakraborty:2016qbw}
S.~Chakraborty and N.~Dadhich, ``{1/r potential in higher dimensions},''
  \href{http://dx.doi.org/10.1140/epjc/s10052-018-5546-1}{{\em Eur. Phys. J. C}
  {\bfseries 78} no.~1, (2018) 81},
  \href{http://arxiv.org/abs/1605.01961}{{\ttfamily arXiv:1605.01961 [gr-qc]}}.

\bibitem{Gannouji:2019gnb}
R.~Gannouji, Y.~Rodríguez~Baez, and N.~Dadhich, ``{Pure Lovelock black holes
  in dimensions $d=3N+1$ are stable},''
  \href{http://dx.doi.org/10.1103/PhysRevD.100.084011}{{\em Phys. Rev. D}
  {\bfseries 100} no.~8, (2019) 084011},
  \href{http://arxiv.org/abs/1907.09503}{{\ttfamily arXiv:1907.09503 [gr-qc]}}.

\bibitem{Chakraborty:2015kva}
S.~Chakraborty and N.~Dadhich, ``{Brown-York quasilocal energy in
  Lanczos-Lovelock gravity and black hole horizons},''
  \href{http://dx.doi.org/10.1007/JHEP12(2015)003}{{\em JHEP} {\bfseries 12}
  (2015) 003}, \href{http://arxiv.org/abs/1509.02156}{{\ttfamily
  arXiv:1509.02156 [gr-qc]}}.

\bibitem{Dadhich:1997ze}
N.~Dadhich, ``{Black hole: Equipartition of matter and potential energy},''
  {\em Curr. Sci.} {\bfseries 76} (1999) 831,
  \href{http://arxiv.org/abs/gr-qc/9705037}{{\ttfamily arXiv:gr-qc/9705037}}.

\bibitem{Dadhich:2016fku}
N.~Dadhich and S.~Chakraborty, ``{Buchdahl compactness limit for a pure
  Lovelock static fluid star},''
  \href{http://dx.doi.org/10.1103/PhysRevD.95.064059}{{\em Phys. Rev. D}
  {\bfseries 95} no.~6, (2017) 064059},
  \href{http://arxiv.org/abs/1606.01330}{{\ttfamily arXiv:1606.01330 [gr-qc]}}.

\bibitem{Dadhich:2012pd}
N.~Dadhich, ``{The gravitational equation in higher dimensions},''
  \href{http://dx.doi.org/10.1007/978-3-319-06761-2\_6}{{\em Springer Proc.
  Phys.} {\bfseries 157} (2014) 43--49},
  \href{http://arxiv.org/abs/1210.3022}{{\ttfamily arXiv:1210.3022 [gr-qc]}}.

\bibitem{Dadhich:2016wtb}
N.~Dadhich, S.~Hansraj, and B.~Chilambwe, ``{Compact objects in pure Lovelock
  theory},'' \href{http://dx.doi.org/10.1142/S0218271817500560}{{\em Int. J.
  Mod. Phys. D} {\bfseries 26} no.~06, (2016) 1750056},
  \href{http://arxiv.org/abs/1607.07095}{{\ttfamily arXiv:1607.07095 [gr-qc]}}.

\bibitem{Molina:2016xeu}
A.~Molina, N.~Dadhich, and A.~Khugaev, ``{Buchdahl-Vaidya-Tikekar model for
  stellar interior in pure Lovelock gravity},''
  \href{http://dx.doi.org/10.1007/s10714-017-2259-y}{{\em Gen. Rel. Grav.}
  {\bfseries 49} no.~7, (2017) 96},
  \href{http://arxiv.org/abs/1607.06229}{{\ttfamily arXiv:1607.06229 [gr-qc]}}.

\bibitem{Glavan:2019inb}
D.~z. Glavan and C.~Lin, ``{Einstein-Gauss-Bonnet gravity in 4-dimensional
  space-time},'' \href{http://dx.doi.org/10.1103/PhysRevLett.124.081301}{{\em
  Phys. Rev. Lett.} {\bfseries 124} no.~8, (2020) 081301},
  \href{http://arxiv.org/abs/1905.03601}{{\ttfamily arXiv:1905.03601 [gr-qc]}}.

\bibitem{Dadhich:2020ukj}
N.~Dadhich, ``{On causal structure of $4D$-Einstein-Gauss-Bonnet black hole},''
  \href{http://arxiv.org/abs/2005.05757}{{\ttfamily arXiv:2005.05757 [gr-qc]}}.

\bibitem{Ai:2020peo}
W.-Y. Ai, ``{A note on the novel 4D Einstein-Gauss-Bonnet gravity},''
  \href{http://arxiv.org/abs/2004.02858}{{\ttfamily arXiv:2004.02858 [gr-qc]}}.

\bibitem{Gurses:2020ofy}
M.~Gurses, T.~C. Sisman, and B.~Tekin, ``{Is there a novel
  Einstein-Gauss-Bonnet theory in four dimensions?},''
  \href{http://arxiv.org/abs/2004.03390}{{\ttfamily arXiv:2004.03390 [gr-qc]}}.

\bibitem{Ge:2020tid}
X.-H. Ge and S.-J. Sin, ``{Causality of black holes in 4-dimensional
  Einstein-Gauss-Bonnet-Maxwell theory},''
  \href{http://arxiv.org/abs/2004.12191}{{\ttfamily arXiv:2004.12191
  [hep-th]}}.

\bibitem{Parattu:2015gga}
K.~Parattu, S.~Chakraborty, B.~R. Majhi, and T.~Padmanabhan, ``{A Boundary Term
  for the Gravitational Action with Null Boundaries},''
  \href{http://dx.doi.org/10.1007/s10714-016-2093-7}{{\em Gen. Rel. Grav.}
  {\bfseries 48} no.~7, (2016) 94},
  \href{http://arxiv.org/abs/1501.01053}{{\ttfamily arXiv:1501.01053 [gr-qc]}}.

\bibitem{Chakraborty:2016yna}
S.~Chakraborty, {\em {Boundary Terms of the Einstein--Hilbert Action}},
  vol.~187, \href{http://dx.doi.org/10.1007/978-3-319-51700-1\_5}{pp.~43--59}.
\newblock 2017.
\newblock \href{http://arxiv.org/abs/1607.05986}{{\ttfamily arXiv:1607.05986
  [gr-qc]}}.

\bibitem{Mishra:2019ged}
A.~K. Mishra and S.~Chakraborty, ``{Strong Cosmic Censorship in higher
  curvature gravity},''
  \href{http://dx.doi.org/10.1103/PhysRevD.101.064041}{{\em Phys. Rev. D}
  {\bfseries 101} no.~6, (2020) 064041},
  \href{http://arxiv.org/abs/1911.09855}{{\ttfamily arXiv:1911.09855 [gr-qc]}}.

\bibitem{Rahman:2018oso}
M.~Rahman, S.~Chakraborty, S.~SenGupta, and A.~A. Sen, ``{Fate of Strong Cosmic
  Censorship Conjecture in Presence of Higher Spacetime Dimensions},''
  \href{http://dx.doi.org/10.1007/JHEP03(2019)178}{{\em JHEP} {\bfseries 03}
  (2019) 178}, \href{http://arxiv.org/abs/1811.08538}{{\ttfamily
  arXiv:1811.08538 [gr-qc]}}.

\bibitem{Dias:2018ynt}
O.~J. Dias, F.~C. Eperon, H.~S. Reall, and J.~E. Santos, ``{Strong cosmic
  censorship in de Sitter space},''
  \href{http://dx.doi.org/10.1103/PhysRevD.97.104060}{{\em Phys. Rev. D}
  {\bfseries 97} no.~10, (2018) 104060},
  \href{http://arxiv.org/abs/1801.09694}{{\ttfamily arXiv:1801.09694 [gr-qc]}}.

\bibitem{Cardoso:2017soq}
V.~Cardoso, J.~L. Costa, K.~Destounis, P.~Hintz, and A.~Jansen, ``{Quasinormal
  modes and Strong Cosmic Censorship},''
  \href{http://dx.doi.org/10.1103/PhysRevLett.120.031103}{{\em Phys. Rev.
  Lett.} {\bfseries 120} no.~3, (2018) 031103},
  \href{http://arxiv.org/abs/1711.10502}{{\ttfamily arXiv:1711.10502 [gr-qc]}}.

\bibitem{Mishra:2020gce}
A.~K. Mishra, ``{Quasinormal modes and Strong Cosmic Censorship in the novel 4D
  Einstein-Gauss-Bonnet gravity},''
  \href{http://arxiv.org/abs/2004.01243}{{\ttfamily arXiv:2004.01243 [gr-qc]}}.

\end{thebibliography}\endgroup

\bibliographystyle{./utphys1}
%%%%%%%%%%%%%%%%%%%%%%%%%%%%%%%%%%%%%%%%%%%%%%%%%%%%%%%%%%%%%%%%%%%%%%%%%%%%%%%%%%%%%%%%%%%%%%%%%%%
%%%%%%%%%%%%%%%%%%%%%%%%%%%%%%%%%%%%%%%%%%%%%%%%%%%%%%%%%%%%%%%%%%%%%%%%%%%%%%%%%%%%%%%%%%%%%%%%%%%
%%%%%%%%%%%%%%%%%%%%%%%%%%%%%%%%%%%%%%%%%%%%%%%%%%%%%%%%%%%%%%%%%%%%%%%%%%%%%%%%%%%%%%%%%%%%%%%%%%%
\end{document}